\documentclass[a4paper,11pt]{article}
\usepackage{a4wide}
\usepackage{amsmath,amssymb}
\usepackage{graphics,graphicx}
\usepackage{changebar}
\usepackage{color}
\usepackage{cancel}
\usepackage[all,cmtip]{xy}

\def\mC{\mathbb{C}}
\def\mP{\mathbb{P}}

\def\mZ{\mathbb{Z}}
\definecolor{color1}{RGB}{255 ,0 ,0 }
\definecolor{color2}{RGB}{0 ,255 ,0 }
\definecolor{color3}{RGB}{0 ,0 ,255 }
\definecolor{color4}{RGB}{255 ,215 ,0 }
\definecolor{color5}{RGB}{0 ,255 ,255 }
\definecolor{color6}{RGB}{255 ,0 ,255 }
\definecolor{color7}{RGB}{255 ,69 ,0 }
\definecolor{color8}{RGB}{20 ,205 ,50 }
\definecolor{color9}{RGB}{70 ,130 ,180 }
\definecolor{color10}{RGB}{186 ,85 ,211 }
\definecolor{color11}{RGB}{139 ,139 ,0 }
\definecolor{color12}{RGB}{210 ,105 ,30 }
\definecolor{color13}{RGB}{188 ,143 ,143 }
\definecolor{color14}{RGB}{139 ,0 ,0 }
\definecolor{color15}{RGB}{34 ,139 ,34 }

\newcommand{\diff}[2]{\textrm{d}^{#1}{#2}}
\newcommand{\e}[1]{\mathrm{e}^{#1}}
\newcommand{\Hyp}[3]{\,{}_3F_2\!\left(\genfrac{}{}{0pt}{0}{#1}{#2} \,; #3\right)}
\renewcommand{\Re}{\mathop{\mathrm{Re}}}
\renewcommand{\Im}{\mathop{\mathrm{Im}}}

\ifx\pdfoutput\undefined
\usepackage
[dvips,
 verbose=false,
 bookmarks=true,
 colorlinks=true,
 linkcolor=webred,
 filecolor=webbrown,
 citecolor=webgreen,
 pagecolor=webblue,
 urlcolor=webblue,
 pdftitle={Hemisphere partition function and analytic continuation},
 pdfauthor={Johanna Knapp, Mauricio Romo, Emanuel Scheidegger},
 pdfsubject={},
 pdfkeywords={},
 bookmarksopen=false,
 pdfpagemode=None,
 pdfview=FitH,
 pdfstartview=FitH,
 extension=pdf]{hyperref}
\else
\usepackage
[pdftex,
 verbose=false,
 bookmarks=true,
 colorlinks=true,
 linkcolor=webred,
 filecolor=webbrown,
 citecolor=webgreen,
 pagecolor=webblue,
 urlcolor=webblue,
 pdftitle={Hemisphere partition function and analytic continuation},
 pdfauthor={Johanna Knapp, Mauricio Romo, Emanuel Scheidegger},
 pdfsubject={},
 pdfkeywords={},
 bookmarksopen=false,
 pdfpagemode=None,
 pdfview=FitH,
 pdfstartview=FitH,
 extension=pdf]{hyperref}
\fi
\definecolor{webred}{rgb}{.8,0,0}
\definecolor{webbrown}{rgb}{.6,0,0}
\definecolor{webgreen}{rgb}{0,0.5,0}
\definecolor{webdkgreen}{rgb}{0,0.3,0}
\definecolor{webblue}{rgb}{0,0,0.5}

\numberwithin{equation}{section}

\begin{document}
\thispagestyle{empty}
\begin{flushright}
TUW-16-04
\end{flushright}
\vspace{1cm}
\begin{center}
{\LARGE\bf Hemisphere Partition Function and Analytic\medskip\\ Continuation to the Conifold Point}
\end{center}
\vspace{8mm}
\begin{center}
{\large Johanna Knapp\footnote{{\tt knapp@hep.itp.tuwien.ac.at}}${}^{\dagger}$, Mauricio Romo\footnote{{\tt mromoj@ias.edu }}${}^{\ast}$ and Emanuel Scheidegger\footnote{{\tt emanuel.scheidegger@math.uni-freiburg.de}}${}^{\ddagger}$}
\end{center}
\vspace{3mm}
\begin{center}
{\em ${}^{\dagger}$ Institute for Theoretical Physics, TU Wien\\ Wiedner Hauptstrasse 8-10, 1040 Vienna, Austria\\
${}^{\ast}$ School of Natural Sciences, Institute for Advanced Study\\ Princeton,
NJ 08540, USA\\
${}^{\ddagger}$ Mathematisches Institut, Albert-Ludwig-Universit\"at\\ Eckerstrasse 1, 79104 Freiburg, Germany
}
\end{center}
\vspace{15mm}
\begin{abstract}
\noindent We show that the hemisphere partition function for certain U(1) gauged linear sigma models (GLSMs) with D-branes is related to a particular set of Mellin-Barnes integrals which can be used for analytic continuation to the singular point in the K\"ahler moduli space of an $h^{1,1}=1$ Calabi-Yau (CY) projective hypersurface. We directly compute the analytic continuation of the full quantum corrected central charge of a basis of geometric D-branes from the large volume to the singular point. In the mirror language this amounts to compute the analytic continuation of a basis of periods on the mirror CY to the conifold point. However, all calculations are done in the GLSM and we do not have to refer to the mirror CY. We apply our methods explicitly to the cubic, quartic and quintic CY hypersurfaces.
\end{abstract}
\newpage
\setcounter{tocdepth}{1}
\tableofcontents
\setcounter{footnote}{0}

\section{Introduction}
Localization techniques in supersymmetric gauge theories in various dimensions pioneered in \cite{Pestun:2007rz} have brought on a plethora of exciting new results. In two dimensions this has led to new methods of computing quantum corrections in string compactifications without having to rely on mirror symmetry. Supersymmetric gauge theories in two dimensions are intimately tied to Calabi-Yau (CY) compactifications in string theory. A remarkable example is the class of $\mathcal{N}=(2,2)$ supersymmetric gauge theories known as gauged linear sigma models (GLSMs) \cite{Witten:1993yc}. 
The low-energy (IR) limit of a GLSM with suitably chosen gauge group and field content corresponds to a superconformal field theory (SCFT) describing a CY compactification of string theory. The space of FI-$\theta$ parameters $t$ of the GLSM can be identified with the space $\mathcal{M}_{K}$ of complexified K\"ahler parameters of the CY~\cite{Morrison:1994fr}. From the point of view of the IR SCFT it corresponds to the space of a particular class of exactly marginal deformations. The parameters $t$ do not get renormalized under the RG flow in the GLSM and therefore we get a family of SCFTs parametrized by $\mathcal{M}_{K}$. This space is naturally divided into regions, called \emph{phases} of the GLSM~\cite{Witten:1993yc}, for which the GLSM has different low energy effective descriptions. Of particular interest are geometric phases whose low-energy description is given in terms of a non-linear sigma model with CY target.

In \cite{Benini:2012ui,Doroud:2012xw}, the exact partition function of a GLSM on a two-sphere $S^{2}$ has been computed using techniques of supersymmetric localization. In \cite{Jockers:2012dk,Gomis:2012wy,Gerchkovitz:2014gta,Gomis:2015yaa} it has been argued that this computes the quantum corrected K\"ahler potential in $\mathcal{M}_{K}$. For GLSMs which have a geometric phase, this information can be used to extract genus $0$ Gromov-Witten invariants without the use of mirror symmetry. 
A natural extension is to consider GLSMs on manifolds with boundaries, such as the disk $D^{2}$. In~\cite{Sugishita:2013jca,Honda:2013uca,Hori:2013ika} localization methods were used to calculate the exact partition function of a GLSM on a hemisphere, i.e. a disk with the round metric. In CY GLSMs with a geometric phase this was conjectured to compute the exact central charge of B-branes in the IR limit including the quantum corrections due to worldsheet instantons. 

Viewing $\mathcal{M}_K$ as parametrizing families of SCFTs in the topological A--model, there is a natural holomorphic vector bundle $\mathcal{H}$ over $\mathcal{M}_K$ whose fiber over $t$ in $\mathcal{M}_K$ is the chiral ring of the corresponding SCFT determined by $t$. Moreover, the bundle $\mathcal{H}$ is equipped with a natural flat connection, the $tt^*$ connection~\cite{Cecotti:1991me}. The central charge of a D-brane is a flat section of this connection~\cite{Ooguri:1996ck}. As the hemisphere partition function of the GLSM is conjectured to compute the central charge in the low energy limit, it is natural to expect that the hemisphere partition function, too, is a flat section of some holomorphic vector bundle with a flat connection obtained from the GLSM. In this paper, we give evidence for such a structure. 

A particularly appealing aspect of the GLSM is that its correlation functions (without D-branes) are rational functions in the algebraic coordinates and therefore can be analytically continued to the entire parameter space $\mathcal{M}_K$. These correlation functions then interpolate between the correlation functions of the SCFTs in the various phases. For the hemisphere partition function we expect a similar picture to hold, although it is not a rational function. Instead, the hemisphere partition function admits a very natural description in terms of a contour integral~\cite{Hori:2013ika}, and is therefore -- when taking into account the grade restriction rule discussed below -- defined as a function over the entire $\mathcal{M}_K$. Since contour integrals are the standard tool to perform analytic continuation, it is a natural question whether they can be used to transport central charges between phases. 

The main goal of this article, then, is to demonstrate that the hemisphere partition function in the GLSM can be understood -- at least locally -- as a multivalued, flat, holomorphic section for an appropriate flat connection, and that it can serve as a tool to analytically continue information about B--type D-branes from one phase to another, or more interestingly, to a phase boundary. The way to achieve this is as follows. First, we mainly focus on GLSMs with gauge group $U(1)$. One of the reasons is that it is clear which integration contour to choose and how to evaluate the contour integral. We make use of the structure of the integrand to derive -- after a change of variable to the algebraic coordinates of the GLSM -- a differential equation for the hemisphere partition function. This derivation is purely within the GLSM, without making reference to any phase. The differential equation is of hypergeometric type, i.e. a linear homogeneous complex differential equation with three regular singularities. The hemisphere partition function for a special set of D-branes is then identified with the Mellin--Barnes integral representation of the hypergeometric functions. Of course, as expected by mirror symmetry, this differential equation is the same as the Picard--Fuchs equation for the periods of the mirror CY. This is also in agreement with the fact that, in the low energy limit, the central charges are mirror to linear combinations of periods of the mirror CY. See \cite{Bizet:2014uua} for a recent application in this context, for the case of CY 4-folds. Our derivation, however, does not involve the mirror at all.

Abelian GLSMs with gauge group $U(1)$ admit a geometric phase whose IR limit corresponds to a CY hypersurface in (weighted) $\mP^N$ and whose parameter space $\mathcal{M}_K$ admits a presentation as $\mP^1$ minus three points. These points are the large volume (or large radius) point $z_{LV}$, the Landau-Ginzburg (LG) or Gepner point $z_{LG}$ and the singular or -- in the language of the mirror -- the conifold point $z_{c}$, where the GLSM develops a non--compact Coulomb branch. They correspond to the regular singularities of the differential equation. Near each of these points there is a natural basis of solutions given in terms of power series whose radius of convergence extends to the closest singularity. Near $z_{LV}$ and $z_{LG}$ these power series can be thought of as the IR limit of the hemisphere partition function in the geometric and Landau--Ginzburg phases, respectively. It is therefore natural to consider the behavior of these bases of solutions, when going around any of these points, or when going from one point to another. This corresponds to parallel transport of central charges of D-branes along various paths in $\mathcal{M}_K$. 

By the work of~\cite{Herbst:2008jq}, D-branes in the GLSM have a remarkable description in terms of matrix factorizations of the GLSM superpotential. Matrix factorizations are well-studied in the context of topological Landau-Ginzburg models where B-branes are described in terms of matrix factorizations of the Landau-Ginzburg potential \cite{Kapustin:2002bi,Brunner:2003dc}. See \cite{Hori:2004zd,Jockers:2007ng,Knapp:2007vc} for reviews. In~\cite{Herbst:2008jq} it was shown that matrix factorizations in the GLSM describe D-branes in the UV not only in the Landau--Ginzburg phase but all over $\mathcal{M}_K$ irrespective of the low-energy description of D-branes in the phases. However, not every D-brane in the GLSM is globally defined on $\mathcal{M}_{K}$: only certain subsets, called grade restricted branes, are.
Only these grade restricted branes can be transported along paths in $\mathcal{M}_K$ in a well-defined way. Since both the branes and the hemisphere partition function can be globally defined, we can in particular study their behavior under monodromy around the singular points. This can be done without making reference to any phase. 

What is a non-trivial task is the analytic continuation of the hemisphere partition function  from one regular singular point to another. The analytic continuation of hypergeometric functions and of D-branes in the GLSM from the large radius to the Landau-Ginzburg point is well-understood~\cite{Norlund:1955ab,Herbst:2009ab}, and we will not address this here. Analytic continuation to the conifold point, however, turns out to be a challenging problem.  
As we will show, the solution is formulated in terms of Mellin--Barnes integrals which in turn have a natural interpretation as hemisphere partition functions. 

The main results presented in this article are:
\begin{itemize}
\item By the use of the hemisphere partition function of the GLSM, we give an interpretation of the analytic continuation in terms of central charges of grade restricted B-branes. From this point of view the computation does not require knowledge of a mirror pair. Specializing to the case of degree $N$ hypersurface in $\mathbb{P}^{N-1}$, we show that the hemisphere partition function for grade restricted B-branes satisfies the hypergeometric differential equation associated to the mirror CY. In this way we can access information about the behavior of B-branes near the singular point and we can directly compute the central charge of a B-brane plus its quantum corrections close to the singularity. Furthermore we use the hemisphere partition function to recalculate the monodromy matrices for the quintic \cite{Candelas:1990rm}.

\item We present mathematical methods to analytically continue a basis of solutions of a particular class of hypergeometric equations which have three regular singular points at $\{0,1,\infty\}$ from the point $0$ to the point $1$. A priori, this is a classical problem on the mirror CY for Mellin-Barnes integral representations of a basis of periods given by the solutions to a Picard-Fuchs equation. This is the point of view taken in a separate note by the third author~\cite{Scheidegger:2016ab}, where this analytic continuation is proven in a more general setting. In the present work, we also propose a second method of analytic continuation and show in examples that its results are in agreement with \cite{Scheidegger:2016ab}. This can lead to non-trivial identities of hypergeometric functions and their derivatives.

\end{itemize}

While the hemisphere partition function is a natural object in physics, in mathematics we need a different approach. From a mathematical point of view one is faced with the issue that neither $\mathcal{M}_K$ nor the bundle $\mathcal{H}$ admit a purely mathematical definition yet. Furthermore, a precise mathematical description for the category of B-branes at every point in $\mathcal{M}_{K}$ is lacking. The equivalence of the category of B-branes at every point in $\mathcal{M}_{K}$ is only conjectural, and moreover, very different descriptions of this category can occur, depending on the IR SCFT at that point. Nevertheless, the result of~\cite{Hori:2013ika} yields an equivalent description of the hemisphere partition function which can be taken as a definition, and hence as a starting point, in mathematics. For abelian GLSMs we make this definition precise. In the IR limit of a geometric phase, we reproduce the expected properties of the central charge. In a geometric phase (i.e. around a large volume point) a mathematical description of the central charge was formulated in \cite{Iritani:2009ab}, based on the ideas of \cite{Hosono:2004jp}.  The problem of relating the central charges along a path joining $z_{LV}$ and $z_{LG}$ has been studied in~\cite{Chiodo:2014ab}. However, from our point of view, these authors worked in the low energy limits of the two phases. The natural setting is the GLSM, in which the hemisphere partition function has the advantage of being defined as a function on the whole space $\mathcal{M}_{K}$ in contrast to, say, the central charge defined in \cite{Iritani:2009ab} which is defined only inside the complexified K\"ahler cone but can be extended outside by analytic continuation.

The article is organized as follows. In section \ref{sec-glsm} we give a lightning review of the GLSM and the hemisphere partition function. We also detail on how convergence of the hemisphere partition function implies a grade restriction rule for abelian one-parameter GLSMs \cite{Herbst:2008jq,Hori:2013ika}. We further show that the hemisphere partition function for a particular grade-restricted set of B-branes satisfies the hypergeometric differential equation. In section \ref{sec-branes} we review the main results of \cite{Herbst:2008jq}, with particular focus on the grade restriction rule. This states that only B-branes whose gauge charges are in a specific subset called ``window'' are globally defined over $\mathcal{M}_K$. We give an algorithmic approach to grade restricting GLSM branes and use it on the GLSM lift of a standard basis of large radius branes. The resulting hemisphere partition functions turn out to be linear combinations of the Mellin-Barnes integrals we know how to analytically continue to the conifold point. In passing, we also recompute the monodromy matrices for the quintic, which becomes an almost trivial calculation using the hemisphere partition function. Section \ref{sec-math} is dedicated to the analytic continuation of a basis of Mellin-Barnes integrals to the conifold point. We propose two methods for analytic continuation. One is based on a generalization of the results due to N{\o}rlund and B\"uhring~\cite{Norlund:1955ab,Buehring:1992ab,Scheidegger:2016ab} which yields the complete result in all cases. The other is based on the application of a specific integral identity involving Gamma functions. We show in examples that the second approach yields the same result, however in quite different-looking expressions. In this way we discover highly non-trivial identities between generalized hypergeometric functions. We apply these methods to the cubic curve, the quartic K3 and the quintic CY threefold. Via the connection to the hemisphere partition function we can use these results to compute the central charge of B-branes near the conifold point. In particular, we reconfirm that the D6-brane on the quintic, corresponding to the structure sheaf, becomes massless at the conifold point. We end our discussion with an outlook on open issues. \\\\

{\bf Acknowledgments:} We would like to thank C. Krattenthaler for providing us with proofs to some of the identities discussed in section \ref{sec-hypersurface}. MR thanks Daniel Pomerleano for discussions. MR acknowledges support from the Institute for Advanced Study and from DOE grant DE-SC0009988.
\section{Gauged Linear Sigma Model and Hemisphere Partition Function}
\label{sec-glsm}
The $\mathcal{N}=(2,2)$ Gauged Linear Sigma Model (GLSM) provides a device to explore the full quantum K\"ahler moduli space of Calabi-Yau (CY) manifold \cite{Witten:1993yc}. Recently, by using the machinery of rigid supersymmetry and localization, the exact, non-perturbative partition function of a GLSM on the disk $D^{2}$ with the round metric (we will also refer to it has hemisphere) has been computed \cite{Sugishita:2013jca,Honda:2013uca,Hori:2013ika}.

In order to define this partition function we need to recall some concepts. (For an even more mathematical description see~\cite{Fan:2015ab}.) We start by defining a GLSM datum. A GLSM datum is a quadruple $(G, W, \rho_{V}, R)$, where $G$ is a compact real Lie group (the gauge group). $\rho_{V}$ is a faithful complex representation of $G$: $\rho_{V}:G\rightarrow GL(V)$ and we set
$m=\mathrm{dim}(V)$. 
$V$ is called the space of chiral fields. $W$ is a holomorphic, $G$-invariant function $W\in \mathrm{Sym}(V^{*})$ (the superpotential). $R$ denotes a representation (the R-symmetry) $R: U(1) \to GL(V)$.
We require that $R$ and $\rho_V$ commute and that $W$ has weight $2$ under the R-symmetry. Moreover we will allow for the weights of the representation $R$ to be rational. However we impose the condition of charge integrality\footnote{The charge integrality condition comes from imposing that all the R-charges of gauge invariant operators reduce (modulo $2$) to the statistics of such an operator, i.e. $(-1)^0$ for bosons and $(-1)^1$ for fermions. This is a condition for the physical theory to be \emph{A-twistable} \cite{Lerche:1989uy}.}:
\begin{equation}
  \label{eq:charge_integrality}
  R(e^{i\pi})=\rho_{V}(J)\text{ \ for some \ } J\in G.
\end{equation}
We denote by $\mathfrak{g}=\mathrm{Lie}(G)$ and $\mathfrak{t}=\mathrm{Lie}(T)$ the Lie algebras of $G$ and of a maximal torus $T$ of $G$, respectively. For future reference we will also use $\{Q_{i}\}_{i=1}^{m}$ to denote the weights of $\rho_{V}$. If we choose a basis $\{\mathfrak{t}^{a}\}_{a=1}^{\mathrm{rk}G}$ for $\mathfrak{t}$ and a basis $\{v_{i}\}_{i=1}^{m}$ of $V$ then $d\rho_{V}(\mathfrak{t}^{a})v_{i}=Q_{i}^{a}v_{i}$. By $\{R_{i}\}_{i=1}^{m}$ we denote the weights of $R$. We will refer to these weights as the gauge charges and the R-charges of the fields $\phi \in V$, respectively. 
The Lie group $G$ can be written as $G=G_{0}\rtimes \pi_{0}(G)$ where $G_{0}$ is the identity component and $\pi_{0}(G)\cong G/G_{0}$ is the group of components of $G$. Then, we define the parameter $t$ as in \cite{Hori:2011pd,Hori:2013gga}:
\begin{align}\label{eqtheta}
e^{t}\in\mathrm{ Hom}(\pi_{1}(G),\mathbb{C}^{*})^{\pi_{0}(G)}.
\end{align}
Since there is an natural adjoint action of $\pi_{0}(G)$ in $G_{0}$, it makes sense to restrict to the $\pi_{0}(G)$ invariant subset in (\ref{eqtheta}). The Lie algebra $\mathfrak{g}$ of a compact Lie group decomposes as $\mathfrak{g} = \mathfrak{s} + \mathfrak{a}$ where $\mathfrak{s}$ is semisimple and $\mathfrak{a}$ is abelian. In particular, $\mathfrak{a} \subset \mathfrak{t}$. We set $s = \dim \mathfrak{a}$.
Given $(G, W,\rho_{V}, R)$ as before, we can define a moment map $\mu:V\rightarrow \mathfrak{g}^{*}$ associated to $\rho_{V}$. We set $t\in \mathfrak{g}^{*}_{\mC}$ such that it factors through the embedding $\mathfrak{a}^{*}_{\mC} \hookrightarrow \mathfrak{g}^{*}_{\mC}$. We choose a basis $t=(t_1,\dots,t_s)$ of $\mathfrak{a}^{*}_{\mC} \cong \mC^s$. Then:
\begin{align}
e^{t}\in\mathrm{ Hom}(\pi_{1}(G),\mathbb{C}^{*})
\end{align}
if we restrict $t_j$ to the cylinder $\mathbb{R}+\mathbb{R}/(2\pi i)\mathbb{Z}$ and $t_j = \zeta_j - i \theta_j$. However the requirement of invariance under $\pi_{0}(G)$ may further restrict $t$ and some of the $\zeta_j$ will be forced to be set to $0$ and some $\theta_{j}$ will be forced to take discrete values making the space of parameters $t$ smaller. In the main examples of this work this does not happen so, we will assume from now on that $\pi_0(G) = \{1\}$ and $\pi_1(G)$ is torsion-free. The parameters $\zeta_j$ and $\theta_j$ are called the Fayet-Illiopoulos (FI) parameters and the $\theta$--angles, respectively.

With all these ingredients at hand we can define the D-term equations given $\phi\in V$:
\begin{align}
\mu_{\phi}(\xi)&=\zeta(\xi)\qquad \text{ \ for all \ } \xi\in \mathfrak{g}^{*}.
\end{align}
This divides the parameter space into chambers and the corresponding low-energy configurations are referred to as phases of the GLSM \cite{Witten:1993yc}. Each phase is characterized by an ideal $I_\zeta \subset \mathrm{Sym}(V^*) \cong \mC[\phi_1,\dots,\phi_m]$, called the irrelevant ideal of the phase.
The F-term equations are
\begin{align}
dW^{-1}(0).
\end{align}
We define the space of classical vacua by the solutions of the F-term equations inside the symplectic quotient of $V$ determined by $\mu$ and $\zeta \in \mathfrak{a}^*$:
\begin{align}
X_{\zeta}&=\{dW^{-1}(0)\}\cap \mu_{\phi}^{-1}(\zeta)/G.
\end{align}
The ideal $I_\zeta$ describes the set of $\phi$ for which the quotient $\mu_{\phi}^{-1}(\zeta)/G$ is ill--defined. The low-energy (IR) behavior of the GLSM is determined largely by $X_{\zeta}$. Finding a model for the IR effective theory is not an easy task in general since the classical vacuum configurations determined by $X_{\zeta}$ can receive quantum corrections, which requires a more in-depth analysis. However we can distinguish a special case which is a weakly coupled geometric phase. This is the case when $X_{\zeta}$ is a smooth projective variety, for given values of $\zeta$. In that region, the IR theory can be well approximated by a non-linear sigma model (NLSM) whose target space is $X_{\zeta}$. Other types of phases are Landau-Ginzburg and hybrid phases. We will identify the FI-$\theta$ parameters $t$ with local coordinates on a space $\mathcal{M}_{K}$. The space $\mathcal{M}_{K}$ is also called \emph{stringy K\"ahler moduli space} and we will describe it in more detail in the next section. 
In the following we will focus on GLSMs which have a weakly coupled geometric phase in which $X_{\zeta}$ is CY. This imposes a further constraint in the GLSM datum, the so-called CY condition, or physically, the cancellation of the axial anomaly. This translates into the condition that $\rho_{V}$ factors through $SL(V)$:
\begin{equation}
\text{ \ CY condition: \ } \rho_{V}:\;G\rightarrow SL(V).
\end{equation}
A quick way to motivate the reason of this definition from a geometric point of view is as follows. If we identify $V$ with $\mC^m$ by choosing a basis $\phi_1,\dots,\phi_m$, then there is a unique natural holomorphic $m$--form on $V$ given by $\diff{}{\phi_1}\wedge\dots\wedge \diff{}{\phi_m}$. The condition that this $m$--form descends to the symplectic quotient $\mu_{\phi}^{-1}(\zeta)/G$ and hence defines a Calabi--Yau structure on $\mu_{\phi}^{-1}(\zeta)/G$ is that the determinant of $\rho_V(g)$ is trivial for all $g \in G$. This Calabi--Yau structure then induces a Calabi--Yau structure on $X_\zeta$.

We will mostly study the concrete case of GLSMs associated with CY hypersurfaces in $\mathbb{P}^{N-1}$. In these cases the GLSM datum is given by~\cite{Witten:1993yc}
\begin{align}
\label{u1glsm}
(U(1),W=pG(x),\rho_{V}:U(1)\rightarrow SL(N+1),R)
\end{align}
where $(p,x_{1},\ldots,x_{N})\in \mathbb{C}^{N+1}$ and the weights of $\rho_{V}$ are $(-N,1,\ldots,1)$. $G(x)$ (sometimes denoted by $G_N(x)$) is a degree $N$ polynomial in $x_1,\dots,x_N$. A set of R-charges consistent with~\eqref{eq:charge_integrality} is given by
\begin{align}
  \label{rcharge}
R(p)=2-2N\varepsilon\qquad R(x_{i})=2\varepsilon\qquad 0 \leq \varepsilon \leq \frac{1}{N}.
\end{align}
The R-charges, as presented in the GLSM datum, take rational values. However, as they appear in the hemisphere partition function $Z_{D^{2}}(\mathcal{B})$, we can safely extend them to be real valued. Moreover, the R-charges of gauge-variant fields cannot be fixed a priori, since they are not physical observables. Therefore assigning real values to them is a sensible thing to do. However, the exact R-charges of gauge invariant operators in the IR make sense physically and are fixed by the low energy theory.

In this case, there are two phases $\zeta \gg 0$ and $\zeta \ll 0$.  The first phase is geometric since $X_{\zeta \gg 0} = \{ G_N = 0 \} \subset \mP^{N-1}$ is a Calabi--Yau hypersurface of degree $N$ in $\mP^{N-1}$. The corresponding irrelevant ideal is $I_{\zeta \gg 0} = (x_1,\dots,x_N)$ which is the irrelevant ideal of the homogeneous coordinate ring of $\mP^{N-1}$. The second phase is non-geometric. It is the Landau--Ginzburg orbifold $X_{\zeta \ll 0} = [\mC^N / \mZ_5]$ with superpotential $G : [\mC^N / \mZ_5] \to \mC$. The corresponding irrelevant ideal is $I_{\zeta \ll 0} = (p)$.

\subsection{Discriminant Locus}
The parameters $t$ defined in the previous section have been identified with coordinates in the stringy K\"ahler moduli space $\mathcal{M}_{K}$. By exponentiating the coordinates $t$ we get coordinates on the algebraic torus $(\mathbb{C}^{*})^s$. 
The manifold $\mathcal{M}_{K}$ is expected to take the form of a partial compactification of $(\mathbb{C}^{*})^s$ where some complex codimension $1$ closed subset $\Delta$ is removed. In the case of $G$ abelian, this compactification comes from toric geometry considerations~\cite{Morrison:1994fr,Cox:1999ab}. For $G$ more general, there is no unified construction of $\mathcal{M}_{K}$.  In general, if the CY condition is satisfied and a mirror CY is known, $\mathcal{M}_{K}$ can be indirectly described by mirror symmetry. In the CY case, the parameters $t$ do not run under RG flow and there is a family of SCFTs in the IR limit for each point in $\mathcal{M}_{K}$. 
The set $\Delta$ is known as the discriminant. The points in $\mathcal{M}_{K}\setminus (\mathbb{C}^{*})^{s}$ correspond to limiting points. Examples of these are Gepner or Landau-Ginzburg (LG) points and large volume (LV) points. On the other hand, at points in the discriminant $\Delta$ there exist non-compact Coulomb branches that render the theory ill defined \cite{Witten:1993yc}. The Coulomb branch will be defined as a subset of $\mathfrak{t}_{\mC}\subset\mathfrak{g}_{\mC}$. It will be helpful in the following to choose a coordinate on the Coulomb branch by denoting
\begin{equation}
\sigma\in \mathfrak{t}_{\mC}\subset\mathfrak{g}_{\mC}.
\end{equation}
For our purposes it will suffice to recall that $\Delta$ can be computed exactly for $G=U(1)$ by a 1-loop computation in the GLSM \cite{Morrison:1994fr,Hori:2003ic}. Being slightly more general, we start by defining the effective twisted superpotential for $G=U(1)^{s}$. Given a GLSM datum $(U(1)^{s}, W,  \rho_{V}, R)$ satisfying the CY condition plus the parameters $t$ we define the effective twisted potential $\widetilde{W}_{\mathrm{eff}}:\mathfrak{u}(1)^{s}\rightarrow \mathbb{C}/(2\pi i \mathbb{Z})$ as
\begin{equation}
\widetilde{W}_{\mathrm{eff}}(\sigma)=-t(\sigma)-\sum_{i=1}^{m}Q_{i}(\sigma)\log(Q_{i}(\sigma))\text{ \ \ }(\mathrm{mod}\: 2\pi i).
\end{equation}
Recall that the $Q_{i}\in\mathfrak{t}^{*}_{\mC} $ are the weights of $\rho_{V}$. In this case, the Coulomb branch corresponds to the points $\sigma$ satisfying $\partial_{\sigma}\widetilde{W}_{\mathrm{eff}}(\sigma)=0$. For a GLSM datum $(U(1), W, \rho_{V}, R)$ satisfying the CY condition the equation $\partial_{\sigma}\widetilde{W}_{\mathrm{eff}}(\sigma)=0$ gives the full discriminant. Moreover the equation $\partial_{\sigma}\widetilde{W}_{\mathrm{eff}}(\sigma)=0$ does not depend on $\sigma$ and we can define $\Delta$ as the values of $t$ satisfying this condition. Then the Coulomb branch has a very simple description:
\begin{equation}
\{\sigma\}=\begin{cases}
    \mathfrak{t}_{\mC}& t\in \partial_{\sigma}\widetilde{W}_{\mathrm{eff}}(\sigma)=0 \\
     \{0\} &\mathrm{ otherwise}.
  \end{cases}
\end{equation}
For the case of a GLSM associated to a CY hypersurface in $\mathbb{P}^{N-1}$, $\Delta$ is a single point:
\begin{equation}
\label{eq:conifold}
\widetilde{W}_{\mathrm{eff}}(\sigma)=-t\sigma+N\sigma\log(-N)\Rightarrow \Delta=\{(\zeta,\theta)=(N\log(N),\pi N)\}.
\end{equation}
At this singular point the GLSM becomes ill-defined. On the mirror CY this corresponds to the conifold point in the complex structure moduli space. Slightly abusing the language we will also refer to the singular point in the GLSM as the conifold point.

For more general $G$ the effective potential on the Coulomb branch only gives part of the discriminant. The remaining components come from mixed Coulomb-Higgs branches \cite{Morrison:1994fr}. Some families of examples for the non-abelian case, where $\Delta$ is studied, can be found in \cite{Hori:2006dk,Jockers:2012zr,Hori:2013gga}.
\subsection{D-branes in the GLSM}
\label{sec:d-branes-glsm}

The GLSM data $(G, W, \rho_{V}, R)$ together with the parameters $t$ can be used to define a GLSM as an actual physical theory. 
To construct this theory we need to specify a Riemann surface $\Sigma$ and a principal $G$-bundle $P_{G}$ over it. Two types of GLSM fields are relevant for our discussion:
\begin{equation}
\begin{aligned}
\text{ \ Chiral: \ }\phi &: \Sigma \rightarrow V\\
\text{ \ Twisted chiral: \ }\sigma &\in\Gamma(P_{G}\times_{Adj}\mathfrak{g}_{\mC})
\end{aligned}
\end{equation}
At a point $p\in\Sigma$, $\sigma(p)\in\mathfrak{g}_{\mC}$. (Note that on the Coulomb branch $\sigma(p)$ is restricted to $\mathfrak{t}_{\mC}$.) For the case of $\Sigma=\mathbb{R}^{2}$, a GLSM can be defined just by the datum $(G, W, \rho_{V}, R)$ and a choice of $t$. The GLSM as an $\mathcal{N}=(2,2)$ supersymmetric gauge theory can be defined in a curved space with enough isometries \cite{Festuccia:2011ws,Closset:2014pda}, in particular the sphere $S^2$ and the hemisphere $D^2$. The action of the GLSM has to be modified and a subgroup of the superconformal symmetry group can be matched with the isometries of the curved space \cite{Pestun:2007rz,Benini:2012ui,Doroud:2012xw}. If $\partial \Sigma\neq \emptyset$, such as in the case $\Sigma=D^{2}$, further information has to be added in order to specify the boundary conditions.  
As in the flat space case, adding a boundary to the GLSM on $S^2$ breaks supersymmetry. In order to preserve some of the original supersymmetry we must add a boundary Lagrangian to the Lagrangian of the GLSM on $S^2$.\\

For this, we need to introduce the boundary datum $\mathcal{B}$. To this aim, we write $\mathrm{Sym}(V^*) = \mC[\phi_1,\dots,\phi_m]$ with $m=\dim V$ and denote this polynomial ring by $S$. Given a GLSM datum $(G, W,\rho_{V}, R)$, a boundary datum is a quadruple $\mathcal{B}=(M,Q,\rho,\mathbf{r}_{*})$ where $M = M^0 \oplus M^1$ is a $\mZ_{2}$--graded free $S$--module and $Q$ is a matrix factorization of $W$, which is a $\mZ_2$ odd map of $\mZ_2$ graded $S$--modules, $Q \in \mathrm{End}_S^1(M)$, that satisfies
\begin{equation}
  \label{eq:MF}
  Q^{2}= W\mathrm{id}_{M}.
\end{equation}
Furthermore, $(M,Q)$ is required to be equivariant with respect to the actions of $\rho_m$ and $R$ on $V$. This is imposed by choosing commuting representations $\rho : G\rightarrow GL(M)$ and $\mathbf{r}_{*} :\mathfrak{u}(1)_{R}\rightarrow \mathfrak{gl}(M)$ on $M$ such that $Q$ has weight $1$ under $\mathbf{r}_*$. Explicitly, their action on $(M,Q)$ is given by
\begin{align}
\label{rrhodef}
\rho(g)^{-1}Q(g\phi)\rho(g)&=Q(\phi)\\
\lambda^{\mathbf{r}_{*}}Q(\lambda^{R}\phi)\lambda^{-\mathbf{r}_{*}}&=\lambda Q(\phi)
\label{gdef}
\end{align}
for all $\lambda\in U(1)_{R}$ and $g\in G$. Moreover, these representations are required to be compatible with the $\mZ_2$ grading on $M$ as follows.
\begin{equation}
e^{\pi i\mathbf{r}_{*}} \rho(J) =\begin{cases}
+1 & \text{ \ on \ } M^{0}\\
-1 & \text{ \ on \ } M^{1}
\end{cases}
\end{equation}
for $J \in G$. We will refer to $\mathcal{B}$ as a B-brane or GLSM brane.

In order to preserve 'B-type' supersymmetry, (the one preserved by B-branes) the boundary action takes the form of a Wilson loop term
\begin{equation}
   \quad\mathrm{tr}_{M}\left[P \exp\left(-\oint_{\partial D^{2}}\phi^{*}\mathcal{A}_{\mathcal{B}} + d\rho (iv + \Re\sigma) \right)\right],
\end{equation}
where $\mathcal{A}_{\mathcal{B}}$ is a connection on $M$ constructed from the boundary data $\mathcal{B}$ and $v$ is the connection on the principal bundle $P_G$. This Wilson line term was originally constructed for abelian GLSMs in~\cite{Herbst:2008jq} and generalized to nonabelian GLSMs in~\cite{Honda:2013uca,Hori:2013ika}. The matrix factorization condition~\eqref{eq:MF} then guarantees that the combined bulk-boundary action is invariant under B-type supersymmetry.

B-branes form a category whose morphisms are determined by the cohomology of $Q$. Given two B-branes $\mathcal{B}_1$ and $\mathcal{B}_2$ the $\mathbb{Z}_2$-graded space of open string states $\Psi$ is \cite{Herbst:2008jq}
\begin{equation}
\label{qcohom}
\mathcal{H}^p(\mathcal{B}_1,\mathcal{B}_2) = H^p_{D}(\mathrm{Hom}_S(M_1,M_2))\qquad p=0,1,
\end{equation}
where
\begin{equation}
  D\Psi=Q_2\Psi-(-1)^{|\Psi|}\Psi Q_1\qquad |\Psi|=0,1.
\end{equation}
The morphisms $\Psi$ are required to be equivariant with respect to the action of $\rho$ and $\bf{r}_*$ as induced by~\eqref{rrhodef} and~\eqref{gdef}. We will call this category $\mathcal{D}_{(G, W,\rho_{V}, R)}$. 

The information encoded in the matrix factorization can be recast into complexes of representations of $G$ and $\mathfrak{u}(1)$, called Wilson line branes~\cite{Herbst:2008jq}. To illustrate this, we first review a natural and general set of matrix factorizations/complexes: the Koszul matrix factorizations~\cite{Buchweitz:1987ab}. By the definition of a GLSM datum, we have $W \in S$. Suppose $W \not = 0$ is given in the form $W = \sum_{\alpha=1}^j a_\alpha b_\alpha$ for polynomials $a_\alpha,b_\alpha \in S$, $\alpha=1,\dots,j$. We collect these polynomials into sequences $a=(a_1,\dots,a_j)$ and $b=(b_1,\dots,b_j)$. Now consider the free $S$-module $E=S^{\oplus j}$ with basis $e_1,\dots,e_j$ and equipped with a inner product.  Given this data, we define $K(a,E) = \left(\bigwedge^\bullet E, \delta\right)$ with twisted differential~\footnote{If $W$ and $a$ are specified as above, then $b$ can be reconstructed. Hence, having fixed $W$ in the GLSM datum, it is sufficient to keep track of $a$ (and $W$).}
\begin{equation}
  \label{eq:delta}
  \delta(w) = \left( \sum_{\alpha=1}^j a_\alpha e_\alpha \right) \wedge w + \left( \sum_{\alpha=1}^j b_\alpha e_\alpha \right) \lrcorner w, \qquad w \in  \sideset{}{^\alpha}\bigwedge E,
\end{equation}
where $\lrcorner$ is the contraction operator. Thinking of the second summand as maps going backwards, we obtain two Koszul complexes, one going from left to right with differential $a\wedge$ and another one going from right to left with differential $b\lrcorner$,
\begin{equation}
  \label{eq:Koszulleftright}
  0 \longrightarrow S\overset{a\wedge}{\underset{b \lrcorner}{\rightleftarrows}} E \overset{a\wedge}{\underset{b \lrcorner}{\rightleftarrows}}  \sideset{}{^2}{\bigwedge} E \overset{a\wedge}{\underset{b \lrcorner}{\rightleftarrows}}  \dots \overset{a\wedge}{\underset{b \lrcorner}{\rightleftarrows}} \sideset{}{^{j-1}}{\bigwedge} E \cong E \overset{a\wedge}{\underset{b \lrcorner}{\rightleftarrows}}  \sideset{}{^j}\bigwedge E \cong S \longleftarrow 0
\end{equation}
where $a\wedge$ and $b\lrcorner$ represent the first and second summand of $\delta$, respectively. We will nevertheless loosely speak of {\em the} Koszul complex when referring to~\eqref{eq:Koszulleftright}. The $S$--module $M$ underlying the Koszul complex is the exterior algebra $\bigwedge^\bullet\! E$. A convenient way to encode the maps $\delta$ as a map of modules is to recall that $\bigwedge^\bullet\! E$ carries the structure of a Clifford algebra. We use a $2^j$-dimensional representation of this Clifford algebra with basis $\eta_\alpha,\bar{\eta}_\beta $, $\alpha,\beta=1,\ldots,j$ satisfying
\begin{align}
\{\eta_\alpha,\bar{\eta}_\beta\}=\delta_{\alpha\beta}\qquad \{\eta_\alpha,\eta_\beta\}=\{\bar{\eta}_\alpha,\bar{\eta}_\beta\}=0.
\end{align}
Choosing a nonzero element  $|0\rangle \in S$ (called the Clifford vacuum) we can build the $S$ module $M$ as $M=\bigoplus_{k=0}^{j} c_{\beta_1\dots\beta_k} \bar{\eta}_{\beta_1}\cdot\ldots\cdot\bar{\eta}_{\beta_k}|0\rangle$, with $c_{\beta_1\dots\beta_k} \in S$. Then
\begin{equation}
\label{koszulmf}
Q=\sum_{\alpha=1}^j a_\alpha\eta_\alpha+b_\alpha\bar{\eta}_\alpha
\end{equation}
is a $2^j\times 2^j$ matrix factorization of $W=\sum_{\alpha=1}^j a_\alpha\cdot b_\alpha$. Note that it can happen that some of the left-arrows (i.e. some of the $b_{\alpha}$) are zero. We will discuss examples of this kind in section \ref{sec-branes}. The Koszul complex $K(a,E)$ is exact with respect to $a\wedge$ except at the $j$th position. In fact, $\mathrm{H}^j( K(a,E) ) = S / (a_1,\dots,a_j)$, and the Koszul complex can be viewed as a resolution of $S / (a_1,\dots,a_j)$.

Up to now, we have only described $M$ and $Q$ and not yet taken into account $\rho$ and $\mathbf{r}_*$. Let $F$ therefore be a free $S$--module and consider representations $\rho : G\rightarrow GL(F)$ and $\mathbf{r}_{*} :\mathfrak{u}(1)_{R}\rightarrow \mathfrak{gl}(F)$. Then define the Koszul matrix factorization $K(a,F)$ as $F\otimes_S K(a,E)$, i.e. we replace $E$ by $F \otimes_S E$ in~\eqref{eq:Koszulleftright}. By repeating the construction above the module $M$ is then $F \otimes_S \bigwedge^\bullet E$. We will use both descriptions, complex or module, interchangeably, depending on what is more convenient. This encodes the full information about the GLSM brane, as opposed to just the matrix factorization~\eqref{eq:Koszulleftright}, and we will refer to it as Koszul brane. 

The Koszul branes do not account for all GLSM branes. More general boundary data $\mathcal{B}$ can be obtained from these by linear algebra operations such as direct sum, tensor, dual, wedge, kernel, cokernel, and homological operations such as shifts and mapping cones.

Finally, we consider our main example (\ref{u1glsm}). 
In this case, $S=\mC[x_1,\dots,x_N,p]$. All representations $\rho$ and $\bf{r}_*$ decompose into direct sums of one-dimensional representations $\mathcal{W}(q_i)_{r_i}$ where $q_i$ and $r_i$ are the weights of (\ref{rrhodef}) and (\ref{gdef}), respectively. Therefore we can choose $F=\bigoplus_{j}\mathcal{W}(q_j)_{r_j}$. The one--dimensional building blocks $\mathcal{W}(q)_r$ are called Wilson line branes~\cite{Herbst:2008jq}. The conditions (\ref{rrhodef}), (\ref{gdef}) determine the gauge and R-charges of the Clifford basis:
\begin{equation}
(q_{\eta_\alpha},r_{\eta_\alpha})=(1,1)\qquad (q_{\bar{\eta}_\alpha},r_{\bar{\eta}_\alpha})=(-1,-1).
\end{equation}
For $x=(x_1,\dots,x_j)$ the resulting Koszul brane $K(x,\mathcal{W}(q)_r)$ then reads
\begin{equation}
\label{koszulbrane}
\xymatrix{\mathcal{W}(q)_{r}\ar@<2pt>[r] & \ar@<2pt>[l]\mathcal{W}(q+1)_{r+1}^{\oplus \binom{j}{1}}\ar@<2pt>[r] & \ar@<2pt>[l]\mathcal{W}(q+2)_{r+2}^{\oplus \binom{j}{2}}\ar@<2pt>[r] & \ar@<2pt>[l]\ldots \ar@<2pt>[r] & \ar@<2pt>[l]\mathcal{W}(q+j)_{r+j}}
\end{equation}
where we have suppressed the maps. The first entry $\mathcal{W}(q)_r$ determines the choice of overall normalization in the definitions (\ref{rrhodef}) and (\ref{gdef}) so that the charges and multiplicities of the Wilson line branes $\mathcal{W}(q_i)_{r_i}$ in the $i$th position of the complex are fully specified by the three integers $q,r,j$. For later use in section~\ref{sec:mellin-barnes} we introduce the abbreviation $K_j = K((x_1,\dots,x_j),\mathcal{W}(0)_0)$, $j=1,\dots,N$, as well as $M_j$ for the corresponding $S$ module.

There are two special Koszul branes associated to the irrelevant ideals $I_{\zeta \gg 0} = (x_1,\dots,x_N)$ and $I_{\zeta \ll 0} = (p)$. Their significance will be discussed in more detail in section~\ref{sec-branes}. The first one is $K_{\zeta \gg 0} = K( I_{\zeta \gg 0} , \mathcal{W}(q)_{r})$ given by
\begin{equation}
\label{lr-empty}
\xymatrix{\mathcal{W}(q)_{r}\ar@<2pt>[r] & \ar@<2pt>[l]\mathcal{W}(q+1)_{r+1}^{\oplus \binom{N}{1}}\ar@<2pt>[r] & \ar@<2pt>[l]\mathcal{W}(q+2)_{r+2}^{\oplus \binom{N}{2}}\ar@<2pt>[r] & \ar@<2pt>[l]\ldots \ar@<2pt>[r] & \ar@<2pt>[l]\mathcal{W}(q+N)_{r+N}}.
\end{equation}
The corresponding matrix factorization has a special form:
\begin{equation}
\label{canonicalmf}
Q=\sum_{i=1}^N x_i{\eta}_i+\frac{1}{N}\frac{\partial W}{\partial x_i}\bar{\eta}_i.
\end{equation}
While Koszul matrix factorizations in general may only exist for particular choices of $W$ (e.g. at the Fermat point), (\ref{canonicalmf}) exists for any generic choice $W$. The second special Koszul brane is $K_{\zeta \ll 0} = K( I_{\zeta \ll 0} , \mathcal{W}(q)_r)$. For later purposes we write the dual matrix factorization $K_{\zeta \ll 0}^*$ as
\begin{equation}
  \label{emptylg}
  \mathcal{W}(q)_r \overset{G\wedge}{\underset{p \lrcorner}{\rightleftarrows}} \mathcal{W}(q+N)_{r+1}
\end{equation}
where we have exchanged the maps $p$ and $G$. The associated $2\times 2$ matrix factorization is $Q=G\eta+p\bar{\eta}$. Note that both matrix factorizations $K_{\zeta \gg 0}$ and $K_{\zeta \ll 0}$ are exact, i.e. have no cohomology.

As mentioned above, a general GLSM branes can be obtained from the matrix factorizations $K_j$, $K_{\zeta \gg 0}$ and $K_{\zeta \ll 0}$ by linear algebra and homological operations. Such a brane then consists of sums of Wilson line branes $\mathcal{W}(q_j)_{r_j}^{\oplus n_j}$ with different gauge charges at a position with fixed R--degree. The multiplicities $n_j$ of these Wilson line branes are determined by the combinatorics of the Clifford algebra. Hence, the GLSM brane is of the form
\begin{equation}
  \label{eq:generalcomplex}
  \dots \leftrightarrows \bigoplus_{i=1}^{L_j} \mathcal{W}(q^{(i)}_j)_{r^{(i)}_j}^{\oplus n^{(i)}_j} \leftrightarrows \bigoplus_{i=1}^{L_{j+1}} \mathcal{W}(q^{(i)}_{j+1})_{r^{(i)}_{j+1}}^{\oplus n^{(i)}_{j+1}} \leftrightarrows \dots
\end{equation}
where $j$ refers to the position in the sequence. It has been conjectured in~\cite{Aspinwall:2007cs} that in this case, the category $\mathcal{D}_{(U(1), pG,\rho_{V}, R)}$ is described by  $\mathrm{DGrS}(W)$, the bounded derived category of finitely generated $S$--modules satisfying~\eqref{eq:MF} and graded by the weight lattices of $G=U(1)$ and the R-symmetry $U(1)$. 

There are also categories of D-branes in the individual phases. In the case of the abelian GLSM it has been conjectured that they can be described as follows~\cite{Herbst:2008jq,Aspinwall:2007cs}. For an irrelevant ideal $I_\zeta \subset S$ we consider the full triangulated subcategory $T_\zeta$ of $\mathrm{DGrS}(W)$ generated by those modules that are annihilated by a power of $I_\zeta$. Then one defines the quotient of triangulated categories $\mathcal{D}_{(G, W,\rho_{V}, R)}^{\zeta} = \mathrm{DGrS}(W)/ T_\zeta$. In a geometric phase, there is an equivalence of categories $\mathrm{DGrS}(W)/ T_\zeta \cong D^b(X_\zeta)$ where $D^b(X_\zeta)$ is the derived category of coherent sheaves on $X_\zeta$. The functors $\mathcal{D}_{(G, W,\rho_{V}, R)} \to \mathcal{D}_{(G, W,\rho_{V}, R)}^{\zeta}$ and $\mathcal{D}_{(G, W,\rho_{V}, R)}^{\zeta} \to \mathcal{D}_{(G, W,\rho_{V}, R)}$ have been explicitly constructed in~\cite{Herbst:2008jq}. We will denote the latter as lifting geometric branes to GLSM branes, or GLSM lift for short. In fact, there is an infinite number of such functors, labeled by $\pi_1(\mathcal{M}_K)$. Some aspects of the latter fact will be further detailed in Section~\ref{sec:analyt-cont-grade}. Explicit GLSM lifts for our main example will be discussed in Section~\ref{sec-branes}.

\subsection{The Hemisphere Partition Function}
\label{sec:hemisph-part-funct}

With all the definitions presented in the previous section, we are ready to state the result of \cite{Hori:2013ika}: the hemisphere partition function $Z_{D^{2}}(\mathcal{B})$ for a GLSM.  
The function $Z_{D^{2}}(\mathcal{B})$ was computed by supersymmetric localization and it depends on $(G, W, \rho_{V}, R)$, the parameter $t$, the boundary datum $\mathcal{B}=(M,Q,\rho,\mathbf{r}_*)$ as well as a choice of integration contour $\gamma \subset \mathfrak{t}_{\mC}$:
\begin{equation}
\label{ZD2}
Z_{D^{2}}(\mathcal{B})=C(r\Lambda)^{\hat{c}/2}\int_{\gamma\subset \mathfrak{t}_{\mC}} d^{l_{G}}\sigma' \prod_{\alpha>0}\alpha(\sigma')\sinh(\pi\alpha(\sigma'))\prod_{i}\Gamma\left(iQ_{i}(\sigma')+\frac{R_{i}}{2}\right)e^{it(\sigma')}f_{\mathcal{B}}(\sigma').
\end{equation}
The integration variable is $\sigma'=r\sigma$ where $r$ is the radius of $D^2$. Since we are only interested in the CY case\footnote{In the non-CY case i.e. when $\rho_{V}$ does not factor through $SL(V)$, the hemisphere partition function depends nontrivially on the dimensionless parameter $r\Lambda$, where $\Lambda$ is the energy scale of the theory. See \cite{Hori:2013ika} for more details.}, where $r$ only enters trivially, we will denote $\sigma'\equiv\sigma$. $C$ is a dimensionless normalization constant, that needs to be fixed and  
$l_{G}=\mathrm{rk}(G)$. The $R_{i}\in \mathbb{Q}$ are the R-charges (weights of $R$ action) of the chiral fields and $Q_{i}\in \mathbb{Z}$ the weights of $\rho_{V}$.  
$\alpha>0$ denotes the positive roots of $G$ and the $t$ are the complexified FI parameters  
\begin{equation}
t=\zeta-i\theta.
\end{equation}
For the CY case, since $\rho_{V}:G\rightarrow SL(V)$, the parameters $t$ do not run with the energy scale $\Lambda$ (i.e. the dependence on $\Lambda$ drops out). The function $f_{\mathcal{B}}(\sigma)$ is defined by
\begin{equation}
\label{fbdef}
f_{\mathcal{B}}(\sigma)=\mathrm{tr}_M\left(e^{i\pi \mathbf{r}_{\ast}}e^{2\pi\rho(\sigma)}\right),
\end{equation}
where $\mathbf{r}_{\ast}$ and $\rho(\sigma)$ have been defined in (\ref{rrhodef}) and (\ref{gdef}), respectively. All the dependence of (\ref{ZD2}) on $\mathcal{B}$ is contained in (\ref{fbdef}). We will refer to $f_{\mathcal{B}}$ as the ``brane factor''. To extract the brane factor we do not require the full information about the associated complex, since the details of the maps do not enter. This reflects the expectation that the brane factor only depends on the Grothendieck group of the category of D-branes in the GLSM. In geometric phases in which the hemisphere partition function reduces to the central charge of the corresponding low energy D-brane, the central charge formula~\cite{Hosono:2004jp} only depends on the K-theory class of the D-brane. 

In the special case of a $U(1)$ GLSM with $\mathcal{B}$ given in terms of a complex of Wilson line branes as in~\eqref{eq:generalcomplex} the brane factor is easily read off:
\begin{equation}
\label{eq:u1fb}
f_{\mathcal{B}}(\sigma)=\sum_j \sum_{i=1}^{L_j} n_{j}^{(i)} e^{i\pi r^{(i)}_j}e^{2\pi q^{(i)}_j\sigma},
\end{equation}
where the sum over $j$ goes over the positions of the complex.

Note that all the poles of the integrand of $Z_{D^{2}}(\mathcal{B})$ are located in the space $\Im(\sigma)\subset \mathfrak{t}_{\mC}$. They are on the complex codimension $1$ hyperplanes:
\begin{equation}
H_{i}:=\{iQ_{i}(\sigma)=-R_{i}/2-k|k\in\mathbb{Z}_{\geq 0}\}\qquad \mathcal{P}:=\bigcup_{i=1}^{\mathrm{dim}V}H_{i}.
\end{equation}
The integration contour $\gamma\subset \mathfrak{t}_{\mC}$ is an important part of the definition of $Z_{D^{2}}(\mathcal{B})$ and necessary if we want a non-perturbative description of it i.e. a description of $Z_{D^{2}}(\mathcal{B})$ which is valid for all values of $t$ on $\mathcal{M}_{K}$.  
The choice of $\gamma$ is determined, among other conditions, by the convergence of the integral.
In the following we will study these contours in much more detail. For this we start by defining admissible contours:  
\\\\
Given a GLSM datum and boundary data $\mathcal{B}$, we define an admissible contour $\gamma \subset \mathfrak{t}_{\mC}$ as a Lagrangian in $\mathfrak{t}_{\mC}\setminus \mathcal{P}$ such that $Z_{D^{2}}(\mathcal{B})$ absolutely converges on $\gamma$ and such that $\gamma$ is a continuous deformation of $\gamma_{\mathbb{R}}=\Re(\mathfrak{t}_{\mC})$ in $\mathfrak{t}_{\mC}\setminus \mathcal{P}$, i.e. a deformation of $\gamma_{\mathbb{R}}$ that avoids the singularities.\\\\
We should remark here that the properties of admissible contours are not yet well understood. They are motivated from the saddle point equations one obtains by supersymmetric localization as explained in \cite{Hori:2013ika}.

\subsection{Analytic Continuation and Grade Restriction}
\label{sec:analyt-cont-grade}
In this section we will discuss the so-called grade restriction rule (GRR) for B-branes \cite{Herbst:2008jq} from the point of view of the hemisphere partition function as done in \cite{Hori:2013ika}. The GRR is a restriction on the boundary data $\mathcal{B}$ that appears naturally as an answer to the question 'Given $\mathcal{B}$, can we find an admissible contour at each point $t\in \mathcal{M}_{K}$?'.  
We will have a more categorical discussion of the GRR in section \ref{sec-branes}, but at this point we can give a purely analytic definition that will fit better in the context of analytic continuation of the hemisphere partition function. 
Let us first study admissible contours in more detail. Start by choosing a basis of $\mathfrak{t}$ and write
\begin{equation}
\sigma^{a}=\sigma^{a}_{1}+i\sigma^{a}_{2} \qquad a=1,\ldots,\mathrm{rk}(G).
\end{equation}
Then, a contour $\gamma$ which is a deformation of $\gamma_{\mathbb{R}}$ can be written as the graph of a function $h:\mathfrak{t}\rightarrow \mathfrak{t}$:
\begin{equation}
\sigma^{a}=\sigma^{a}_{1}+ih^{a}(\sigma_{1})
\end{equation}
with some additional conditions so that it avoids $\mathcal{P}$. Explicitly, there are $\mathrm{dim} V$ conditions given by $-Q_{i}(h(\sigma_{1}))+R_{i}/2>0$ in the region $Q_{i}(\sigma_{1})=Q_{i}^{a}\sigma_{1}^{a}=0$. In addition we also have the Lagrangian condition $\sum_{a}d\sigma^{a}_{1}\wedge d h^{a}(\sigma_{1})=0$, but that will be trivially satisfied in the one-parameter models that we study in this paper.  Let us denote the integrand of (\ref{ZD2}) by $F_{\mathcal{B}}(\sigma)$.  Asymptotically $F_{\mathcal{B}}(\sigma)$ in (\ref{ZD2}) takes the form
\begin{equation}\label{asymptF}
|F_{\mathcal{B}}(\sigma)|\sim P(\sigma) e^{-A_{\mathcal{B}}(\sigma)} \text{ \ as \ }|\sigma|\rightarrow \infty
\end{equation}
where $P(\sigma)$ is polynomial in $\sigma$. An explicit expression for $A_{\mathcal{B}}(\sigma)$ in (\ref{asymptF}) can be found using the Stirling approximation:
\begin{align}
\Gamma(z)\sim e^{z\log z-\frac{1}{2}\log z-z}\qquad |z|\rightarrow\infty,|\mathrm{Arg}(z)|<\pi.
\end{align}
Then, given $\gamma$, determined by $h$ as above, the condition for $\gamma$ to be admissible is that $h^{*}A_{\mathcal{B}}(\sigma)\rightarrow \infty$ as $|\sigma_{1}|\rightarrow\infty$. This motivates the following definition of grade restricted branes.
\\\\
Given a GLSM datum $(G, \rho_{V}, W, R)$, recall that the associated parameters $t_{I}=\zeta_{I}-i\theta_{I}$ can be used as coordinates in $\mathcal{M}_{K}\cong (\mathbb{C}^{*})^{s}\backslash\Delta$ by setting $z_{I}=\exp(-t_{I})\in \mathcal{M}_{K}$. Set $\mathrm{Arg}:(\mathbb{C}^{*})^{s}\rightarrow \mathbf{T}^{s}:=(\mathbb{R}/2\pi\mathbb{Z})^{s}$ the mapping $\mathrm{Arg}(z)=\mathrm{arg}(z)$ and $p:\mathbb{R}^{s}\rightarrow\mathbf{T}^{s}$ its universal covering. Given $\Delta\subset (\mathbb{C}^{*})^{s}$ considering the coamoeba of $\Delta$, i.e. the image of $\Delta$ under the map $\mathrm{Arg}$: $C:=\mathrm{Arg}(\Delta)$. Therefore the space $\Theta:=p^{-1}(\mathbf{T}^{s}\setminus C)\subset \mathbb{R}^{s}$ will have a chamber structure. Consider $\{\theta_{I}\}_{I=1}^{s}$ as elements on the universal cover of $\mathbf{T}^{s}$, then fix a chamber $T\subset \Theta$.  
 We say that 
$\mathcal{B}$ is grade restricted if there exists an admissible contour $\gamma_{t}$ for every point $t\in \mathbb{R}^{s}\times T$.
\\\\
In the concrete case of GLSMs associated with CY hypersurfaces in $\mathbb{P}^{N-1}$ described by the GLSM datum (\ref{u1glsm})  
the hemisphere partition function is
\begin{align}
\label{hypzd}
Z_{D^{2}}(\mathcal{B})=C(r\Lambda)^{\hat{c}/2}\int_{\gamma} d\sigma\Gamma\left(-iN\sigma+1-N\varepsilon\right)\Gamma\left(i\sigma+\varepsilon\right)^{N}e^{it\sigma}\sum_j \sum_{i=1}^{L_j} n_{j}^{(i)} e^{i\pi( r^{(i)}_j+2\varepsilon)}e^{2\pi q^{(i)}_j\sigma},
\end{align}
where we have explicitly written the function $f_{\mathcal{B}}$ in terms of the weights $(q_{i},r_{j})$ of $(\rho,\mathbf{r}_{*})$. The factor $e^{2\pi i\varepsilon}$ is due to~\eqref{rcharge} and~\eqref{gdef}. The integration contour $\gamma$ should be an admissible contour. In order to determine which contours are allowed we need to start by computing the function $A_{\mathcal{B}}$ in (\ref{asymptF}). For $G=U(1)$, we can focus on a particular weight $q:=q_{j}$ of $\rho$ since the analysis of all the other terms will be analogous. Then we write $A_{q}$ instead of $A_{\mathcal{B}}$ with
\begin{align}
A_{q}=(\zeta-N\log N)\sigma_{2}+\left(N\pi -\mathrm{sgn}(\sigma_{1})(\theta+2\pi q)\right)|\sigma_{1}|.
\end{align}
The convergence of $Z_{D^{2}}(\mathcal{B})$ is determined, in general, by $\zeta,\theta$ and $q$. However, note that if $\zeta\neq N\log N$ we can always choose the following $h$ to define $\gamma$
\begin{align}
\sigma_{2}=h(\sigma_{1})=\mathrm{sgn}(\zeta-N\log N)(\sigma_{1})^{2},
\end{align}
so that $A_q$ becomes
\begin{align}
\lim_{|\sigma_{1}|\rightarrow\infty}\Big(|\zeta-N\log N|(\sigma_{1})^{2}+\left(N\pi -\mathrm{sgn}(\sigma_{1})(\theta+2\pi q)\right)|\sigma_{1}|\Big)\sim |\sigma_{1}|^{2} \rightarrow\infty.
\end{align}
Therefore, if $\zeta\neq N\log N$, there always exists a $\gamma$, for all $\mathcal{B}$ such that $Z_{D^{2}}(\mathcal{B})$ is absolutely convergent. For $\zeta= N\log N$ the dependence on $\sigma_{2}$ in $A_{q}$ drops out and we get a condition on the weights of $\rho$:
\begin{align}
  \label{GRRhyp}
  -\frac{N}{2}<\frac{\theta}{2\pi}+q<\frac{N}{2}.
\end{align}
This imposes a condition on the allowed charges $q$ of a brane for a given length $2\pi$ interval of $\theta$. The condition is called the grade restriction rule. The set of allowed values for $q$, together with $\theta$ is referred to as the ``(charge) window'' \cite{Herbst:2008jq}. If this condition is satisfied, we can take any function $h$ satisfying
\begin{align}
-1/N+\varepsilon <h(\sigma_{1}=0)<\varepsilon.
\end{align}
In particular, we can take $h\equiv 0$ i.e. $\gamma=\gamma_{\mathbb{R}}$. In order to determine the grade restricted branes in this class of examples, we have to specify the values of $\theta$. Recall that the conifold point is located at
\begin{align}
t_{c}=N\log N-i\pi N.
\end{align}
In the universal covering of $\mathcal{M}_{K}$, $\theta$ takes values in
\begin{align}
\theta\in \mathbb{R}\setminus(N\pi+2\pi\mathbb{Z}).
\end{align}
Therefore, the different chambers are determined by an integer $n$. Call them $T_{n}$, so
\begin{align}
\label{chamber}
 T_{n}=(\pi(N+2n),\pi(N+2+2n))\qquad n\in\mathbb{Z}.
\end{align}
Thus, we have shown the following: For the GLSM (\ref{u1glsm}), given a chamber $T_{n}$ and a B-brane $\mathcal{B}=(M,Q,\rho,\mathbf{r}_{*})$ there exists an admissible contour $\gamma_{t}$ for $\mathcal{B}$ at all points $(\zeta, \theta)\in \mathbb{R}\times T_{n}$ if and only if all the weights of $\rho$ satisfy (\ref{GRRhyp}). In particular, $\gamma_{t}=\gamma_{\mathbb{R}}$ is an admissible contour.
\\\\
Hence, we have shown that the B-branes satisfying (\ref{GRRhyp}) are grade restricted. 
\subsection{Hemisphere Partition Function and Mellin-Barnes Integrals}
\label{sec:mellin-barnes}

Next, we work out some of the consequences of this result for the hemisphere partition function~(\ref{hypzd}) in the case of the GLSM associated to  a CY hypersurface of degree $N$ in $\mathbb{P}^{N-1}$. We choose the chamber $T_n$ with $n = -N$ and $\theta \in T_n$. Then,~\eqref{GRRhyp} is solved by the weights $q=0,\dots,N-1$. Now consider the modules $M_q$ associated to the Koszul complexes $K_q$ introduced after (\ref{koszulbrane}) for $q=1,\dots,N-1$. Note, that they are automatically grade-restricted to the chosen charge window. Then,~\eqref{eq:u1fb} yields
\begin{equation}
  \label{eq:1}
  \mathrm{tr}_{M_q}\left(e^{i\pi \mathbf{r}_{\ast}}e^{2\pi\rho(\sigma)}\right) = (1-e^{2\pi \sigma})^q, \quad q=1,\dots, N-1.
\end{equation}
For a generic GLSM brane, this is typically not the case, which also means that it is not well-defined outside a given phase. Two prominent examples where this is not the case are the Koszul branes (\ref{lr-empty}) and (\ref{emptylg}). Further note that the brane factors of a general GLSM brane will not be of the form (\ref{eq:1}). For instance, if we multiply (\ref{eq:1}) by an integer, the resulting GLSM brane is no longer represented by a simple Koszul brane of the form (\ref{koszulbrane}).  All these issues will be addressed in examples in section \ref{sec-branes}.

Hence we have completely specified the brane datum $\mathcal{B}_q = (M_{N-q},Q,\rho,\mathbf{r}_*)$, $q=1,\dots,N-1$ (note the reverse order). The hemisphere partition functions~\eqref{hypzd} for $\mathcal{B}_q$ then becomes
\begin{equation}
  \label{eq:2}
  Z_{D^{2}}(\mathcal{B}_q)=iC(r\Lambda)^{\hat{c}/2}\int_{-i\infty}^{i\infty} ds\,\Gamma\left(Ns+1\right)\Gamma\left(-s\right)^{N}e^{-t(s+\varepsilon)} (1-e^{2\pi i s})^{N-q}
\end{equation}
where we have set $s=-i\sigma + \varepsilon$ where $\varepsilon$ was introduced in (\ref{rcharge}). The admissible contour $\gamma_{\mathbb{R}}$ then becomes the imaginary axis. Using the reflection formula (\ref{reflection}) and the multiplication formula (\ref{multiplication}) for the Gamma functions this can be rewritten as (where we have chosen $\varepsilon = 0$)
\begin{equation}
  \label{eq:yhemisphere}
  Z_{D^{2}}(\mathcal{B}_q)=iC(r\Lambda)^{\hat{c}/2} \frac{(2\pi i)^NN^{\frac{1}{2}}}{(2\pi)^{\frac{N-1}{2}}}\int_{-i\infty}^{i\infty} ds\, \frac{\prod_{j=1}^{N-1}\Gamma\left(s+\frac{j}{N}\right)}{\Gamma(s+1)^{N-1}}\frac{\left(N^{N}e^{-t} e^{-i\pi N}\right)^s}{(1-e^{2\pi i s})^{q}}
\end{equation}
In the following we are going to derive some properties of these integrals. First we change to the algebraic coordinate of the GLSM~\cite{Morrison:1994fr}
\begin{equation}
  z=e^{-i\pi N} N^{N}e^{-t}.
  \label{eq:4}
\end{equation}
We denote the integral without the prefactor by $y^*_q(z)$
\begin{equation}
  \label{eq:ydef}
  y^*_q(z) = \int_{-i\infty}^{i\infty} ds\, \frac{\prod_{j=1}^{N-1}\Gamma\left(s+\frac{j}{N}\right)}{\Gamma(s+1)^{N-1}}\frac{z^s}{(1-e^{2\pi i s})^{q}}, \qquad q=1,\dots, N-1
\end{equation}
We are now going to show that $y^*_q(z), q=1,\dots, N-1$ satisfies a linear homogeneous differential equation of order $N-1$ in $z$. For this purpose, we set $g_N(s) = \prod_{j=1}^{N-1}\frac{\Gamma\left(s+\frac{j}{N}\right)}{\Gamma(s+1)}$. We first note that
\begin{equation}
  \label{eq:6}
  g_N(s+1) = \frac{\prod_{j=1}^{N-1} \left(s+\frac{j}{N}\right)}{(s+1)^{N-1}} g_N(s).
\end{equation}
If we set $\theta = z\frac{d}{dz}$, then we also observe that
\begin{equation}
  \label{eq:8}
  \theta y^*_q(z) = \int_{-i\infty}^{i\infty} ds\, s g_N(s) \frac{z^s }{(1-e^{2\pi i s})^{q}}
\end{equation}
and hence
    \begin{align}
    \label{eq:7}
      \theta^{N-1} y^*_q(z) &= \int_{-i\infty}^{i\infty} ds\, s^{N-1} g_N(s) \frac{z^s }{(1-e^{2\pi i s})^{q}}\\
    \label{eq:10}
     \prod_{j=1}^{N-1} \left(\theta + \tfrac{j}{N}\right) y^*_q(z) &=
      \int_{-i\infty}^{i\infty} ds\, \prod_{j=1}^{N-1} \left(s +
        \tfrac{j}{N}\right) g_N(s) \frac{z^s}{(1-e^{2\pi i s})^{q}}.
    \end{align}
Now, by~\eqref{eq:6} the equation~\eqref{eq:10} becomes
\begin{equation}
  \label{eq:9}
  z\prod_{j=1}^{N-1} \left(\theta + \tfrac{j}{N}\right) y^*_q(z) = \int_{-i\infty}^{i\infty} ds\, (s+1)^{N+1} g_N(s+1) \frac{z^{s+1}}{(1-e^{2\pi i s})^{q}}.
\end{equation}
The integrands of the left hand side of~\eqref{eq:7} and~\eqref{eq:9} have the same poles, hence the integrals are equal, and we conclude that $y^*_q(z)$ satisfies the differential equation
\begin{equation}
  \label{eq:11}
   \left(\theta^{N-1} - z\prod_{j=1}^{N-1} \left(\theta + \tfrac{j}{N}\right) \right) y^*_q(z) = 0, \quad q=1,\dots,N-1
\end{equation}
This is a generalized hypergeometric differential equation. The solutions $y^*_1(z),\dots,y^*_{N-1}(z)$ are linearly independent. This can be seen as follows. Evaluating the integrals $y^*_{q}(z)$ by closing the contour to the right with a semi-circle and applying the residue theorem, they are of the form
\begin{equation}
  \label{eq:12}
  y^*_q(z) = \sum_{k=0}^\infty \left( g_N^{(q-1)}(k) + \binom{q-1}{1} g_N^{(q-2)}(k) \log z + \dots + g_N(k) (\log z)^{q-1}\right) z^k
\end{equation}
where $g_N(0), g^{(1)}_N(0), \dots, g_N^{(q-1)}(0)$ do not all vanish simultaneously.
Hence, there are nonzero constants $B_q \in \mC$ such that
\begin{equation}
  \label{eq:13}
  \lim_{z\to 0} \frac{y^*_q(z)}{(\log z)^{q-1}} = B_q,\quad q=1,\dots,N-1.
\end{equation}
Since these constants $B_q$ are non-vanishing, the $y^*_q(z)$ are linearly independent. We will study the analytic properties of the functions $y^*_q(z)$ in more detail in section~\ref{sec-math}.

A couple of remarks are in order.
The differential equation~(\ref{eq:11}) has regular singularities at $z=0,\infty$, and $1$. By~\eqref{eq:4}, these correspond to the points $t = \infty, -\infty$ and $t_c=N\log N - i\pi$, respectively. These are precisely the large radius limit, the Landau-Ginzburg point, and by~\eqref{eq:conifold} the singular point where the GLSM develops a non-compact Coulomb branch.

The differential equation (\ref{eq:11}) is the Picard-Fuchs equation satisfied by the periods of the mirror hypersurface. In view of mirror symmetry, this is of course not surprising, in particular since (\ref{eq:ydef}) is nothing but the Mellin-Barnes representation of these periods. Note however, that we have derived this hypergeometric differential equation without any reference to the mirror. 

Note that the Koszul complexes $K_q$ we started with do not refer to any specfic phase, i.e. they are genuine GLSM branes and not viewed as lifts of low-energy D-branes in some phase. Therefore the derivation of the differential equation is irrespective of the phase. If in~\eqref{eq:2} we had chosen to bring $\Gamma(Ns+1)$ into the denominator instead of $\Gamma(-s)^N$, we would have obtained 
\begin{equation}
  \label{eq:5}
  \left(\prod_{j=1}^{N-1} \left(\theta + \tfrac{j}{N}\right) - \tfrac{1}{z}\theta^{N-1}\right) \overline{y}^*_q(z) = 0,
\end{equation}
for
\begin{equation}
  \label{eq:17}
  \overline{y}^*_q(z)  = \int_{-i\infty}^{i\infty} ds\, \frac {\Gamma(s)^{N-1}}{\prod_{j=1}^{N-1}\Gamma\left(s+\frac{j}{N}\right)}\frac{z^s}{(1-e^{2\pi i( s+\frac{q}{5})})}, \qquad q=1,\dots, N-1
\end{equation}
This is the same differential equation as~\eqref{eq:11} and the $\overline{y}^*_q(z)$ are linear combinations of the $y^*_q(z)$.

The hemisphere partition function (\ref{eq:2}) is defined over the whole parameter space. To compute the IR limit of $Z_{D^2}$ in the two phases we have to close the integration contour either to the right ($\zeta\gg0$) or to the left ($\zeta\ll0$) and evaluate the residue integral. This leads to convergent expressions in either phase which we will denote by $Z_{D^2}^{\zeta\gg0}$ and $Z_{D^2}^{\zeta\ll0}$. While the evaluation of the hemisphere partition function in one of the phases just amounts to computing residues it is not obvious how one can extract a convergent expansion around the singular point.

\section{GLSM branes and Grade Restriction}
\label{sec-branes}
In this section we discuss the D-brane interpretation of the Mellin-Barnes integrals in (\ref{eq:2}). So far, these integrals have been studied only for the GLSM branes $\mathcal{B}_q$. We will relate a special set of GLSM branes to specific combinations of these Mellin-Barnes integrals. These GLSM branes are obtained from the lift $\mathcal{D}_{(G, W,\rho_{V}, R)}^{\zeta\gg0} \to \mathcal{D}_{(G,W,\rho_{V}, R)}$ of branes in the geometric phase. In such a phase the Grothendieck group of $\mathcal{D}_{(G, W,\rho_{V}, R)}^{\zeta\gg0}$ is expected to agree with the K-theory group $K(X_{\zeta \gg 0})$. The special set is then the lift of a basis of $K(X_{\zeta \gg 0})$. Since not all of these branes will be grade restricted, we first have to grade restrict them before we can make the connection to (\ref{eq:2}). Otherwise it is not possible to sensibly talk about analytic continuation to or beyond a phase boundary. We will first explain how to do this by recalling the essentials of \cite{Herbst:2008jq}, then apply this to a suitably chosen set of GLSM branes. The associated hemisphere partition functions can then be related to certain combinations of the $Z_{D^2}(\mathcal{B}_q)$ whose analytic continuation to the conifold point will then be discussed in section \ref{sec-math}. Since we have all tools available, we also recompute the monodromy matrices of the quintic directly in the GLSM by using the hemisphere partition function.
\subsection{Grade Restriction}
According to the seminal paper \cite{Herbst:2008jq} one can only transport grade-restricted branes past the conifold singularity from one phase to another in a well-defined way. In fact, as soon as we approach a phase boundary, grade restriction is required. One can show in examples that non-grade restricted branes lead to divergence of the hemisphere partition function near the conifold point. Grade restriction means that the gauge charges of the Wilson line branes in the GLSM have to be restricted to a certain window as defined in (\ref{GRRhyp}). Due to the periodicity of the theta-angle there is an infinite number of windows of the same width, each corresponding a specific path past the conifold point in the FI-theta parameter space inside a chamber $T_n$ as defined in (\ref{chamber}). Without grade restriction, the Mellin-Barnes integrals coming from a hemisphere partition function are not well-defined because they are not globally defined throughout the parameter space.  For example, B-branes which have a simple geometric interpretation such as the structure sheaf $\mathcal{O}_X\in\mathcal{D}_{(G, W,\rho_{V}, R)}^{\zeta \gg 0}$ in the large radius phase, do not lift to grade restricted GLSM branes. In order to study the behavior of such branes close to and beyond a phase boundary they have to be grade restricted first. Since this procedure can become quite tedious, we will propose an algorithmic method to grade restrict GLSM branes and extract the brane factor without doing involved calculations.

Grade restriction replaces a given GLSM brane by a different one which is the same in the IR but whose gauge charges fit into the desired window. This can be done by systematically binding ``empty'' branes via tachyon condensation. The empty branes are those branes which reduce to ``nothing'' in the IR. This means in particular in the respective phase the K-theory class and the hemisphere partition function of such a brane are zero. For every phase there is a set of empty branes which can be used to grade restrict. Having fixed a charge window, grade restricting a GLSM-brane is a two-step process.
\begin{enumerate}
\item Produce a bound state between a non-grade-restricted brane $\mathcal{B}$ and an empty brane $\mathcal{B}_E$ by turning on boundary-changing $\mathbb{Z}_{2}$-odd open string state $\Psi\in \mathcal{H}^1(\mathcal{B},\mathcal{B}_E)$ (the ``tachyon'') as in (\ref{qcohom}). In the homological language this amounts to computing the mapping cone $\mathrm{Cone}(\Psi:\mathcal{B}\rightarrow\mathcal{B}_E)$.  The open string state must have the specific property that there is the identity map between the non-grade-restricted Wilson line branes appearing in $\mathcal{B}$ and $\mathcal{B}_E$. Typically, one has to bind more than one empty brane to $\mathcal{B}_E$ with different gauge and R charges to achieve grade restriction.
\item Remove trivial brane-antibrane pairs by replacing the identity map by a specific composition of other maps. This removes that Wilson line brane that is not in the window.
\end{enumerate}
Step $1$ is the crucial one because we need to construct a very specific open string state stretching between two D-branes. Step $2$ is more of a cosmetic nature. In the language of matrix factorizations this amounts to elementary row and column manipulations to single out blocks of the form $\left(\begin{array}{cc}0&1\\W&0\end{array}\right)$ which can be removed from the matrix factorization, as they describe trivial brane-antibrane pairs. This step is however not strictly necessary since the brane is already grade restricted after the first step, so it may be skipped entirely or performed only partially.

In the following we restrict ourselves to the main example of the $U(1)$ GLSM (\ref{u1glsm}) associated to CY hypersurfaces of degree $N$ in $\mathbb{P}^{N-1}$. In this case there are two sets of empty branes $\mathcal{B}_E\in \mathcal{D}_{(G,W,\rho_{V}, R)}$, one corresponding to each phase. We denote them by $\mathcal{B}_E^{\zeta\gg0}$ and $\mathcal{B}_{E}^{\zeta\ll0}$, respectively. These have already been introduced in section \ref{sec:d-branes-glsm}. If we start off in the large radius phase ($\zeta\gg0$) the empty brane is given by the Koszul brane $K_{\zeta \gg 0}$ in (\ref{lr-empty}). Even though we refer to this as the empty brane, (\ref{lr-empty}) actually consists of an infinite number of GLSM branes, labeled by the choice of $q$ and $r$. Note that the empty brane is not grade restricted since by~\eqref{GRRhyp} the allowed window of gauge charges $q$ consists of $N$ consecutive integers. Therefore, at least one of the Wilson line branes in (\ref{lr-empty}) cannot by in any given window. This also shows that the notion of an empty brane cannot be globally defined but only makes sense in connection with a phase. Indeed, (\ref{lr-empty}) only reduces to ``nothing'' in the large radius phase, while it corresponds to a non-trivial B-brane in the Landau-Ginzburg phase. With the associated brane factor 
\begin{align}
f_{\mathcal{B}_E^{\zeta\gg0}}=e^{2\pi q\sigma}e^{i\pi r}(1-e^{2\pi\sigma})^N
\end{align}
this can be verified explicitly by evaluation of the hemisphere partition function in both phases which yields $Z_{D^2}^{\zeta\gg0}(\mathcal{B}_E^{\zeta\gg0})=0$ while $Z_{D^2}^{\zeta\ll0}(\mathcal{B}_E^{\zeta\gg0})\neq0$. Similarly, there is also an empty brane associated to the Landau-Ginzburg phase. This is given by the Koszul brane (\ref{emptylg}) whose brane factor is 
\begin{align}
f_{\mathcal{B}_E^{\zeta\ll0}}=e^{2\pi q\sigma}e^{i\pi r}(1-e^{2\pi N\sigma}).
\end{align}
To illustrate the procedure of grade restriction we repeat a discussion of \cite{Herbst:2008jq} for $N=3$. i.e. the GLSM associated to the cubic curve. Consider the GLSM matrix factorization corresponding to the Koszul brane $K = K_{\zeta \ll 0}^*$ in~\eqref{emptylg}:
\begin{equation}
\label{strmf}
\widehat{\mathcal{W}}_{X}:\quad Q=\left(\begin{array}{cc}0&G_3\\p&0\end{array}\right)\quad\rightarrow\quad \xymatrix{\mathcal{W}(0)_0 \ar@<2pt>[r]^{G_3}&\ar@<2pt>[l]^p \mathcal{W}(3)_1}
\end{equation}
Note that the matrix factorization $Q$ is not enough to fully specify the GLSM brane $\mathcal{B}=(M,Q,\rho,\mathbf{r}_{*})$, while the representation in terms of the Wilson line branes encodes the full information of $\mathcal{B}$. It includes in particular the matrices $\rho$ and $\mathbf{r}_{*}$ in (\ref{rrhodef})  and (\ref{gdef}) with a specific choice of normalization corresponding to overall shifts. In the following we will often only specify $Q$ without explicitly giving $\rho$ and $\mathbf{r}_{*}$, as they can be reconstructed from the Wilson line brane representation. 
This brane (\ref{strmf}) is not grade restricted and we will denote non-grade-restricted branes by $\widehat{\mathcal{W}}$. In the large radius phase $\zeta \gg 0$, this reduces to the structure sheaf on the cubic given by $G_3(x_1,x_2,x_3)=0$ and described by the complex $ \xymatrix{\mathcal{O}(0) \ar[r]^{G_3}& \mathcal{O}(3)}$. The window we would like to grade restrict to is $q\in\{0,1,2\}$, which corresponds to $\theta\in (-3\pi,-\pi)$. The brane factor is
\begin{equation}
f_{\widehat{\mathcal{W}}_X}=1-e^{6\pi\sigma}.
\end{equation}
Inserting this into the definition of the hemisphere partition function and evaluating it in the large radius phase one recovers the expected charges\footnote{In order to obtain this result one has to implement a $\theta$-angle shift between the UV and the IR theory: $\theta^{IR}=\theta^{UV}+N\pi$ \cite{Herbst:2008jq}.}:
\begin{equation}
Z_{D^2}^{\zeta\gg0}(\widehat{\mathcal{W}_X})=3\varpi_1=3\Phi_{0,12},
\end{equation}
where the $\varpi$ and $\Phi_0$ are bases of solutions of the hypergeometric equation near $z=0$ whose series expansions are defined in appendix \ref{app-pf}. Here we have made a specific choice for the normalization constant $C$. Since this is nothing but the empty brane in the Landau-Ginzburg phase one also computes $Z_{D^2}^{\zeta\ll0}(\widehat{\mathcal{W}_X})=0$. This is certainly not the correct analytic continuation of the structure sheaf to the Landau-Ginzburg point. This shows why we have to grade restrict. To put the brane into desired window we have to get rid of the Wilson line brane $\mathcal{W}(3)_1$. This can be achieved by binding an empty brane in the following way. We choose the empty brane given by the Koszul brane $K' = K_{\zeta \gg 0} \otimes \mathcal{W}(0)_{-1}$, and a morphism $\Psi \in H^1_{D}(\mathrm{Hom}( K , K'))$ as follows:
\begin{equation}
\label{step1}
\xymatrix{&\mathcal{W}(0)_0\ar@<2pt>[r] \ar[rd]^{\varphi}& \ar@<2pt>[l]\mathcal{W}(3)_1\ar[rd]^{1}\\
\mathcal{W}(0)_{-1}\ar@<2pt>[r] & \ar@<2pt>[l] \mathcal{W}(1)_0^{\oplus 3}\ar@<2pt>[r]& \ar@<2pt>[l] \mathcal{W}(2)_1^{\oplus 3}\ar@<2pt>[r] &\ar@<2pt>[l] \mathcal{W}(3)_2.
}
\end{equation}
The map $\varphi$ can be determined explicitly (see \cite{Herbst:2008jq}) but we actually do not need its explicit form to compute the brane factor. Removing trivial brane-antibrane factors this reduces to
\begin{equation}
\label{step2}
\xymatrix{&\mathcal{W}(0)_0 \ar@<2pt>[rd]&\\
\mathcal{W}(0)_{-1}\ar@<2pt>[r] & \ar@<2pt>[l] \mathcal{W}(1)^{\oplus 3}_0\ar@<2pt>[r]& \ar@<2pt>[l] \mathcal{W}(2)^{\oplus 3}_1.\ar@<2pt>[lu]
}
\end{equation}
This new GLSM brane is a bound state $\mathrm{Cone}(\Psi:K\to K')$ and is clearly grade restricted. We give it the name $\mathcal{W}_X$. By~\eqref{eq:u1fb}, its brane factor is
\begin{equation}
  f_{\mathcal{W}_X}=3 e^{2\pi\sigma}(1-e^{2\pi\sigma}).
\end{equation}
One can easily convince oneself that, evaluated in the large radius phase, the hemisphere partition function gives the same result for both the grade-restricted and the non-grade-restricted  brane factors. Using the grade-restricted brane $Z_{D^2}^{\zeta\ll0}(\mathcal{W}_X)$ will no longer be zero, as it should be. This means that both branes $\widehat{\mathcal{W}}_X$ and $\mathcal{W}_X$ are GLSM lifts of the structure sheaf $\mathcal{O}_X$, but only $\mathcal{W}_X$ is globally defined over $\mathcal{M}_K$. In terms of the basis $y_q^*(z)$ in~\eqref{eq:ydef}, the hemisphere partition function is
\begin{equation}
\label{cubicd2}
Z_{D^2}(\mathcal{W}_X)=-\frac{3\sqrt{3}}{2\pi}(y_2^{\ast}-y_1^{\ast}).
\end{equation}
We would like to emphasize again that removing the trivial brane-antibrane pair is not necessary. Both (\ref{step1}) and (\ref{step2}) give the same brane factors. Only the gauge charges of the Wilson line branes and their relative positions enter. 

If we want to consider more complicated branes the procedure of grade restriction, in particular the calculation of the maps specifying the tachyon, can become complicated. Even the binding of only two empty branes can become very tedious, as demonstrated for instance for a D4-brane on the quintic in \cite{Herbst:2008jq}.  
Typically the calculations are even more involved than explained there. Given that most of the information is irrelevant if we are just interested in the brane factor we would like to shortcut this calculation. Instead of explicitly computing the maps corresponding to the desired open string states we simply {\em assume} that they exist and position the Wilson line branes describing the empty brane accordingly. In this way we skip a tedious calculation but still keep the information that is relevant for extracting the brane factor. Of course one needs an independent check that one has done the right thing. This is easily achieved by verifying that the resulting grade restricted brane factor yields the same result for the hemisphere partition function in the large radius phase as the non-grade-restricted one. Let us illustrate this procedure by considering a D0-brane on the cubic. For the generic cubic the D0 brane is the intersection of a divisor $h=\sum_{i=1}^3\alpha_ix_i$ with the cubic hypersurface equation $G_3(x)=0$. This can be lifted to the Koszul brane $K( (h,-G_3), \mathcal{W}(0)_+)$:
\begin{equation}
\label{cubicd0complex}
\widehat{\mathcal{W}}_{D0}:\quad  \xymatrix{\mathcal{W}(0)_+\ar@<2pt>[r]^{f_1} & \ar@<2pt>[l]^{g_1}{\begin{array}{c}\mathcal{W}(1)_-\\\oplus\\\mathcal{W}(3)_-\end{array}}\ar@<2pt>[r]^{f_2} & \ar@<2pt>[l]^{g_2}\mathcal{W}(4)_+
}.
\end{equation}
The explicit maps are
\begin{equation}
 f_1=\left(\begin{array}{c}h\\-G_3\end{array}\right)\quad g_1=(0,-p) \qquad f_2=(G_3,h) \quad g_2=\left(\begin{array}{c}p\\0\end{array}\right).
\end{equation}
We need not specify the R-charges of the Wilson line branes because for computing the brane factor it only matters if the charge is even or odd, which we indicate by a subscript $\pm$. The explicit matrix factorization is
\begin{equation}
Q=\left(\begin{array}{cccc}
0&h&-G_3&0\\
0&0&0&G_3\\
-p&0&0&h\\
0&p&0&0
\end{array}\right)=h\eta_1+G_3\eta_2+p\bar{\eta}_2.
\end{equation}
The matrices $\rho$ and $\mathbf{r}_{*}$ can be reconstructed from (\ref{cubicd0complex}). The brane factor is
\begin{equation}
f_{\widehat{\mathcal{W}}_{D0}}=1-e^{2\pi\sigma}-e^{6\pi\sigma}+e^{8\pi\sigma}.
\end{equation}
Grade restriction to the standard window $\{0,1,2\}$ requires binding five empty branes. We indicate the procedure in the following table.
\begin{equation}
\label{cubicd0grr}
\mathcal{W}_{D0}:\qquad \begin{array}{ccccc|c}
-&+&-&+&-&\#\\
\hline
&\mathcal{W}(0)&\mathcal{W}(1)&&&\\
&&{\color{green}\mathcal{W}(3)}&{\color{red}\mathcal{W}(4)}&&\\
\hline
&\mathcal{W}(1)&\mathcal{W}(2)^{\oplus 3}&{\color{blue}\mathcal{W}(3)^{\oplus 3}}&{\color{red}\mathcal{W}(4)}&1\\
\mathcal{W}(0)&\mathcal{W}(1)^{\oplus 3}&\mathcal{W}(2)^{\oplus 3}&{\color{green}\mathcal{W}(3)}&&1\\
&\mathcal{W}(0)&\mathcal{W}(1)^{\oplus 3}&\mathcal{W}(2)^{\oplus 3}&{\color{blue}\mathcal{W}(3)}&3
\end{array}
\end{equation}
The first two lines capture the information of the non-grade restricted brane (\ref{cubicd0complex}) with all the maps omitted. The columns indicate the R-degree. In order to grade restrict the brane have to bind a combination of empty branes such that the Wilson line branes $\mathcal{W}(3)$ and $\mathcal{W}(4)$ are removed. The first step is to remove $\mathcal{W}(4)$ by binding an the empty brane as described in (\ref{step1}). Since the tachyon we turn on is a $\mathbb{Z}_2$-odd state we have to position the empty brane appropriately such that the maps between the Wilson line branes go from even/odd to odd/even R-degree. Only the $\mathbb{Z}_2$-grading and not the integer R-grading is relevant for the brane factor. Therefore the positions of the Wilson line branes may be wrong up to an even number of shifts to the left or to the right. Since we are only interested in computing the brane factor here, we can afford ``mistakes'' of this kind. In order for $\mathcal{W}(4)$ to be removed the open string state has to be such that there is the identity map between the two $\mathcal{W}(4)$. We indicate that by giving the pairs of Wilson line branes the same color. There may be further ``accidental'' identity maps, which might cancel more terms than indicated in the table. This can terminate the procedure sooner, but for the sake of systematics, we do not concern ourselves with those. Keeping them only means that step $2$ indicated above is not fully performed. We proceed analogously to remove $\mathcal{W}(3)$. However, we are not done yet, because by binding the first empty brane we were forced to introduce Wilson line branes $\mathcal{W}(3)^{\oplus 3}$ which is also not in the window. To remove this we have to bind three more empty branes as indicated in the table: the last column keeps track of how many copies of the brane we need. After this last step we are done and can easily compute the brane factor from (\ref{cubicd0grr}) by using~\eqref{eq:u1fb}:
\begin{equation}
f_{\mathcal{W}_{D0}}=3-6e^{2\pi\sigma}+e^{4\pi\sigma}=3(1-e^{2\pi\sigma})^2.
\end{equation}
If we evaluate the hemisphere partition function in the large radius phase we get $Z_{D^2}^{\zeta\gg0}(\mathcal{W}_{D0})=3\varpi_0=3\Phi_{0,11}$.
The relation to the $y^{\ast}$-basis is
\begin{equation}
Z_{D^2}(\mathcal{W}_{D0})=\frac{3\sqrt{3}}{2\pi}y_1^{\ast}.
\end{equation}
This brane does not have the minimal charge of a point-like object of the cubic. Indeed, the way we constructed the brane, we end up with three points on the hypersurface. The object with minimal charge can be obtained from a GLSM lift of a permutation-type matrix factorization at the Fermat point \cite{Ashok:2004zb,Brunner:2004mt,Brunner:2005fv,Enger:2005jk,Govindarajan:2005im}.  
Given the Fermat cubic $\sum_{i=1}^3x_i^3$ we consider the Koszul brane $K((x_1-x_2,x_3), \mathcal{W}(0)_+)$
\begin{equation}
\mathcal{W}_\mathrm{pt}:\quad \xymatrix{\mathcal{W}(0)_+\ar@<2pt>[r] & \ar@<2pt>[l]\mathcal{W}(1)_-^{\oplus 2}\ar@<2pt>[r] & \ar@<2pt>[l]\mathcal{W}(2)_+},
\end{equation}
corresponding to the matrix factorization:
\begin{align}
\label{cubic-permmf}
Q=(x_1-x_2)\eta_1+x_3\eta_2+p(x_1^2-x_1x_2+x_2^2)\bar{\eta}_1+px_3^2\bar{\eta}_2.
\end{align}
The brane is automatically grade restricted and its brane factor is
\begin{equation}
f_{\mathcal{W}_\mathrm{pt}}=(1-e^{2\pi\sigma})^2.
\end{equation}
The hemisphere partition function evaluated in the large radius phase is $Z_{D^2}^{\zeta\gg0}(\mathcal{W}_{\mathrm{pt}})=\varpi_0=\Phi_{0,11}$, which is indeed one third of the value of the ``geometric'' D0. This is in agreement with the literature on Landau-Ginzburg branes and D-branes in Gepner models where it has been shown that the permutation branes or ``short'' branes generate the charge lattice \cite{Brunner:2005fv,Enger:2005jk,Govindarajan:2005im}.  
The matrix factorization (\ref{cubic-permmf}) is obtained as a GLSM lift of a matrix factorization in the Landau-Ginzburg phase in which the $p$-field has been fixed to $p=1$. Going from the Landau-Ginzburg phase to the GLSM one has to add in the $p$-field. How to to this rigorously has been demonstrated in \cite{Herbst:2008jq}. Here we just mention the following. There seem to be ambiguities on how to include the $p$-field and the result are different GLSM branes. So it looks like one Landau-Ginzburg matrix factorization yields several different GLSM branes. Of course this cannot be true. The resolution to this apparent puzzle is that a Landau-Ginzburg brane is specified not only by the matrix factorization but also by two additional matrices, typically denoted by $(\gamma,R)$ which encode the orbifold and R-charges of the brane \cite{Walcher:2004tx}. By choosing (\ref{cubic-permmf}) as the lift we have simply picked a Landau-Ginzburg brane with a specific orbifold charge. The identification with the Mellin-Barnes basis gives
\begin{equation}
\label{cubicd0}
Z_{D^2}(\mathcal{W}_\mathrm{pt})=\frac{\sqrt{3}}{2\pi}y_1^{\ast}
\end{equation}
as expected.

The procedure of grade restriction outlined in this section works equally well for other GLSMs, in particular also non-abelian ones. We present a further examples of the grade restriction of GLSM branes on the quartic and quintic in appendix \ref{app-grr}.

\subsection{D-branes on the Quintic}
In section \ref{sec-math} we will discuss the analytic continuation of a specific basis of Mellin-Barnes integrals to the conifold point. We would like to understand which geometric objects this corresponds to in the large radius phase. In order to achieve this we proceed as follows. We first choose a basis of $K(X_{\zeta \gg 0})$ of large radius branes which we characterize by their central charges $Z_{D^2}^{\zeta\gg0}$. Next we find an lift $\mathcal{D}_{(G, W,\rho_{V}, R)}^{\zeta\gg0} \to \mathcal{D}_{(G,W,\rho_{V}, R)}$  of these branes to the GLSM and grade restrict to the charge window $q\in \{0,1,2,3,4\}$ if necessary. Comparing with (\ref{eq:yhemisphere}) we can relate the hemisphere partition function of the large radius brane to a linear combination of the Mellin-Barnes integrals $y_q^{\ast}$ (\ref{eq:ydef}) whose analytic continuation to the conifold point we can do. 
In this way we can directly study how certain D-branes behave near the conifold point, in particular their masses including all quantum corrections. We only focus on the quintic in this section. A basis of branes on the cubic has already been discussed. The completely analogous discussion for the quartic has been relegated to appendix \ref{app-grr}.

We start with a suitable set of D-branes in the geometric phase by picking a basis of $K(X_{\zeta \gg 0})$ given by the classes of $(\mathcal{O}_{\mathrm{pt}},\mathcal{O}_{\ell},\mathcal{O}_{H},\mathcal{O}_X)$, which are the sheaves of a point, of a line $\ell$, of a hyperplane $H$ and the structure sheaf of a one-parameter Calabi--Yau threefold $X$. We will study the analytic continuation of the corresponding central charges
\begin{align}
Z_{D^2}^{\zeta\gg0}(\mathcal{O}_{\mathrm{pt}})&=\varpi_0=\Phi_{0,11}\label{quintd0}\\
Z_{D^2}^{\zeta\gg0}(\mathcal{O}_{\ell})&=\varpi_1=\Phi_{0,12}\label{quintd2}\\ 
Z_{D^2}^{\zeta\gg0}(\mathcal{O}_{H})&=\frac{H^3}{2}\varpi_2-\frac{H^3}{2}\varpi_1+\left(\frac{c_2\cdot H}{24}+\frac{H^3}{6}\right)\varpi_0\label{quintd4}\\
Z_{D^2}^{\zeta\gg0}(\mathcal{O}_{X})&=\frac{H^3}{6}\varpi_3+\frac{c_2\cdot H}{24}\varpi_1+\frac{\zeta(3)c_3}{(2\pi i)^3}\varpi_0=-\Phi_{0,14}\label{quintd6},
\end{align}
The bases $\Phi_0$ and $\varpi$ of solutions to the hypergeometric differential equation can be found in appendix \ref{app-pf}. For the case of the quintic the topological numbers are
\begin{equation}
H^3=5\qquad c_2\cdot H=50\qquad c_3=-200.
\end{equation}
We have already discussed a GLSM lift $\mathcal{W}_X$ of the structure sheaf $\mathcal{O}_X$ in the context of the cubic where we found the GLSM brane (\ref{step2}). In the quintic case the grade restricted GLSM brane $\mathcal{W}_X$ is represented by
\begin{equation}
\label{grox}
\xymatrix{\mathcal{W}(0)_+\ar@<2pt>[r] & \ar@<2pt>[l]\mathcal{W}(1)_-^{\oplus 5}\ar@<2pt>[r] & \ar@<2pt>[l]\mathcal{W}(2)_+^{\oplus 10}\ar@<2pt>[r] & \ar@<2pt>[l]\mathcal{W}(3)_-^{\oplus 10}\ar@<2pt>[r]& \ar@<2pt>[l]\mathcal{W}(4)_+^{\oplus 5}\ar@<2pt>[r] & \ar@<2pt>[l]\mathcal{W}(0)_-
}.
\end{equation}
We extract the following brane factor
\begin{equation}
f_{\mathcal{W}_X}=-5e^{2\pi\sigma}+10e^{4\pi\sigma}-10e^{6\pi\sigma}+5e^{8\pi\sigma}.
\end{equation}
This brane is clearly in the desired charge window and a quick calculation shows that the hemisphere partition function indeed reproduces the correct central charge (\ref{quintd6}) in the large radius phase. In terms of the $y_q^{\ast}$ basis we have the correspondence
\begin{equation}
\label{quinticd6}
Z_{D^2}(\mathcal{W}_X)=5(2\pi i)(y^{\ast}_1-2y^{\ast}_2+2y^{\ast}_3-y^{\ast}_4).
\end{equation}
We will see in section \ref{sec-math} that it is straightforward to analytically continue this particular combination of Mellin-Barnes integrals to the conifold point, where we will confirm directly the known result that this brane becomes massless.

The GLSM brane $\mathcal{W}_H$ associated to $\mathcal{O}_H$ which produces the central charge (\ref{quintd4}) has already been discussed in \cite{Herbst:2008jq}. We start with the Koszul brane $K( (h,-G_5), \mathcal{W}(-1)_-)$ where the linear polynomial $h=0$ is the defining equation of the hyperplane $H$:
\begin{equation}
\widehat{\mathcal{W}}_H:\quad \xymatrix{\mathcal{W}(-1)_+\ar@<2pt>[r] & \ar@<2pt>[l]{\begin{array}{c}\mathcal{W}(0)_-\\\oplus\\\mathcal{W}(4)_-\end{array}}\ar@<2pt>[r] & \ar@<2pt>[l]\mathcal{W}(5)_+
}.
\end{equation}
The structure of the maps is the same as for the D0 brane on the cubic (\ref{cubicd0}). The brane is not grade restricted. The brane factor is
\begin{equation}
f_{\widehat{\mathcal{W}}_H}=-e^{-2\pi\sigma}+1+e^{8\pi\sigma}-e^{10\pi\sigma},
\end{equation}
and the hemisphere partition function gives the desired central charge (\ref{quintd4}). Grade restriction yields a GLSM brane $\mathcal{W}_H$ as follows:
\begin{equation}
\label{groh}
\begin{array}{ccccccc|c}
+&-&+&-&+&-&+&\#\\
\hline
&&{\color{color2}\mathcal{W}(-1)}&\mathcal{W}(0)&&&&-\\
&&&\mathcal{W}(4)&{\color{color1}\mathcal{W}(5)}&&&-\\
\hline
\mathcal{W}(0)&\mathcal{W}(1)^{\oplus 5}&\mathcal{W}(2)^{\oplus 10}&\mathcal{W}(3)^{\oplus 10}&\mathcal{W}(4)^{\oplus 5}&{\color{color1}\mathcal{W}(5)}&&1\\
&{\color{color2}\mathcal{W}(-1)}&\mathcal{W}(0)^{\oplus 5}&\mathcal{W}(1)^{\oplus 10}&\mathcal{W}(2)^{\oplus 10}&\mathcal{W}(3)^{\oplus 5}&\mathcal{W}(4)&1
\end{array}
\end{equation}
Comparing with the result in \cite{Herbst:2008jq}, we find agreement concerning all the information that is relevant to extract the hemisphere partition function.
The brane factor of $\mathcal{W}_H$ is
\begin{equation}
\label{ohfb}
f_{\mathcal{W}_H}=5-15 e^{2\pi\sigma}+20e^{4\pi\sigma}-15e^{6\pi\sigma}+5e^{8\pi\sigma},
\end{equation}
which also gives (\ref{quintd4}) in the large radius phase. The hemisphere partition function can be written as
\begin{equation}
\label{quinticd4}
Z_{D^2}(\mathcal{W}_H)=\frac{5\sqrt{5}}{5\pi^2}(y^{\ast}_1-y^{\ast}_2+y^{\ast}_3).
\end{equation}
When we discuss the analytic continuation to the conifold point we will see that the following combination of Mellin-Barnes integrals is favored: $Z_{D^2}(\mathcal{W}_{D4})\sim 5(y_3^*-2y_2^*+y_1^*)$. Comparing with (\ref{eq:yhemisphere}) the corresponding brane factor would be
\begin{equation}
f_{\mathcal{W}_{D4}}=5( e^{4\pi\sigma}-2e^{6\pi\sigma}+e^{8\pi \sigma}).
\end{equation}
The hemisphere partition function, evaluated in the large radius phase is in this case
\begin{equation}
Z_{D^2}^{\zeta\gg0}(\mathcal{W}_{D4})=\left(\frac{5}{2}\varpi_2+\frac{5}{2}\varpi_1-\frac{25}{12}\varpi_0\right)=\Phi_{0,13}.
\end{equation}
The brane factor does not come from a Koszul brane. It rather indicates that the corresponding brane is a bound state of the D4 brane (\ref{groh}) with a D2 and a D0 discussed in appendix \ref{app-grr}. Indeed the brane factors of these three branes sum up as follows:
\begin{align}
f_{\mathcal{W}_{D4}}&=(\ref{ohfb})+(\ref{geomd2})+(\ref{geomd0})\nonumber\\
&=5(1-e^{2\pi\sigma})^2(1-e^{2\pi\sigma}+e^{4\pi\sigma})-5e^{2\pi\sigma}(1-e^{2\pi\sigma})^3+5(1-e^{2\pi\sigma})^4.
\end{align}

Moving on to the GLSM lift of $\mathcal{O}_l$ we note that the following Koszul brane $K( (f_1,f_2,x_5) , \mathcal{W}(1)_-)$ yields a GLSM brane $\mathcal{W}_l$ that is automatically grade-restricted and leads to the desired central charge (\ref{quintd2}) in the large radius phase:
\begin{equation}
\label{grol}
\mathcal{W}_l:\quad \xymatrix{\mathcal{W}(1)_-\ar@<2pt>[r] & \ar@<2pt>[l]\mathcal{W}(2)^{\oplus 3}_+\ar@<2pt>[r] & \ar@<2pt>[l]\mathcal{W}(3)^{\oplus 3}_-\ar@<2pt>[r] & \ar@<2pt>[l]\mathcal{W}(4)_+}.
\end{equation}
The corresponding matrix factorization of the Fermat quintic is well-known in the literature on Landau-Ginzburg branes \cite{Baumgartl:2007an}. It is the matrix factorization of a permutation brane. With
\begin{align}
f_1&=x_1+x_2\label{permf1def}\\
f_2&=x_3+x_4\\
g_1&=x_1^4-x_1^3x_2+x_1^2x_2^2-x_1x_2^3+x_2^4\label{permg1def}\\
g_2&=x_3^4-x_3^3x_4+x_3^2x_4^2-x_3x_4^3+x_4^4
\end{align}
the associated matrix factorization
\begin{equation}
Q=f_1\eta_1+f_2\eta_2+x_5\eta_3+p g_1\bar{\eta}_1+p g_2\bar{\eta}_2+px_5\bar{\eta}_3.
\end{equation}
The brane factor is
\begin{equation}
f_{\mathcal{W}_l}=-e^{2\pi \sigma}+3e^{4\pi\sigma}-3e^{6\pi\sigma}+e^{8\pi\sigma}.
\end{equation}
which yields
\begin{equation}
\label{quinticd2}
Z_{D^2}(\mathcal{W}_l)=-\frac{\sqrt{5}}{4\pi^2}(y_2^{\ast}-y_1^{\ast}).
\end{equation}
This is another combination of Mellin-Barnes integrals we will be able to analytically continue to the conifold point in section \ref{sec-math}. 
Note that we could also have constructed a D2-brane as a complete intersection of two linear hyperplanes $h_1$ and $h_2$ with the quintic equation. This is a straight-forward generalization of the construction of the D4-brane above. We give the details in appendix \ref{app-grr}. The corresponding GLSM brane is not grade restricted. Grade restriction of such a brane is already quite involved and the algorithm described above turns out to be very useful. The central charge is five times the charge of the permutation-type matrix factorization. 

Finally we come to the GLSM lift $\mathcal{W}_\mathrm{pt}$ of the D0-brane $\mathcal{O}_{\mathrm{pt}}$. The corresponding hemisphere partition function is produced by the Koszul brane $K( (f_1,x_3,x_4,x_5) , \mathcal{W}(0)_+)$ which is automatically grade-restricted:
\begin{equation}
\label{gropt}
\mathcal{W}_{\mathrm{pt}}:\quad \xymatrix{\mathcal{W}(0)_+\ar@<2pt>[r] & \ar@<2pt>[l]\mathcal{W}(1)^{\oplus 4}_-\ar@<2pt>[r] & \ar@<2pt>[l]\mathcal{W}(2)^{\oplus 6}_+\ar@<2pt>[r] & \ar@<2pt>[l]\mathcal{W}(3)^{\oplus 4}_-\ar@<2pt>[r] & \ar@<2pt>[l]\mathcal{W}(4)_+}.
\end{equation}
The corresponding matrix factorization is a permutation-type factorization of the Fermat quintic:
\begin{equation}
Q=f_1\eta_1+x_3\eta_2+x_4\eta_3+x_5\eta_4+pg_1\bar{\eta}_1+px_3^4\bar{\eta}_2+px_4^4\bar{\eta}_3+px_5^4\bar{\eta}_4,
\end{equation}
where $f_1$ and $g_1$ are the same as in (\ref{permf1def}) and (\ref{permg1def}). The brane factor is
\begin{equation}
f_{\mathcal{W}_{\mathrm{pt}}}=1-4e^{2\pi\sigma}+6e^{4\pi\sigma}-4e^{6\pi\sigma}+e^{8\pi\sigma}=(1-e^{2\pi\sigma})^4.
\end{equation}
With that one easily confirms
\begin{equation}
\label{quinticd0}
Z_{D^2}(\mathcal{W}_{\mathrm{pt}}) =\frac{\sqrt{5}}{4\pi^2}y_{1}^{\ast},
\end{equation}
which we will also analytically continue to the conifold point in section \ref{sec-math}. Let us add a few extra comments. The hemisphere partition function for this brane, evaluated at the large radius limit, is the same for every integer shift of the GLSM brane. This reflects the fact that the D0-brane is invariant under large radius monodromy. Alternatively, we could also construct a D0-brane an an intersection of three linear hyperplanes $h_1,h_2,h_3$ with the generic quintic which we present in appendix \ref{app-grr}. The central charge in the large radius phase is five times the one of the permutation brane. This is consistent with the fact that such a construction actually describes five points on the quintic and not a single one.

This concludes our discussion on the analytic continuation of a specific basis of large radius branes to the conifold point. A completely analogous analysis for the quartic K3 can be found in appendix \ref{app-grr}.
\subsection{Hemisphere Partition Function and Monodromy}
After the preceding discussion, computing monodromy matrices with the help of the hemisphere partition function is almost trivial. It can be done without referring to a particular phase, i.e. without choosing a reference point. Consider a GLSM brane $\mathcal{B}$ and a set of branes such as $\mathcal{W}=(\mathcal{W}_X, \mathcal{W}_H, \mathcal{W}_l, \mathcal{W}_{\mathrm{pt}})$ in (\ref{grox}), (\ref{groh}), (\ref{grol}) and (\ref{gropt}), which is obtained from a lift of a basis of $K(X)$. By a slight abuse of language, we also call $\mathcal{W}$ a basis. We assume that $\mathcal{B}$ and $\mathcal{W}$ are grade restricted to a specific charge window such that the brane factor $f_{\mathcal{B}}$ can be expressed in terms of linear combinations of the brane factors of the basis $\mathcal{W}$. The monodromy operation generates a new brane $\tilde{\mathcal{B}}$, which is typically not grade restricted to the selected window anymore. After grade restricting back to the original window we get a brane $\mathcal{B}'$ whose brane factor $f_{\mathcal{B}'}$ can be expressed in terms of the brane factors of the basis $\mathcal{W}$ and we can read off the monodromy. Below, we will apply this procedure to the basis $\mathcal{W}$ of branes on the quintic which will lead to relations $Z_{D^2}(\mathcal{W}')=M Z_{D^2}(\mathcal{W})$, where $M$ is the monodromy matrix.

\subsubsection{Large Radius Monodromy}
The monodromy around the large radius point corresponds to a shift in the $\theta$-angle $\theta\rightarrow\theta+ 2\pi$ which affects the term $e^{it\sigma}$ in the hemisphere partition function: $e^{it\sigma}\rightarrow e^{it\sigma}e^{2\pi\sigma}$. This shift has to be absorbed in the brane factor of the new brane $\tilde{\mathcal{B}}$ such that under large radius monodromy:
\begin{align}
f_{\mathcal{B}}\rightarrow f_{\tilde{\mathcal{B}}}=e^{2\pi\sigma}f_{\mathcal{B}}.
\end{align}
The additional factor can be interpreted as a change in the overall normalization of $\rho$ in (\ref{gdef}), whereas the matrix factorization $Q$ remains unchanged. Writing the GLSM brane $\mathcal{B}$ in terms of Wilson line branes $\mathcal{W}(q_i)_{r_j}$ the large radius monodromy amounts to a global shift where the gauge charge of each constituent brane is shifted by $+1$: $\mathcal{W}(q_i)_{r_j}\rightarrow\mathcal{W}(q_i+1)_{r_j}$, i.e. we tensor the whole GLSM brane with $\mathcal{W}(1)$. In case we consider a brane $\mathcal{B}$ grade restricted to a window associated to the chamber $T_n$, the brane $\tilde{\mathcal{B}}$ after large radius monodromy will be grade restricted to a window associated to the chamber $T_{n+1}$. To extract the monodromy matrix we grade restrict $\tilde{\mathcal{B}}$ back to $T_n$ using the empty brane in the large radius phase (\ref{lr-empty}). This yields $\mathcal{B}'$ whose brane factor we can express in terms of the basis $\mathcal{W}$.

Applying this procedure to our basis $\mathcal{W}$ we can read off the large radius monodromy matrix $M_{LR}$ from the relation $Z_{D^2}(\mathcal{W}')=M_{LR}Z_{D^2}(\mathcal{W})$. We obtain the following result:
\begin{align}
 \left(\begin{array}{c}
{Z}_{D^2}(\mathcal{W}'_\mathrm{pt})\\
{Z}_{D^2}(\mathcal{W}'_{l})\\
{Z}_{D^2}(\mathcal{W}'_{H})\\
{Z}_{D^2}(\mathcal{W}'_{X})\\
\end{array}\right)=\left(\begin{array}{cccc}1&0&0&0\\1&1&0&0\\0&5&1&0\\0&5&1&1\end{array}\right)\left(\begin{array}{c}
Z_{D^2}(\mathcal{W}_\mathrm{pt})\\
Z_{D^2}(\mathcal{W}_{l})\\
Z_{D^2}(\mathcal{W}_{H})\\
Z_{D^2}(\mathcal{W}_{X})\\
\end{array}\right).
\end{align}
This relation can be read off simply by comparing brane factors, without evaluating the hemisphere partition function in a phase. We could, however, also compare the central charges in the large radius phase $Z_{D^2}^{\zeta\gg0}(\mathcal{W}')=M_{LR}Z_{D^2}^{\zeta\gg0}(\mathcal{W})$, which gives the same result. This amounts to choosing a reference point in the large radius phase. Grade restriction is not an issue then because we do not cross a phases boundary. Therefore we get the same result, irrespective of whether we perform the monodromy operation on the grade restricted or non-grade-restricted GLSM branes, which simplifies the calculation.
\subsubsection{Conifold Monodromy}
Conifold monodromies of GLSM branes have been discussed in \cite{Herbst:2008jq}. Without loss of generality we assume that we start in the large radius phase. We would like to transport a given brane in a loop around the conifold point which, in our examples, is at $\zeta={N^N}$ and $\theta=\pi\mod 2\pi$. Computing the conifold monodromy matrix $M_C$ is a three-step process. First we choose a brane $\mathcal{B}$ and a basis $\mathcal{W}$ and grade restrict it to a given charge window. In contrast to the large radius monodromy this is strictly necessary since the path inevitably crosses a phase boundary. Grade restriction selects a path within a specific chamber $T_n$ corresponding to $\theta$-angle interval of width $2\pi$ between two specific copies of the conifold singularity. This makes sure that the brane is well-defined along the full length of this path, all the way down to the Landau-Ginzburg point. The next step is to bring the brane back along a path corresponding to the adjacent charge window such that the conifold singularity in encircled. Since the ``turning-point'' of the path has to be in the Landau-Ginzburg phase in order for the conifold point to be encircled, we have to grade restrict to the adjacent window using the empty brane (\ref{emptylg}) associated to the Landau-Ginzburg phase. The resulting brane is $\mathcal{\tilde{B}}$. In order to make the comparison, the last step is to grade restrict back to the original window. This yields $\mathcal{B}'$.

Let us apply this to our choice of basis $\mathcal{W}$ for the quintic which is grade restricted with respect to the charge window $q\in \{0,1,2,3,4\}$ corresponding to the chamber $T_{-5}=(-5\pi,-3\pi)$. In the second step we use the empty brane (\ref{emptylg}) to grade restrict to the adjacent window $q\in\{1,2,3,4,5\}$. Since this corresponds to the chamber  $T_{-6}=(-7\pi,-5\pi)$, we have $\theta\rightarrow\theta-2\pi$. This amounts to choosing a clockwise path around the conifold point. This is not a problem, but we have to be aware that we compute the inverse $M_C^{-1}$ of the conifold monodromy matrix. Due to the simple structure of the empty brane (\ref{emptylg}), step two simply amounts to replacing any $\mathcal{W}(0)_{\pm}$ by a $\mathcal{W}(5)_{\pm}$. This yields $\tilde{\mathcal{W}}$. In the last step we have to use the empty brane (\ref{lr-empty}) of the large radius phase to grade restrict back to the original window. This yields $\mathcal{W}'$. By comparing the respective brane factors we arrive at a relation $Z_{D^2}(\mathcal{W}')=M^{-1}_{C}Z_{D^2}(\mathcal{W})$ with
\begin{align}
\left(\begin{array}{c}
{Z}_{D^2}(\mathcal{W}'_\mathrm{pt})\\
{Z}_{D^2}(\mathcal{W}'_{l})\\
{Z}_{D^2}(\mathcal{W}'_{H})\\
{Z}_{D^2}(\mathcal{W}'_{X})\\
\end{array}\right)=\left(\begin{array}{cccc}1&0&0&1\\0&1&0&0\\0&0&1&5\\0&0&0&1\end{array}\right)\left(\begin{array}{c}
Z_{D^2}(\mathcal{W}_\mathrm{pt})\\
Z_{D^2}(\mathcal{W}_{l})\\
Z_{D^2}(\mathcal{W}_{H})\\
Z_{D^2}(\mathcal{W}_{X})\\
\end{array}\right).
\end{align}
Since we have not evaluated to hemisphere partition function in any phase, the choice of a reference point does not matter. We could have started with the same basis $\mathcal{W}$ and could have assumed a reference point in the Landau-Ginzburg phase. In the second step one would then have to use the empty brane in the large radius phase (\ref{lr-empty}) to grade restrict to the window $q\in\{1,2,3,4,5\}$. For the third step we are back in the Landau-Ginzburg phase and therefore have to use the corresponding empty brane (\ref{emptylg}) to grade restrict to the original window. Since this describes a counter-clockwise path around the singular point, this procedure will compute $M_C$, rather than its inverse.

Depending on the reference point we could also obtain $M_C$ or its inverse by comparing central charges in the phases. In this case one can omit the third step. In the Landau-Ginzburg phase this should reproduce known results for the monodromy of Landau-Ginzburg branes \cite{Jockers:2006sm,Brunner:2008fa}.
\subsubsection{Landau-Ginzburg Monodromy}
To summarize, we have computed the following monodromy matrices:
\begin{align}
M_C=\left(\begin{array}{rrrr}1&0&0&-1\\ 0&1&0&0 \\ 0&0&1&-5 \\ 0&0&0&1\end{array}\right)\qquad
M_{LR}=\left(\begin{array}{cccc}1&0&0&0\\1&1&0&0\\0&5&1&0\\0&5&1&1\end{array}\right).
\end{align}
Using $M_{LG}\cdot M_{LR}\cdot M_C=1$, the monodromy matrix around the Landau-Ginzburg point is
\begin{align}
M_{LG}=M_C^{-1}\cdot M_{LR}^{-1} =\left(\begin{array}{rrrr} 1&0&-1&1 \\ -1&1&0&0 \\ 5&-5&-4&5 \\ 0&0&-1&1 \end{array}\right).
\end{align}
One can check that $M_{LG}^5=1$, as expected. Our results are in agreement with \cite{Candelas:1990rm}.
\section{Analytic Continuation to the Conifold Point}
\label{sec-math}
In this section we study analytic properties of the hemisphere partition function. We present various methods to analytically continue this function to the conifold point. We would like to emphasize that grade restriction in the GLSM is crucial to make contact with the discussion in this section. The proofs of the results can be found in general textbooks such as~\cite{Erdelyi:1953ab,Slater:1966ab} and the articles~\cite{Norlund:1955ab, Buehring:1992ab, Scheidegger:2016ab}. These references show that the methods presented here are can be applied in a more general context.

Recall from section~\ref{sec:mellin-barnes} that the hemisphere partition function for branes in the GLSM associated to CY hypersurfaces of degree $N$ in $\mathbb{P}^{N-1}$  satisfies a generalized hypergeometric differential equation of order $n=N-1$ after a change of variables. Such a differential equation takes the general form
\begin{equation}
\label{hypeq}
 \left( \theta\, \prod_{j=1}^{n-1} \left( \theta - \gamma_j\right) - z \prod_{j=1}^n \left(\theta - \alpha_j\right) \right) y(z) = 0 ,
\end{equation}
where $\alpha_1,\dots,\alpha_n,\gamma_1,\dots,\gamma_{n-1} \in \mathbb{C}$ and $\theta = z\frac{\diff{}{}}{\diff{}{z}}$. For convenience we also introduce a $\gamma_n\equiv 0$.
For the case of one-parameter Calabi-Yau hypersurfaces in $\mathbb{P}^{n}$ we see from~\eqref{eq:11} that
\begin{equation}
  \label{eq:14}
  \alpha_i=\frac{i}{n+1},\qquad \gamma_i=0,\qquad i=1,\ldots,n\\
\end{equation}
This differential equation has regular singularities at $z=0,1,\infty$ with exponents
\begin{equation}
    \begin{array}[]{c|cccccc}
    z= 0 & \gamma_1 & \gamma_2 & \gamma_3 & \dots & \gamma_{n-1} & 0\\
    z= 1 & 0 & 1 & 2 & \dots & n-2 & \beta_n\\
    z = \infty & \alpha_1 & \alpha_2 & \alpha_3 & \dots & \alpha_{n-1}
    & \alpha_n
  \end{array}
\end{equation}
where
\begin{equation}
\beta_n = n - 1 - \sum_{i=1}^n (\alpha_i + \gamma_i).
\end{equation}
For $q\in \{2,\dots,n\}$ let $E$ be a maximal subset of exponents $\{\gamma_{i_1},\dots,\gamma_{i_q}\}$ at $z=0$ with the property that $\lambda
  - \mu \in \mZ$ for all $\lambda,\mu \in E$. Then $E$ is called to be
  in resonance or resonant. Most results in the mathematics literature on analytic continuation of solutions to the hypergeometric differential equation only deal with the non-resonant case. For CY hypersurfaces however, by~\eqref{eq:14}, the whole set of exponents $\gamma_i$ is in resonance. This is the case we are interested in, and we will frequently state the results only for this case.
Before we discuss the solutions to~\eqref{hypeq} in the resonant case in more detail, we review the Frobenius method to solve ordinary linear differential equations. The most compact way to present this method is to consider the equivalent system of $n$ first order equations
\begin{equation}
  \label{eq:15}
  \theta\, Y(z) = A(z) Y(z)
\end{equation}
where $Y(z) = (y(z),\theta y(z),\dots,\theta^{n-1} (z))^t$ and $A(z)$ is an $n\times n$ matrix of holomorphic functions. For~\eqref{hypeq} the matrix $A$ becomes
\begin{equation}
  A(z) =
  \begin{pmatrix}
    0 & 1 & 0  & \dots & 0\\
    0 & 0 & 1  & \dots & 0\\
    \vdots & \vdots & \vdots & \ddots & \vdots \\
    0 & 0 & 0 & \dots & 1\\
    a_{n1} & a_{n2} & a_{n3} & \dots & a_{nn} \\
  \end{pmatrix}
\end{equation}
where
\begin{equation}
  a_{ni} = \frac{e_{n+1-i}(\gamma_1,\dots,\gamma_{n-1}) - z\,
    e_{n+1-i}(\alpha_1,\dots,\alpha_n)}{1-z},
\end{equation}
and $e_i(x_1,\dots,x_k)$ is the elementary symmetric polynomial of degree $i$. Since there are $n$ linearly independent solutions $y_i(z)$ we have $n$ linearly independent solution vectors $Y_i(z)$ which form the columns of the fundamental matrix of~\eqref{eq:15}:  $\Phi(z) =
\begin{pmatrix}
  Y_1(z) & \dots & Y_n(z)
\end{pmatrix}
$. The matrix $\Phi(z)$ is only determined up to multiplication by an element in $GL(n,\mC)$.

Now the general solution of~\eqref{eq:15} can be described as follows. There are a constant $n\times n$ matrix $R$ and a single-valued, holomorphic $n \times n$ matrix $S(z)$ such that $\Phi(z) = z^R S(z)$. Under certain conditions on the eigenvalues of $A$, $R$ can be taken to be $A(0)$.

While everything so far has been explained in a neighborhood of $z=0$, we can of course change the variable in the differential equation to go to a neighborhood of any other point in $\mathbb{P}^1$. Of particular interest are the other two regular singularities, $1$ and $\infty$. We denote fundamental matrices near $z=z_i$ by $\Phi_i(w_i) = w_i^{R_i} S_i(w_i)$ for $i=0,1,\infty$, where $w_0 = z, w_1 = 1-z, w_\infty = \frac{1}{z}$. Analytic continuation of the solutions from $w_i$ to $w_j$ then implies the existence of constant matrices $M_{ij}$ such that $\Phi_i(w_i) = \Phi_j(w_j(w_i))M_{ji}$. The goal here is to determine the matrices $M_{ij}$ given the fundamental matrices in some basis.

For the hypergeometric differential equation the matrices $M_{ij}$ have been determined in the literature for all $i,j$ in the non-resonant case. In the resonant case, the matrices $M_{ij}$ are known for all $i,j$ if $n=2$. For $n>2$ only the matrix $M_{0\infty}$ has been fully determined determined so far.

The main idea to determine these matrices is to use the fact that the solutions of (\ref{hypeq}) have a Mellin--Barnes integral representation $y_j^*(z)$ given by
\begin{align}
\label{yjdef}
y_j^*(z)=\int \diff{}{t} \, z^t \prod_{k=1}^n
  \frac{\Gamma(\alpha_k+t)}{\Gamma(1-\gamma_k+t)}  \frac{1}{\left(1-\e{2\pi
      i(t-\gamma_1)}\right)^j} \qquad j=1,\dots,n.
\end{align}
 Closing the contour to the right and invoking the residue theorem, yields a fundamental matrix whose entries are a polynomial in $\log z$ and a power series in $z$. By comparing coefficients, it is straightforward to relate the basis of Mellin-Barnes integrals to the fundamental matrix $\Phi(z)$ obtained from the Frobenius method. In the examples of our interest, these relations are given in appendix~\ref{app-pf}. Since we can equally well close the contour to the left, we obtain a basis of solutions near $z=\infty$, since the residue theorem now yields a series in $\frac{1}{z}$. At the same time, this computation yields the change of basis to the solutions near $z=0$, i.e. it yields the matrix $M_{0\infty}$.

This method does not work so straightforwardly when going from $z=0$ to $z=1$. A solution to this problem is due to N\o rlund~\cite{Norlund:1955ab} in the case where no exponents are in resonance. The generalization of his method to the resonant case has recently been developed in~\cite{Scheidegger:2016ab}. We will review the results of this method in the remainder of this section.

For this purpose, we first consider different types of solutions to~\eqref{hypeq} and express them in terms of (\ref{yjdef}). The first type is the (generalized) hypergeometric function ${}_nF_{n-1}(z)$
\begin{equation}
  \label{eq:nFn-1}
  \begin{aligned}
    y_1^*(z) &= \int \frac{\diff{}{t}}{2\pi i}\, \e{-i\pi t} z^t
    \prod_{j=1}^n \frac{\Gamma(\alpha_j+t)}{\Gamma(1-\gamma_j+t)}
    \Gamma(-t)\Gamma(1+t) \\
    &= \prod_{j=1}^n
  \frac{\Gamma(\alpha_j+\gamma_1)}{\Gamma(1-\gamma_j+\gamma_1)} {}_nF_{n-1}\left(\genfrac{}{}{0pt}{}{\alpha_1,\dots,
      \alpha_n}{1-\gamma_1, \dots, 1-\gamma_{n-1}};z\right)
  \end{aligned}
\end{equation}
where we have set $\gamma_n=0$. This is a holomorphic function for $|\arg z| < \pi$. In particular it does not need to be holomorphic at $z=1$.

The second type of solutions to~\eqref{hypeq} is given in terms of the Meijer G--function~\cite{Erdelyi:1953ab}:
\begin{equation}
\label{eq:meijerg}
 G^{p,n}_{n,n}\left(\genfrac{}{}{0pt}{}{1-\alpha_1,\dots,
      1-\alpha_n}{\gamma_1, \dots, \gamma_{n}};z\right) =
  \int \frac{\diff{}{t}}{2\pi i} z^t \prod_{j=1}^n
  \frac{\Gamma(\alpha_j+t)}{\Gamma(1-\gamma_j+t)} \prod_{h=1}^p \Gamma(\gamma_h-t)\Gamma(1-\gamma_h+t).
\end{equation}
for $1 \leq p \leq n$. This integral converges for $|\arg z| < p\,\pi$. We introduce the shorthand notation
\begin{equation}
  \label{eq:Gp}
 G_p(z) :=  G_p\left(\genfrac{}{}{0pt}{}{\alpha_1,\dots,
      \alpha_n}{\gamma_1, \dots, \gamma_{n}};z\right) := G^{p,n}_{n,n}\left(\genfrac{}{}{0pt}{}{1-\alpha_1,\dots,
      1-\alpha_n}{\gamma_1, \dots, \gamma_{n}};(-1)^{p-2}z\right).
\end{equation}
In the case that the whole set of exponents $\gamma_i$ is in resonance, the $G_p(z), 1 \leq p \leq n$, form a basis of solutions to~\eqref{hypeq}~\cite{Scheidegger:2016ab}. Moreover, they are related to the basis (\ref{yjdef}) by
\begin{equation}
\label{meijer-yrel}
    G_p(z) = \e{-2\pi i \gamma_1}\e{-i\pi\sum_{j=1}^p\gamma_{j}} (2\pi i)^{p-1} \sum_{j=1}^{p} (-1)^{p-j} \binom{p-1}{p-j} y_j^*(z)
\end{equation}
in particular, $G_1(z) = \e{-i\pi\gamma_1} y_1^*(z)$. The essential property of the $G_p(z)$ is that the convergence condition entails that for $p>1$ they define holomorphic functions in a neighborhood around $z=1$. In this neighborhood they therefore yield $n-1$ linearly independent solutions to~\eqref{hypeq}. In fact, these solutions correspond to the exponents $0, 1, \dots, n-2$.

This brings us to the third type of solutions, namely the special solution near $z=1$ corresponding to the remaining exponent $\beta_n$:
\begin{align}
\xi_n(z) := \xi_n\left(\genfrac{}{}{0pt}{}{\alpha_1,\dots,
    \alpha_n}{\gamma_1, \dots, \gamma_{n}};z\right) := z^{\gamma_1}
(1-z)^{\beta_n} \sum_{k=0}^\infty c_k\,(1-z)^k.
\end{align}
The coefficients $c_k$ are determined recursively by the differential
operator. Explicit formulas can be found in~\cite{Norlund:1955ab}. In this reference, also the analytic continuation of $\xi_n(z)$ to a neighborhood of $z=0$ is determined. Again assuming that the whole set of exponents $\gamma_i$ is in resonance, we have
\begin{equation}
   \label{eq:xi_n}
   \xi_n(z) = \frac{\Gamma(\beta_n+1)}{2\pi i} \sum_{j=1}^q
   \frac{(-1)^j}{(q-j)!} \psi^{(q-j)}(\e{2\pi i \gamma_1}) \e{-2\pi i j \gamma_1} y_j^*(z) ,
  \end{equation}
where
\begin{align}
    \psi(x) = \e{-i\pi \beta_n} \frac{\prod_{k=1}^n (x-\e{-2\pi i
        \alpha_k})}{\prod_{k=q+1}^n (x-\e{2\pi i \gamma_k})} .
\end{align}
If in $\xi_n(z)$ we interchange $\alpha_j$ and $\gamma_j$, $j=1,\dots,n$,
and replace $z$ by $\frac{1}{z}$, we obtain a solution which we denote by $\bar \xi_n(z)$ and differs from $\xi_n(z)$ by a factor $\e{\pm i\pi \beta_n}$. This solution will play an auxiliary role in the discussion of $G_p(z)$ in Section~\ref{sec:Norlund}. 

Returning to the problem of the analytic continuation from $z=0$ to $z=1$, the fact that the $G_p(z)$, $p>1$, are holomorphic at $z=1$, together with~\eqref{eq:xi_n} already gives a partial answer. It remains to determine the series expansions of the $G_p(z)$, $p>1$, at $z=1$, as well as to compute the analytic continuation of $G_1(z)$ to $z=1$. The solution to the latter problem by B\"uhring in~\cite{Buehring:1992ab} will be reviewed in Section~\ref{sec:buehring}. The former has been recently solved in~\cite{Scheidegger:2016ab}. This will be discussed in Section~\ref{sec:Norlund}.

Before we go into details, we explain the strategy to obtain these results. The idea is to find recurrences for all the solutions. The integral representation of a solution to an order $n$ equation is written in terms of an integral representation of a solution to an
order $n-1$ equation. By repeated application of these recurrences, the problem of analytic
continuation is reduced to solutions of an order $2$ equation. As mentioned above, the order $2$ case is completely understood. In fact, we have
\begin{equation}
  \label{eq:2f1cont}
\frac{\Gamma(a)\Gamma(b)}{\Gamma(c)} {}_2F_1 \left(\genfrac{}{}{0pt}{0}{a,b}{c};z\right) = \frac{1}{\Gamma(c-a)\Gamma(c-b)} G_2 \left(\genfrac{}{}{0pt}{0}{a,b}{c-a-b,0};1-z\right).
\end{equation}
We refer to~\cite{Norlund:1963ab} for details, where the function $\frac{(-1)^{c-1}\Gamma(c)}{\Gamma(a)\Gamma(b)} G_2 \left(\genfrac{}{}{0pt}{0}{a,b}{1-c,0};z\right)$ was denoted $g(a,b,c;z)$.
\subsection{Holomorphic Solution: B\"uhring's Method}
\label{sec:buehring}
For the first type of solutions, the (generalized) hypergeometric function ${}_{n}F_{n-1}$, B\"uhring proved in~\cite{Buehring:1992ab} the following recurrence (which we state here in terms of Mellin--Barnes integrals instead of power series~\cite{Scheidegger:2016ab}):
\begin{align}
\label{buehring}
  \begin{aligned}
& \,{}_n F_{n-1}\left(\genfrac{}{}{0pt}{}{\alpha_1,\dots, \alpha_n}{1-\gamma_1, \dots, 1-\gamma_{n-1}};z\right) \\
& = \frac{\Gamma(1-\gamma_{n-2})\Gamma(1-\gamma_{n-1})}{\Gamma(\alpha_n)\Gamma(1-\gamma_{n-1}-\alpha_{n})\Gamma(1-\gamma_{n-2}-\alpha_{n})}\\
& \phantom{=} \cdot \int \frac{\diff{}{t}}{2\pi i} \,\e{\pm\pi i t}\frac{\Gamma(-t)\Gamma(1-\gamma_{n-1}-\alpha_{n}+t)\Gamma(1-\gamma_{n-2}-\alpha_{n}+t)}{\Gamma(2-\gamma_{n-1}-\gamma_{n-2} - \alpha_n + t)}
 \\
& \phantom{=} \cdot \,{}_{n-1} F_{n-2}\left(\genfrac{}{}{0pt}{}{\alpha_1,\dots, \alpha_{n-1}}{1-\gamma_1, \dots, 1-\gamma_{n-3}, 2-\gamma_{n-1}-\gamma_{n-2} - \alpha_n + t};z\right)
\end{aligned}
\end{align}
By repeated application of this recurrence one can express ${}_nF_{n-1}$ in terms of ${}_2F_1$ and then use (\ref{eq:2f1cont}). The series expansion at $z=1$ is then obtained by evaluating the contour integrals with the residue theorem. The closure of the contour is determined by the convergence of the integral.

In $\beta_n\not \in\mZ$ (as for instance for the quartic) B\"uhring finds the following series expansion:
\begin{equation}
  \label{eq:Buehring_noninteger}
    \begin{aligned}
      &
      \frac{\Gamma(\alpha_1)\cdots\Gamma(\alpha_n)}{\Gamma(1-\gamma_1) \cdots\Gamma(1-\gamma_{n-1})}  \,{}_nF_{n-1} \left(\genfrac{}{}{0pt}{}{\alpha_1,\dots,\alpha_n}{1-\gamma_1,\dots,
          1-\gamma_{n-1}};z\right) \\
      & =\sum_{m=0}^\infty g_m(0)(1-z)^m + (1-z)^{\beta_n} \sum_{m=0}^\infty
      g_m(\beta_n)(1-z)^m
    \end{aligned}
  \end{equation}
  where
  \begin{equation}
  \label{eq:gm_ell}
  \begin{aligned}
    g_m(\ell) &= (-1)^m  \frac{\Gamma(\alpha_1+\ell+m)\Gamma(\alpha_2+\ell+m)\Gamma(\beta_n-2\ell-m)}{\Gamma(\beta_n+\alpha_1)\Gamma(\beta_n+\alpha_2)\Gamma(m+1)}\\
    &\phantom{=} \cdot \sum_{k=0}^\infty
    \frac{(\beta_n-\ell-m)_k}{(\alpha_1+\beta_n)_k(\alpha_2+\beta_n)_k}A^{(n)}(k)
  \end{aligned}
\end{equation}
with
\begin{equation}
\label{eq:An_k}
\begin{aligned}
A^{(n)}(k)=&\sum_{k_2=0}^k\frac{(n-1-\gamma_n-\gamma_{n-1}+\ldots-\gamma_2-\alpha_{n+1}-\alpha_n-\ldots -\alpha_3+k_2)_{k-k_2}(1-\gamma_1-\alpha_3)_{k-k_2}}{(k-k_2)!}\\
&\cdot \sum_{k_3=0}^{k_2}\frac{(n-2-\gamma_n-\gamma_{n-1}+\ldots-\gamma_3-\alpha_{n+1}-\alpha_n-\ldots-\alpha_4+k_3)_{k_2-k_3}(1-\gamma_2-\alpha_4)_{k_2-k_3}}{(k_2-k_3)!}\\
&\cdot \ldots\\
&\cdot \sum_{k_{n-1}=0}^{k_{n-2}}\frac{(2-\gamma_n-\gamma_{n-1}-\alpha_{n+1}-\alpha_n+k_{n-1})_{k_{n-2}-k_{n-1}}(1-\gamma_{n-2}-\alpha_n)_{k_{n-2}-k_{n-1}}}{(k_{n-2}-k_{n-1})!}\\
&\cdot\frac{(1-\gamma_n-\alpha_{n+1})_{k_{n-1}}(1-\gamma_{n-1}-\alpha_{n+1})_{k_{n-1}}}{k_{n-1}!}
\end{aligned}
\end{equation}
and where $(\ldots)_n$ is the Pochhammer symbol. The series in $g_m(\ell)$ terminates when $\ell =\beta_n$, while for $\ell=0$ we need the conditions $\Re(\alpha_j+m) > 0$, $j=3,\dots,n$, for convergence.

For the case $\beta_n\in\mathbb{Z}$ (corresponding to the quintic in our examples) the analytic continuation formula for the holomorphic solution reads
\begin{equation}
  \label{eq:Buehring_integer}
  \begin{aligned}
    & \frac{\Gamma(\alpha_1)\cdots\Gamma(\alpha_n)}{\Gamma(1-\gamma_1)\cdots\Gamma(1-\gamma_{n-1})}
    \,{}_n F_{n-1}\left(\genfrac{}{}{0pt}{}{\alpha_1,\dots,\alpha_n}{1-\gamma_1,\dots,
        1-\gamma_{n-1}};z\right) \\
    & =\sum_{m=0}^{\beta_n-1} l_m(1-z)^m + (1-z)^{\beta_n} \sum_{m=0}^\infty
    \left( w_m + q_m\log(1-z) \right)(1-z)^m
  \end{aligned}
\end{equation}
where $l_m=g_m(0)$, $q_m=g_m(\beta_n)$ and
\begin{equation}
  \label{eq:wm}
\begin{aligned}
 w_m&=  (-1)^{\beta_n}\frac{(\alpha_1+\beta_n)_m(\alpha_2+\beta_n)_m}{\Gamma(\beta_n+m+1)\Gamma(m+1)}\sum_{k=0}^m
        \frac{(-m)_k}{(\alpha_1+\beta_n)_k(\alpha_2+\beta_n)_k}A_n(k)\\
  &\phantom{=} \cdot \left( \psi(1+m-k) + \psi(1+\beta_n+m) - \psi(\alpha_1+\beta_n+m)
    - \psi(\alpha_2+\beta_n+m) \right)\\
  &\phantom{=} + (-1)^{\beta_n+m}\frac{(\alpha_1+\beta_n)_m(\alpha_2+\beta_n)_m}{\Gamma(\beta_n+m+1)}\sum_{k=m+1}^\infty
        \frac{\Gamma(k-m)}{(\alpha_1+\beta_n)_k(\alpha_2+\beta_n)_k}A^{(n)}(k),
\end{aligned}
\end{equation}
where $\psi$ is the digamma function.   The convergence of the series in $l_m$ requires the
conditions $\Re(\alpha_j+m) > 0$, $j=3,\dots,n$, while the convergence of
the series in $w_m$ requires the conditions $\Re(\alpha_j+\beta_n+m) > 0$, $j=3,\dots,n$.
 In section \ref{sec-hypersurface} we will explicitly apply these formulas to one-parameter Calabi-Yau hypersurfaces in $\mathbb{P}^n$.
\subsection{Logarithmic solutions: Generalization of N{\o}rlund's Method}
\label{sec:Norlund}
The analytic continuation for the solutions of the second type, i.e. those containing logarithms has recently been derived in~\cite{Scheidegger:2016ab}. There a method of analytic continuation due to N{\o}rlund~\cite{Norlund:1955ab} has been generalized to the resonant case. We refer to these references for the proofs and state here only the relevant results. One of the basic reasons why the analytic continuation can be performed is the property of the functions $G_p(z)$ in~(\ref{eq:meijerg}) mentioned above: they are holomorphic at $z=1$ for $p>1$. Moreover, they admit a rather simple series expansion at $z=1$:
\begin{equation}
  \label{eq:Gp_series}
    \begin{aligned}
    &G_p
    \left(\genfrac{}{}{0pt}{0}{\alpha_{1},\dots,\alpha_n}{\gamma_{1},\dots,\gamma_n};z\right) = z^{\gamma_q} \sum_{m=0}^\infty \frac{1}{m!} G_p
    \left(\genfrac{}{}{0pt}{0}{\alpha_{1},\dots,\alpha_n}{\gamma_{1},\dots,\gamma_{q-1},\gamma_q+m,\gamma_{q+1},\dots,\gamma_n};1\right) (1-z)^m
  \end{aligned}
\end{equation}
for any $p>1$. In order to determine the coefficients in this expansion, we again make use of a recurrence. To state it we need an auxiliary function
\begin{align}
  \widetilde G_p(z) := \widetilde
  G_p\left(\genfrac{}{}{0pt}{}{\alpha_1,\dots, \alpha_p}{\gamma_1,
      \dots, \gamma_{p}};z\right) :=
  G^{p,p}_{p,p}\left(\genfrac{}{}{0pt}{}{1-\alpha_1,\dots,
      1-\alpha_p}{\gamma_1, \dots, \gamma_{p}};(-1)^{p-2}z\right).
\end{align}
Then one can show~\cite{Scheidegger:2016ab} that for any $p\geq 1$ there exists the following integral representation of $G_p(z)$
\begin{align}
      G_p \left(\genfrac{}{}{0pt}{0}{\alpha_{1},\dots,\alpha_n}{\gamma_{1},\dots,\gamma_n};z\right) = \frac{1}{\Gamma(\beta_n-\beta_p)}
    \int_0^z \frac{\diff{}{t}}{t}\, \widetilde G_p \left(\genfrac{}{}{0pt}{0}{\alpha_{1},\dots,\alpha_p}{\gamma_{1},\dots,\gamma_p};t\right)\bar\xi_{n-p}\left(\genfrac{}{}{0pt}{0}{\alpha_{p+1},\dots,\alpha_n}{\gamma_{p+1},\dots,\gamma_n};\frac{z}{t}\right).
\end{align}
if $\Re \beta_n > \Re\beta_p$,
 $\Re(\alpha_s+\gamma_j)>0$, $j=1\dots,p$, $s=p+1,\dots,n$.
Moreover, the auxiliary function $\widetilde G_p(z)$ satisfies the recurrence for $p>2$
\begin{align}
    \begin{aligned}
      \widetilde G_p \left(\genfrac{}{}{0pt}{0}{\alpha_{1},\dots,\alpha_p}{\gamma_{1},\dots,\gamma_p};z\right) &=\Gamma(\alpha_{p-1}+\gamma_p)\Gamma(\alpha_p+\gamma_p) \\
      &\phantom{=} \cdot \int \frac{\diff{}{s}}{2 \pi i}\e{\pm \pi i
        s}\frac{\Gamma(\alpha_{p-1}+s)\Gamma(\alpha_p+s)}{\Gamma(\alpha_{p-1}+\alpha_p+\gamma_p+s)}
      \widetilde G_{p-1}\left(\genfrac{}{}{0pt}{0}{\alpha_{1},\dots,\alpha_{p-2},-s}{\gamma_{1},\dots,\gamma_{p-1}};z\right).
    \end{aligned}
\end{align}
By repeated application of this recurrence, the argument is again reduced to the case $n=2$ where $\widetilde G_2(z) = G_2(z)$ and we can apply~\eqref{eq:2f1cont}. The $t$ integral can be solved by using the integral representation of ${}_2F_1$ and special properties of integrals of $\bar \xi_{n-p}$ due to~\cite{Norlund:1955ab}. The result is then the following series expansion for $G_p(z)$, $p=2$, at $z=1$~\cite{Scheidegger:2016ab}. If $|z-1| < 1$, $\Re \beta_n > \Re\beta_p$,
 $\Re(\alpha_s+\gamma_j)>0$, $j=1\dots,p$, $s=p+1,\dots,n$, $\alpha_{p}+\gamma_p,
    \alpha_s+\gamma_{s+1} \not \in \mZ_{\leq 0}$, $s=2,\dots,p-1$ then
  \begin{align}
  \begin{aligned}
    G_2(z) &= \sum_{m=0}^\infty
  \frac{\Gamma(\alpha_1+\gamma_2+m)\Gamma(\alpha_2+\gamma_2+m)}{\Gamma(m+1)}
    \\
   &\phantom{=} \cdot \int \frac{\diff{}{v}}{2\pi i} \e{-i\pi v} \frac{\Gamma(\alpha_1+\gamma_1+v) \Gamma(\alpha_2+\gamma_1+v)  \Gamma(-v)}{\Gamma(\alpha_1+\alpha_2+\gamma_1+\gamma_2+m+v)} \\
   &\phantom{=} \cdot \int
    \frac{\diff{}{u}}{2\pi i} \e{-i\pi u}
    \frac{\Gamma(-v+u)\Gamma(-u)}{\Gamma(-v)} \prod_{s=3}^n
    \frac{\Gamma(\alpha_s+\gamma_1+u)}{\Gamma(1-\gamma_s+\gamma_1+u)} (1-z)^m.
  \end{aligned}
  \end{align}
  If $p>2$ then the expansion is slightly more involved:
  \begin{align}
    \begin{aligned}
    G_p(z) &= \sum_{m=0}^\infty \Gamma(\alpha_1+\gamma_2) \int \frac{\diff{}{v}}{2\pi i} \e{-i\pi v} \Gamma(\alpha_1+\gamma_1+v) \Gamma(-v) \\
   &\phantom{=} \cdot \int \frac{\diff{}{s}}{2 \pi i} \frac{B_{p,m}(s)}{\Gamma(m+1)} \frac{ \Gamma(\gamma_2-s) \Gamma(\gamma_1+v-s) }{\Gamma(\alpha_1+\gamma_1+\gamma_2+v-s)}\\
   &\phantom{=} \cdot \int
    \frac{\diff{}{u}}{2\pi i} \e{-i\pi u}
    \frac{\Gamma(-v+u)\Gamma(-u)}{\Gamma(-v)} \prod_{s=p+1}^n
    \frac{\Gamma(\alpha_s+\gamma_1+u)}{\Gamma(1-\gamma_s+\gamma_1+u)} (1-z)^m
  \end{aligned}
  \end{align}
  where $B_{p,m}(s) = B_p(s)|_{\gamma_p \to \gamma_p+m}$ with
  \begin{align}
  \begin{aligned}
\label{bpdef}
    B_{p}(s) &= \Gamma(\alpha_p+\gamma_p) \Gamma(\alpha_{p-1}+\gamma_p) \Gamma(\alpha_{p-2}+\gamma_{p-1})\dots\Gamma(\alpha_2+\gamma_3) \\
   &\phantom{=} \cdot\int \frac{\diff{}{s_{p-2}}}{2\pi i}\, \e{-i\pi s_{p-2}} \frac{\Gamma(\alpha_p+s_{p-2})\Gamma(\alpha_{p-1}+s_{p-2})\Gamma(\gamma_{p-1}+s_{p-2})}{\Gamma(\alpha_p+\alpha_{p-1}+\gamma_p+s_{p-2})} \\
   &\phantom{=} \cdot\int \frac{\diff{}{s_{p-3}}}{2\pi i}\, \e{-i\pi s_{p-3}} \frac{\Gamma(\alpha_{p-2}+s_{p-3})\Gamma(\gamma_{p-2}+s_{p-3})\Gamma(-s_{p-2}+s_{p-3})}{\Gamma(\alpha_{p-2}+\gamma_{p-1}-s_{p-2}+s_{p-3})} \\
   &\phantom{=}\cdot \phantom{int} \dots\\
   &\phantom{=} \cdot\int \frac{\diff{}{s_{2}}}{2\pi i}\, \e{-i\pi s_{2}} \frac{\Gamma(\alpha_{3}+s_{2})\Gamma(\gamma_{3}+s_{2})\Gamma(-s_{3}+s_{2})}{\Gamma(\alpha_{3}+\gamma_{4}-s_{3}+s_{2})} \\
  &\phantom{=} \cdot \e{-i\pi s} \frac{\Gamma(\alpha_2+s)\Gamma(-s_2+s)}{\Gamma(\alpha_2+\gamma_3-s_2+s)}.\\
  \end{aligned}
  \end{align}
These residue integrals can be evaluated by closing the integration contours in either direction. In the examples discussed in section \ref{sec-hypersurface} we will make a convenient choice. For $p=2,3$ we will end up with series expansions of the following form:
\begin{align}
 G_2(z)& = z^{\gamma_1} \sum_{m=0}^\infty h_m\, (1-z)^m\\
 G_3(z)& = z^{\gamma_1} \sum_{m=0}^\infty k_m\, (1-z)^m,
\end{align}
where we refer to \cite{Scheidegger:2016ab} for the explicit expressions of the coefficients $h_m,k_m$ as they arise after evaluating (\ref{bpdef}). 

With that, we have collected all the necessary mathematical methods to analytically continue a basis of solutions of (\ref{hypeq}) to the singular point $z=1$.

\subsection{An Alternative Approach}
\label{sec-method2}
We now present an alternative approach to the analytic continuation to the hemisphere partition function. In contrast to the previous methods this approach always leads to a double integral. However, evaluating these integrals is not always straightforward. Despite these difficulties we have successfully implemented this alternative method for several examples of branes on the cubic and the quartic.

Let us first explain the general idea. The hemisphere partition functions that we encounter in our main examples, upon grade restriction, are Mellin-Barnes integrals of the form
\begin{align}
I_{q}(z):=\int_{-i\infty}^{i\infty} ds K_{q}(s)z^{s}\qquad q=1,\ldots N-1
\end{align}
where
\begin{align}
K_{q}(s):=\frac{e^{i\pi (q-N)s}\Gamma(-s)^{N-q}\prod_{j=1}^{N-1}\Gamma(s+\frac{j}{N})}{\Gamma(s+1)^{q-1}}
\end{align}
and the contour is defined such that it passes through the left of the pole at $s=0$. In order to investigate the behavior of this integral around $z=1$ we simply use the identity (\ref{identityMB}) to shift $z\rightarrow 1-z$, i.e. we write $z^{s}$ as:
\begin{align}
z^{s}=\frac{1}{2\pi i}\frac{1}{\Gamma(-s)}\int_{-i\infty}^{i\infty}du\Gamma(u-s)\Gamma(-u)(z-1)^{u}.
\end{align}
Then the Mellin-Barnes integral becomes (up to an irrelevant constant)
\begin{align}
\label{u-integral}
I_{q}(z)\sim \int_{-i\infty}^{i\infty} du \Gamma(-u) F_{q}(u)(z-1)^{u},
\end{align}
where $F_{q}(u)$ takes the form
\begin{eqnarray}
F_{q}(u) &=& \int_{-i\infty}^{i\infty}ds\frac{e^{-i\pi (N-q)s}\Gamma(u-s)\Gamma(-s)^{N-q-1}\prod_{j=1}^{N-1}\Gamma(s+\frac{j}{N})}{\Gamma(s+1)^{q-1}}\nonumber\\
&=& G^{N-q,N-1}_{N-1,N-1}\left({1-\frac{1}{N},\ldots,1-\frac{N-1}{N} \atop u,0,\ldots,0};e^{-i\pi(N-q)}\right).
\end{eqnarray}
In principle we cannot draw any conclusions about convergence of $F_{q}(u)$. Since we are interested in the solution in the region $|z-1|\ll 1$, we can perform the integral (\ref{u-integral}) by the residue method, closing the contour to the right, provided that $\Gamma(-u)F_{q}(u)$ does not grow faster than $(z-1)^{u}$ as $|u|\rightarrow\infty$ and $|\mathrm{Arg}(u)|<\frac{\pi}{2}$. Assuming that this is the case, we would like to see what the poles of $F_{q}(u)$ in the region $\Re(u)>0$ are. In order to do this, first we notice that by repeated application of the identity
\begin{eqnarray}
\label{exp-trick}
e^{-2\pi i s}=1+\frac{2\pi i e^{-i\pi s }}{\Gamma(-s)\Gamma(s+1)}
\end{eqnarray}
we can bring $F_{q}(u)$ to the form:
\begin{align}
\label{eq:Fqdecomp}
F_{q}(u)=\sum_{L=0}^{N-q-1}C_{L}\int_{-i\infty}^{i\infty}ds\frac{e^{-i\pi \alpha_{L}s}\Gamma(u-s)\Gamma(-s)^{N-q-1-L}\prod_{j=1}^{N-1}\Gamma(s+\frac{j}{N})}{\Gamma(s+1)^{q-1+L}}=\sum_{L=0}^{N-q-1}C_{L}F^{(L)}_{q}(u)
\end{align}
where $C_{L}$ is a numerical constant, that is irrelevant for this analysis and
\begin{align}
\alpha_{L}=0\text{ \ or \ }1\qquad \alpha_{L}+L\text{ \ is even/odd if \ }N-q \text{ \ is even/odd \ }.
\end{align}
Let us analyze the convergence of the integrals $F^{(L)}_{q}(u)$. First consider the asymptotics of the integrand. Using the formula:
\begin{align}
\lim_{|y|\rightarrow\infty}|\Gamma(x+iy)|=\sqrt{2\pi}e^{-|y|\pi/2}|y|^{x-\frac{1}{2}}
\end{align}
we obtain
\begin{align}\label{integrandF}
\lim_{\Im(s)=\pm i\infty}\left[\mathrm{Integrand}(F^{(L)}_{q}(u))\right]\sim e^{\pi |\Im(s)|(\pm\alpha_{L}-N+q+L)}|\Im(s)|^{\Re(u)-\frac{N-1}{2}}.
\end{align}
Since $\pm\alpha_{L}-N+q+L\leq 0$ the integral $F^{(L)}_{q}(u)$ is absolutely convergent unless $\alpha_{L}=1$ and $L=N-q-1$. If $\pm\alpha_{L}-N+q+L< 0$ then, the exponential term dominates and the integral $F^{(L)}_{q}(u)$ is a well defined function for all $u$ such that $\Re(u)\geq 0$ since there are no poles hitting the contour. Indeed we can even take $\Re(u)< 0$ and continuously deform the contour to avoid the poles. The first pole for $\Re(u)< 0$ will be at $u=-\frac{1}{N}$. However, we are not interested in the behavior for $\Re(u)< 0$. Alternatively, we can compare $F_{q}^{(L)}(u)$ with (\ref{eq:meijerg}):
\begin{eqnarray}
F^{(L)}_{q}(u) =G^{N-q-L,N-1}_{N-1,N-1}\left({1-\frac{1}{N},\ldots,1-\frac{N-1}{N} \atop u,0,\ldots,0};e^{-i\pi\alpha_{L}}\right).
\end{eqnarray}
From this we also can conclude that $F^{(L)}_{q}(u)$ is convergent except for $L=N-q-1$ where further analysis is necessary, as we are going to do next.\\

Coming back to the case of $\Re(u)\geq 0$, we concluded that the only integral that can potentially be divergent for $u$ in this range is $F^{(L)}_{q}(u)$ with $\alpha_{L}=1$ and $L=N-q-1$. In such a case (\ref{integrandF}) shows that $F^{(L=N-q-1)}_{q}(u)$ diverges when
\begin{align}
\Re(u)\geq\frac{N-3}{2}.
\end{align}
We will now show that $F^{(L=N-q-1)}_{q}(u)$ can be analytically continued to a function with simple poles at $\Re(u)=\frac{N-3}{2}+k$ for all $k\in \mathbb{Z}_{\geq 0}$. Let us start by writing
\begin{align}
F^{(N-q-1)}_{q}(u)=\int_{-i\infty}^{i\infty}ds\frac{e^{-i\pi s}\Gamma(u-s)\prod_{j=1}^{N-1}\Gamma(s+\frac{j}{N})}{\Gamma(s+1)^{N-2}}.
\end{align}
Then, assume $0<\Re(u)<\frac{N-3}{2}$, so the integral in $s$ is convergent and perform it by taking the poles at $s=u+\mathbb{Z}_{\geq 0}$. Then we get
\begin{align}\label{specialFq}
F^{(N-q-1)}_{q}(u)=(2\pi i)\frac{e^{-i\pi u}\prod_{j=1}^{N-1}\Gamma(u+\frac{j}{N})}{\Gamma(u+1)^{N-2}}{}_{N-1}F_{N-2}\left({\frac{1}{N}+u,\ldots,\frac{N-1}{N}+u\atop 1+u,\ldots,1+u};1\right).
\end{align}
Note that the hypergeometric function appearing in (\ref{specialFq}) has balance $-u+\frac{N-3}{2}$ and hence satisfies all the assumptions of Theorem 1 of~\cite{Buehring:1992ab}. Therefore we can write
\begin{align}\label{specialFq2}
F^{(N-q-1)}_{q}(u)=(2\pi i)\frac{e^{-i\pi u}\Gamma(\frac{1}{N}+u)\Gamma(\frac{2}{N}+u)\Gamma(-u+\frac{N-3}{2})}{\Gamma(\frac{N-3}{2}+\frac{1}{N})\Gamma(\frac{N-3}{2}+\frac{2}{N})}
\sum_{k=0}^{\infty}\frac{(-u+\frac{N-3}{2})_{k}}{(\frac{N-3}{2}+\frac{1}{N})_{k}(\frac{N-3}{2}+\frac{2}{N})_{k}}A^{(N-2)}(k)
\end{align}
where the coefficients $A^{(N-2)}(k)$ are defined in~\eqref{eq:An_k}. Note that the series in (\ref{specialFq2}) is convergent, according to theorem 2 in~\cite{Buehring:1992ab} provided that $\Re(u)+\frac{j}{N}+k>0$ for $j=3,\ldots,N-1$ and for all $k\in\mathbb{Z}_{\geq 0}$. This is clearly satisfied in our case. Moreover, equation (\ref{specialFq2}) is precisely equation (4.6) in \cite{Norlund:1955ab}. In \cite{Norlund:1955ab}, there are further remarks regarding these expressions. The most important is that (\ref{specialFq2}) provides an analytic continuation of the left-hand side from $0<\Re(u)<\frac{N-3}{2}$ to $u\in \mathbb{C}\setminus \{\frac{N-3}{2}+k\}_{k\in\mathbb{Z}_{\geq 0}}$, since the right-hand side of (\ref{specialFq2}) is defined in this range of $u$, with first order poles at $u\in \{\frac{N-3}{2}+k\}_{k\in\mathbb{Z}_{\geq 0}}$. This concludes our analysis of $F^{(N-q-1)}_{q}(u)$. Before we end this section, we give a list of interesting and useful consequences of the properties of the functions $F_{q}(u)$:
\begin{itemize}
  \item The odd/even properties of $\alpha_{L}$ and $L$ implies that the functions $F^{(L)}_{q}(u)$ are always expressed as linear combinations of generalized hypergeometric functions of type ${}_{N-1}F_{N-2}$ at unit argument and their derivatives with respect to its parameters (not with respect to the argument).
  \item We argued that only $F^{(N-q-1)}_{q}(u)$ has poles in the region $\Re(u)\geq 0$. Moreover we argued that these poles are simple and located at $u\in\{\frac{N-3}{3}+k\}_{k\in\mathbb{Z}_{\geq 0}}$. As a consequence, if $\{\Phi_{1,a}(y)\}$ is a basis of solutions for the Picard-Fuchs equation around the conifold point $y=1-z$, then if $N$ is even, none of the solutions contain a $\log y$ term but they may have fractional powers of $y$. If $N$ is odd, they contain at most one power $\log y$ and no fractional powers of $y$. The same conclusion can be reached by looking at the matrix of exponents of the Picard-Fuchs operator at $y$.
  \item We will see that, if we write $F^{(L)}_{q}(u)$ in terms of hypergeometric functions, it is not apparent that $F^{(L)}_{q}(k)$ are finite, leading to interesting relations between hypergeometric functions and their derivatives.
\end{itemize}
\section{Application to Calabi-Yau Hypersurfaces in $\mathbb{P}^N$}
\label{sec-hypersurface}
In this section we apply our results of analytic continuation to CY hypersurfaces in $\mathbb{P}^N$ for $N=3,4,5$. By making a connection with the discussion of GLSM branes in section \ref{sec-branes}, we are able to study the behavior of D-branes near the conifold point. For convenience, we introduce the variable $y=1-z$ so that the conifold point is at $y=0$. The series expansions of the solutions of the hypergeometric differential equations around the conifold point are given in appendix \ref{app-pf}.
\subsection{The Cubic}
\subsubsection{Method 1}
The analytic continuation of the a D0 and a D2 brane to the conifold point is almost trivial. In (\ref{cubicd0}) we have identified the D0 brane with $y_{1}^{\ast}$. By making use of (\ref{eq:2f1cont}) and (\ref{meijer-yrel}) we get
\begin{equation}
y_1^{\ast}(z)=\frac{1}{\Gamma\left(\frac{1}{3}\right)\Gamma\left(\frac{2}{3}\right)}G_2\left(\begin{array}{c}\frac{1}{3},\frac{2}{3}\\0,0\end{array};y\right)
=\frac{2\pi i}{\Gamma\left(\frac{1}{3}\right)\Gamma\left(\frac{2}{3}\right)}\left(y_2^{\ast}(y)-y_1^{\ast}(y)\right).
\end{equation}
To get the analytic continuation of the D2 brane (\ref{cubicd2}) we simply replace $z\leftrightarrow y$. This yields the full analytic continuation matrix $\Phi_0=\Phi_1M_{10}$ with
\begin{equation}
M_{10}=\left(\begin{array}{cc}0&-\frac{3\Gamma\left(\frac{1}{3}\right)\Gamma\left(\frac{2}{3}\right)}{2\pi i}  \\\frac{2\pi i}{\Gamma\left(\frac{1}{3}\right)\Gamma\left(\frac{2}{3}\right)}&0 \end{array}\right).
\end{equation}
\subsubsection{Method 2}
We first discuss the analytic continuation of a D0-brane on the cubic. Its hemisphere partition function has been linked to the Mellin-Barnes integral
\begin{align}
y_1^{\ast}&=\int_{-i\infty}^{i\infty}ds\frac{\Gamma\left(\frac{1}{3}+s\right)\Gamma\left(\frac{2}{3}+s\right)}{(1-e^{2\pi i s})\Gamma(1+s)^{2}}z^{s}\nonumber\\
&=\int_{-i\infty}^{i\infty} \frac{ds}{2\pi i}\frac{\Gamma\left(\frac{1}{3}+s\right)\Gamma\left(\frac{2}{3}+s\right)\Gamma(-s)}{\Gamma(1+s)}z^{s}e^{-i\pi s}.
\end{align}
Replacing $z$ by $y=1-z$ using (\ref{identityMB}) 
the $s$-integral can be evaluated by using Jantzen's additional identity (\ref{jantzen}) \cite{Jantzen:2012cb}. The resulting $u$-integral looks like a hemisphere partition function with $z$ replaced by $y$ for the grade restricted structure sheaf for the cubic which we could relate to the combination $y_2^{\ast}-y_1^{\ast}$ of our basis of Mellin-Barnes integrals.
\begin{align}
\label{dual1}
y_1^{\ast}(z)&=\frac{1}{(2\pi i){\Gamma\left(\frac{1}{3}\right)\Gamma\left(\frac{2}{3}\right)}}\int_{-i\infty}^{i\infty}du\Gamma\left(\frac{1}{3}+u\right)\Gamma\left(\frac{2}{3}+u\right)\Gamma(-u)^{2}y^{u}e^{-2\pi i u}\nonumber\\
&=\frac{2\pi i}{\Gamma\left(\frac{1}{3}\right)\Gamma\left(\frac{2}{3}\right)}
(y_2^{\ast}(y)-y_1^{\ast}(y)).
\end{align}
If we start with the D2-brane in the large radius phase, the $s$-integral reduces to the first Barnes lemma (\ref{barnes1}). The remaining $u$-integral then looks like a hemisphere partition function for a D0-brane:
\begin{align}
\label{dual2}
y_2^{\ast}(z)-y_1^{\ast}(z)
&=
\int_{-i\infty}^{i\infty} \frac{ds}{(2\pi i)^{2}}\Gamma\left(\frac{1}{3}+s\right)\Gamma\left(\frac{2}{3}+s\right)\Gamma(-s)^{2}z^{s}\nonumber\\
&=\frac{\Gamma\left(\frac{1}{3}\right)\Gamma\left(\frac{2}{3}\right)}{(2\pi i)^{2}}\int_{-i\infty}^{i\infty} du\frac{\Gamma\left(\frac{1}{3}+u\right)\Gamma\left(\frac{2}{3}+u\right)\Gamma(-u)}{\Gamma(1+u)}e^{-i\pi u}y^{u}\nonumber\\
&=
\frac{\Gamma\left(\frac{1}{3}\right)\Gamma\left(\frac{2}{3}\right)}{(2\pi i)}y_1^{\ast}(y).
\end{align}
This is indeed the same result as obtained from method 1. It is interesting that the result near the conifold point again looks like a hemisphere partition function for the cubic with a different normalization and the variable $z$ replaced by $y$. Furthermore the brane factors for the D0 and D2-branes get exchanged. This behavior is very special to the cubic curve\footnote{This statement also holds for other elliptic curves that are described in terms of hypersurfaces in weighted $\mathbb{P}^2$.}. The reason is that the monodromy behavior around the conifold point is the same as around the large radius point. Furthermore the hypergeometric differential operator in the $z$-variable and the $y$-variable are the same. 
\subsection{The Quartic}
\subsubsection{Method 1}
For the mirror quartic we fix $n=3$ and $\beta_3=\frac{1}{2}$. B\"uhring's method for the analytic continuation of the holomorphic solution yields
\begin{equation}
\Gamma(\tfrac{1}{4}) \Gamma(\tfrac{1}{2}) \Gamma(\tfrac{3}{4})
\Phi_{0,11} = g_0(0) \Phi_{1,11} + g_1(0) \Phi_{1,12} +
g_0(\tfrac{1}{2}) \Phi_{1,13}
\end{equation}
with
\begin{align}
    g_0(0) &= \frac{\Gamma(\frac{1}{4})
      \Gamma(\frac{1}{2})^2}{\Gamma(\frac{3}{4})}
    \Hyp{\frac{1}{4},\frac{1}{2},\frac{1}{4}}{\frac{3}{4},1}{1}
    = \frac{\Gamma(\frac{1}{2})\Gamma(\frac{1}{8}) \Gamma(\frac{3}{8})}{2\,\Gamma(\frac{5}{8}) \Gamma(\frac{7}{8})}\\
    g_1(0) &=
    -\frac{\Gamma(\frac{5}{4})\Gamma(\frac{3}{2})\Gamma(-\frac{1}{2})}{\Gamma(
      \frac{3}{4})}
    \Hyp{-\frac{1}{2},\frac{1}{4},\frac{1}{4}}{\frac{3}{4},1}{1}= 2\,\Gamma(\tfrac{1}{2})
    \left( \frac{3\,\Gamma(\tfrac{1}{8})\Gamma(\frac{3}{8})}{64\, \Gamma(\frac{5}{8}) \Gamma(\tfrac{7}{8})} + \frac{\Gamma(\tfrac{5}{8})\Gamma(\frac{7}{8})}{\Gamma(\frac{1}{8}) \Gamma(\tfrac{3}{8})}\right)\\
    g_0(\tfrac{1}{2}) &= -2\,\Gamma(\tfrac{1}{2})
 \end{align}
This describes the analytic continuation of the central charge of the D0 brane (\ref{quarticd0}).

The method of~\cite{Scheidegger:2016ab} for analytic continuation of the solution $\Phi_{0,12}$ yields
\begin{equation}
\label{d2meth1}
 - 2 \pi i \Gamma(\tfrac{1}{4}) \Gamma(\tfrac{1}{2})
\Gamma(\tfrac{3}{4}) \Phi_{0,12} = h_0 \Phi_{1,11} + h_1 \Phi_{1,12}.
\end{equation}
The left-hand side has been identified as the large-radius value of the hemisphere partition function for the D2-brane (\ref{quarticd2}). The coefficients on the right-hand side are
\begin{align}
    h_0 &=\Gamma(\tfrac{1}{4})^2\Gamma(\tfrac{1}{2})^2
    \Hyp{\frac{1}{4},\frac{1}{2},\frac{1}{4}}{\frac{3}{4},1}{1} =
    \frac{\Gamma(\frac{1}{4})\Gamma(\frac{1}{2})\Gamma(\frac{3}{4})\Gamma(\frac{1}{8})\Gamma(\frac{3}{8})}{2\,\Gamma(\frac{5}{8})\Gamma(\frac{7}{8})}\\
    h_1 &= \tfrac{1}{6}
    \Gamma(\tfrac{1}{4})^2\Gamma(\tfrac{1}{2})^2\Hyp{\frac{1}{4},\frac{1}{2},\frac{1}{4}}{\frac{7}{4},1}{1}= 2\,\Gamma(\tfrac{1}{4})\Gamma(\tfrac{1}{2}) \Gamma(\tfrac{3}{4})
    \left(
      \frac{3\,\Gamma(\frac{1}{8})\Gamma(\frac{3}{8})}{64\,\Gamma(\frac{5}{8})\Gamma(\frac{7}{8})}
      -
      \frac{\Gamma(\frac{5}{8})\Gamma(\frac{7}{8})}{\Gamma(\frac{1}{8})\Gamma(\frac{3}{8})}
    \right).
  \end{align}
For the evaluation of the ${}_3F_2$ at $1$ we used identities due to Dixon (\ref{dixon}) \cite{Slater:1966ab} and Lavoie (\ref{lavoie}), (\ref{lavoie2}) \cite{Lavoie:1994ab}. For $\Phi_{0,13}$ we find from~\eqref{eq:xi_n}
\begin{equation}
\Phi_{1,13}(y) = -\frac{\Gamma(\frac{3}{2})}{\pi} \left( 2\,y^*_1(z) -3\, y^*_2(z)
      +2\,y^*_3(z) \right),
\end{equation}
which yields
\begin{equation}
\Phi_{0,13}(z) = -\frac{1}{\Gamma(\frac{1}{4})\Gamma(\frac{3}{4})}\Phi_{1,13}(y).
\end{equation}
This is the analytic continuation of the hemisphere partition function of the structure sheaf (\ref{quarticd4}) to the conifold point. Since $\Phi_{1,13}(0)=0$ we immediately see that the corresponding D4 brane becomes massless at the conifold point. Collecting all the information we have computed the full analytic continuation matrix
\begin{equation}
    M_{10} =
    \begin{pmatrix}
      \frac{A}{2\,\sqrt{2}\pi} & -\frac{A}{4\pi i}  & 0\\
      \frac{2}{\sqrt{2}\pi}
      \left(\frac{3\,A}{64} + \frac{1}{A}\right) &  -\frac{1}{\pi i}
      \left(\frac{3\,A}{64} - \frac{1}{A}\right) & 0 \\
      -\frac{2}{\sqrt{2}\pi} & 0 & -\frac{1}{\sqrt{2}\pi}
    \end{pmatrix}
\end{equation}
 where $A=\frac{\Gamma(\frac{1}{8})\Gamma(\frac{3}{8})}{\Gamma(\frac{5}{8}) \Gamma(\frac{7}{8})}$.

\subsubsection{Method 2}
The basis $y^{\ast}$ of Mellin-Barnes integral has the following form for the quartic:
\begin{align}
y_j^{\ast}(z)=\frac{1}{(2\pi i)^{j}}\int_{-i\infty}^{i\infty} ds e^{-j\pi i s }\frac{\Gamma\left(s+\frac{1}{4}\right)\Gamma\left(s+\frac{1}{2}\right)\Gamma\left(s+\frac{3}{4}\right)\Gamma(-s)^{j}}{\Gamma(1+s)^{3-j}}z^{s}\qquad j=1,2,3.
\end{align}
Apart from (\ref{identityMB}) we also we make repeated use of the identity (\ref{exp-trick}) 
in order to carefully treat the factor $e^{-j\pi i s }$. The D0-brane corresponds to $j=1$. Applying (\ref{identityMB}) we get
\begin{align}
\label{quartic1}
y_1^{\ast}(z)=\frac{1}{(2\pi i)^{2}}\int_{-i\infty}^{i\infty}duds e^{-i\pi (s+u)}\frac{\Gamma\left(s+\frac{1}{4}\right)\Gamma\left(s+\frac{1}{2}\right)\Gamma\left(s+\frac{3}{4}\right)\Gamma(u-s)\Gamma(-u)}{\Gamma(1+s)^{2}}y^{u}.
\end{align}
We can evaluate the $s$-integral in (\ref{quartic1}) by closing the contour either way. We choose to close the contour to the right and pick up poles\footnote{This means we choose a clockwise orientation of the contour.} at $s=u+k$. Evaluating the residue we get
\begin{equation}
y_1^{\ast}=\frac{1}{(2\pi i)}\int du\Gamma(-u)e^{-2i\pi u}\frac{\Gamma\left(u+\frac{1}{4}\right)\Gamma\left(u+\frac{1}{2}\right)\Gamma\left(u+\frac{3}{4}\right)}{\Gamma(u+1)^{2}}{}_{3}F_{2}\left({u+\frac{1}{4},u+\frac{1}{2},u+\frac{3}{4}\atop u+1,u+1};1\right)y^{u}.
\end{equation}
In order to get an expansion around the conifold point we have to evaluate the contour integral for $\Re(u)>0$. In this region the generalized hypergeometric function in the integral is clearly divergent and we need to regularize. In this case we are lucky because it turns out that all the divergence comes from poles at $u\in \frac{1}{2}+\mathbb{Z}_{\geq 0}$. We can use the Thomae relation (\ref{thomae1}) to write
\begin{equation}
{}_3F_2\left(\begin{array}{c}\frac{1}{4}+u,\frac{1}{2}+u,\frac{3}{4}+u\\1+u,1+u \end{array};1\right)=\frac{\Gamma(1+u)\Gamma\left(\frac{1}{2}-u\right)}{\Gamma\left(\frac{5}{4}\right)\Gamma\left(\frac{1}{4}\right)}{}_3F_2\left(\begin{array}{c}\frac{3}{4},\frac{1}{2},\frac{3}{4}+u\\\frac{5}{4},1+u \end{array};1\right).
\end{equation}
The remaining $u$-integral only has first order poles at $u=k$ and $u=k+\frac{1}{2}$ for $k\in\mathbb{Z}_{\geq 0}$ and can be evaluated using the residue theorem. The result is
\begin{align}
y_1^{\ast}(z)=&\frac{\pi}{\Gamma\left(\frac{5}{4}\right)\Gamma\left(\frac{1}{4}\right)}\bigg[\sum_{k=0}^{\infty}y^k\frac{\Gamma\left(k+\frac{1}{4}\right)\Gamma\left(k+\frac{3}{4}\right)}{\Gamma(1+k)^2}{}_3F_2\left(\begin{array}{c}\frac{3}{4},\frac{1}{2},\frac{3}{4}+k\\\frac{5}{4},1+k \end{array};1\right)\nonumber\\
&-\sum_{k=0}^{\infty}y^{k+\frac{1}{2}}\frac{\Gamma\left(k+\frac{3}{4}\right)\Gamma\left(k+\frac{5}{4}\right)}{\Gamma\left(k+\frac{3}{2}\right)^2}{}_3F_2\left(\begin{array}{c}\frac{3}{4},\frac{1}{2},\frac{5}{4}+k\\\frac{5}{4},\frac{3}{2}+k \end{array};1\right)
\bigg].
\end{align}
Expanding in $k$ and comparing with the periods at the conifold point in appendix \ref{app-quartic-periods} we get
\begin{align}
y_1^{\ast}(z)=&\pi\bigg[\frac{\Gamma\left(\frac{3}{4}\right)}{\Gamma\left(\frac{5}{4}\right)}{}_3F_2\left(\begin{array}{c}\frac{3}{4},\frac{1}{2},\frac{3}{4}\\\frac{5}{4},1\end{array};1\right)\Phi_{1,11}+\frac{\Gamma\left(\frac{7}{4}\right)}{\Gamma\left(\frac{1}{4}\right)}{}_3F_2\left(\begin{array}{c}\frac{3}{4},\frac{1}{2},\frac{7}{4}\\\frac{5}{4},2\end{array};1\right)\Phi_{1,12}\nonumber\\
&-\frac{\Gamma\left(\frac{3}{4}\right)}{\Gamma\left(\frac{1}{4}\right)\Gamma\left(\frac{3}{2}\right)^2}{}_3F_2\left(\begin{array}{c}\frac{3}{4},\frac{1}{2},\frac{5}{4}\\\frac{5}{4},\frac{3}{2}\end{array};1\right)\Phi_{1,13} \bigg]
\end{align}
By repeated application of the identities collected in appendix \ref{app-identities}, in particular (\ref{dixon}) and (\ref{lavoie}), one can show that this  perfectly agrees with the result of method 1.

Next, we consider a D2-brane which we can relate to the combination $y_2^{\ast}-y_1^{\ast}$ by making use of (\ref{exp-trick}):
\begin{equation}
y_2^{\ast}(z)-y_1^{\ast}(z)=\frac{1}{(2\pi i)^{3}}\int duds e^{-i\pi u} \frac{\Gamma\left(s+\frac{1}{4}\right)\Gamma\left(s+\frac{1}{2}\right)\Gamma\left(s+\frac{3}{4}\right)\Gamma(-s)\Gamma(u-s)\Gamma(-u)}{\Gamma(s+1)}y^{u}.
\end{equation}
Applying the analysis of Section~\ref{sec-method2}, in particular the decomposition~\eqref{eq:Fqdecomp}, this equation can be written as
\begin{equation}
  \label{eq:3}
  \int_{-i\infty}^{i\infty} du \Gamma(-u) F_{2}(u)(z-1)^{u} - \int_{-i\infty}^{i\infty} du \Gamma(-u) F^{(1)}_{2}(u)(z-1)^{u} = \int_{-i\infty}^{i\infty} du \Gamma(-u) F^{(0)}_{2}(u)(z-1)^{u}.
\end{equation}
We argued that $F^{(0)}_{2}(u)$ is convergent for $\Re u \geq 0$, hence the integral on the right hand side will define a holomorphic function of $z-1$. To determine this function, we perform in a first step the integral over $s$ while keeping $u$ an arbitrary parameter. The integral we have to evaluate is
\begin{equation}
I_1=\int\frac{ds}{2\pi i}\frac{\Gamma\left(s+\frac{1}{4}\right)\Gamma\left(s+\frac{1}{2}\right)\Gamma\left(s+\frac{3}{4}\right)\Gamma(-s)\Gamma(u-s)}{\Gamma(1+s)}.
\end{equation}
If we close the contour to the right there are first order poles at $s=k$ and $s=u+k$ with $k\in\mathbb{Z}_{\geq 0}$. The result is
\begin{align}
I_1=&-\frac{\Gamma(-u)\Gamma(1+u)\Gamma\left(\frac{1}{4}\right)\Gamma\left(\frac{1}{2}\right)\Gamma\left(\frac{3}{4}\right)}{\Gamma(1-u)} {}_3 F_2\left(\begin{array}{c}\frac{1}{4},\frac{1}{2},\frac{3}{4}\\ 1,1-u\end{array};1\right)\nonumber\\
&+ \frac{\Gamma(-u)\Gamma\left(u+\frac{1}{4}\right)\Gamma\left(u+\frac{1}{2}\right)\Gamma\left(u+\frac{3}{4}\right)}{\Gamma(1+u)}{}_3 F_2\left(\begin{array}{c}u+\frac{1}{4},u+\frac{1}{2},u+\frac{3}{4}\\ 1+u,1+u\end{array};1\right).
\end{align}
One can check that both terms are divergent, but we know that the divergences must cancel. As in the D0-case we can use the Thomae relations to factor out the poles. Specifically, we use (\ref{thomae2}) for the first term and (\ref{thomae1}) for the second term. The hemisphere partition function then becomes
\begin{align}
y_2^{\ast}(z)-y_1^{\ast}(z)=&
\frac{1}{(2\pi i)^2}\int du y^u\nonumber\\
&\cdot\bigg[-e^{-i\pi u}\frac{\Gamma(-u)^2\Gamma\left(\frac{1}{4}\right)\Gamma\left(\frac{3}{4}\right)\Gamma\left(\frac{1}{2}-u\right)\Gamma\left(1+u\right)}{\Gamma\left(\frac{3}{4}-u\right)\Gamma\left(\frac{5}{4}-u\right)} {}_3F_2\left(\begin{array}{c}\frac{1}{2},\frac{1}{2}-u,\frac{1}{2}-u\\\frac{3}{4}-u,\frac{5}{4}-u \end{array};1\right)\nonumber\\
&+e^{-i\pi u} \frac{\Gamma\left(u+\frac{1}{2}\right)\Gamma\left(u+\frac{1}{4}\right)\Gamma\left(u+\frac{3}{4}\right)\Gamma(-u)^2\Gamma\left(\frac{1}{2}-u\right)}{\Gamma\left(\frac{1}{4}\right)\Gamma\left(\frac{5}{4}\right)}{}_3F_2\left(\begin{array}{c}\frac{3}{4},\frac{1}{2},\frac{3}{4}+u\\\frac{5}{4},1+u \end{array};1\right)
\bigg].\nonumber\\
\end{align}
There are second order poles at $u=k$ and first order poles at $u=k+\frac{1}{2}$ for $k\in\mathbb{Z}$. To deal with the double poles we introduce a small shift $u=k-\epsilon$. Then can rewrite the integral as follows
\begin{align}
\label{d2meth2}
&\sum_{k=0}^{\infty}\bigg[-\oint \frac{d\epsilon}{(2\pi i)^2}y^{k-\epsilon}e^{i\pi\epsilon}\frac{\Gamma\left(\frac{1}{4}\right)\Gamma\left(\frac{3}{4}\right)\Gamma\left(\frac{1}{4}+k-\epsilon\right)\Gamma\left(-\frac{1}{4}+k-\epsilon\right)}{\Gamma\left(1+k-\epsilon\right)\Gamma\left(\frac{1}{2}+k-\epsilon\right)}\nonumber\\
&\cdot\pi\frac{\sin\pi\left(\frac{3}{4}+\epsilon\right)\sin\pi\left(\frac{5}{4}+\epsilon\right)}{\sin^2\pi\epsilon\sin\pi\left(\frac{1}{2}+\epsilon\right)}{}_3F_2\left(\begin{array}{c}\frac{1}{2},\frac{1}{2}-k+\epsilon,\frac{1}{2}-k+\epsilon\\\frac{3}{4}-k+\epsilon,\frac{5}{4}-k+\epsilon \end{array};1\right)\nonumber\\
&+\oint \frac{d\epsilon}{(2\pi i)^2}y^{k-\epsilon}e^{i\pi\epsilon}\frac{\Gamma\left(\frac{1}{4}+k-\epsilon\right)\Gamma\left(\frac{3}{4}+k-\epsilon\right)}{\Gamma\left(\frac{1}{4}\right)\Gamma\left(\frac{5}{4}\right)\Gamma(1+k-\epsilon)^2}\frac{\pi^3}{\sin^2\pi\epsilon\sin\pi\left(\frac{1}{2}+\epsilon\right)}{}_3F_2\left(\begin{array}{c}\frac{3}{4},\frac{1}{2},\frac{3}{4}+k-\epsilon\\\frac{5}{4},1+k-\epsilon \end{array};1\right)\bigg]\nonumber\\
&+\frac{1}{(2\pi i)}\sum_{k=0}^{\infty}\bigg[-e^{-\frac{i\pi}{2}}y^{k+\frac{1}{2}}\frac{\Gamma\left(\frac{1}{4}\right)\Gamma\left(\frac{3}{4}\right)\Gamma\left(\frac{3}{4}+k\right)\Gamma\left(\frac{1}{4}+k\right)}{\Gamma\left(\frac{3}{2}+k\right)\Gamma(1+k)}\frac{\sin\frac{\pi}{4}\sin\frac{3\pi}{4}}{\sin^2\frac{-\pi}{2}}{}_3F_2\left(\begin{array}{c}\frac{1}{2},-k,-k\\\frac{1}{4}-k,\frac{3}{4}-k \end{array};1\right)\nonumber\\
&+e^{\frac{i\pi}{2}}y^{k+\frac{1}{2}}\frac{\Gamma\left(\frac{3}{4}+k\right)\Gamma\left(\frac{5}{4}+k\right)}{\Gamma\left(\frac{1}{4}\right)\Gamma\left(\frac{5}{4}\right)\Gamma\left(\frac{3}{2}+k\right)^2}\frac{\pi^2}{\sin^2\frac{-\pi}{2}}{}_3F_2\left(\begin{array}{c}\frac{3}{4},\frac{1}{2},\frac{5}{4}+k\\\frac{5}{4},\frac{3}{2}+k \end{array};1\right)
\bigg]
\end{align}
Evaluating the poles from the first two summands will yield a term holomorphic in $y$ and a term proportional to $\log y$. We have argued that the integral must define a holomorphic function, hence the coefficient of the term proportional to $\log y$ must vanish. The explicit calculation yields the identity
\begin{equation}
  \begin{aligned}
    \label{newid2}
\, _3F_2\left(\begin{array}{c}\frac{1}{2},\frac{3}{4},k+\frac{3}{4}\\\frac{5}{4},k+1\end{array};1\right)&=-\frac{\Gamma \left(\frac{5}{4}\right) \Gamma \left(k-\frac{1}{4}\right) \Gamma (k+1) }{\Gamma\left(\frac{3}{4}\right) \Gamma \left(k+\frac{1}{2}\right)
   \Gamma \left(k+\frac{3}{4}\right)}\,
   _3F_2\left(\begin{array}{c}\frac{1}{2},\frac{1}{2}-k,\frac{1}{2}-k\\\frac{3}{4}-k,\frac{5}{4}-k\end{array};1\right)\quad k\in\mathbb{Z}_{\geq0}.
  \end{aligned}
\end{equation}
Similarly, the third and fourth summand in~\eqref{d2meth2} must vanish for the same reason as they yield an expression proportional to $\sqrt{y}$ and leads to the identity
\begin{equation}
\begin{aligned}
\label{newid1}
\, _3F_2\left(\begin{array}{c}\frac{1}{2},\frac{3}{4},k+\frac{5}{4}\\\frac{5}{4},k+\frac{3}{2}\end{array};1\right)&=\frac{ \Gamma \left(\frac{5}{4}\right) \Gamma \left(k+\frac{1}{4}\right) \Gamma \left(k+\frac{3}{2}\right)}{\Gamma\left(\frac{3}{4}\right) \Gamma (k+1) \Gamma \left(k+\frac{5}{4}\right)}\,   _3F_2\left(\begin{array}{c}\frac{1}{2},-k,-k\\\frac{1}{4}-k,\frac{3}{4}-k\end{array};1\right)\quad k\in\mathbb{Z}_{\geq0}\\
\end{aligned}
\end{equation}
A direct proof of these two identities has been given by C. Krattenthaler \cite{krattenthaler} using techniques explained in~\cite{Krattenthaler:2006ab}. It involves resummations and repeated application of the identities (\ref{thomae1}), (\ref{thomae2}) and (\ref{id3}).

Defining two functions $f_1(k,\epsilon)$ and $f_2(k,\epsilon)$ by writing the terms with second order poles in (\ref{d2meth2}) as $\oint \frac{d\epsilon}{\sin^2\pi\epsilon}y^{k-\epsilon}(f_1(k,\epsilon)+f_2(k,\epsilon))$ the final result is
\begin{equation}
y_2^{\ast}(z)-y_1^{\ast}(z)=(2\pi i)\sum_{k=0}^{\infty}y^k\frac{1}{\pi^2}\left.\left(\frac{df_1(k,\epsilon)}{d\epsilon}+\frac{df_2(k,\epsilon)}{d\epsilon}\right)\right\vert_{\epsilon=0}.
\end{equation}
This is convergent and can be expanded in terms of the periods $\Phi_{1,11}$ and $\Phi_{1,12}$. The coefficients correspond to the terms for $k=0$ and $k=1$ in the above sum. Even though the expressions look completely different, comparison with (\ref{d2meth1}) shows that the results agree numerically. We have not managed to prove this non-trivial identity analytically and leave this as an open conjecture.

The D4-brane is slightly more complicated. With (\ref{exp-trick}) one finds that it is convenient to evaluate the following combination
\begin{equation}
y_3^{\ast}(z)-y_2^{\ast}(z)=\frac{1}{(2\pi i)^4}\int duds e^{-i\pi(u+s)}\Gamma\left(s+\frac{1}{4}\right)\Gamma\left(s+\frac{1}{2}\right)\Gamma\left(s+\frac{3}{4}\right)\Gamma(-u)\Gamma(-s)^2\Gamma(u-s).
\end{equation}
The $s$-integral now has second order poles for positive integer $s$:
\begin{equation}
I_2=\int\frac{ds}{2\pi i}\Gamma\left(s+\frac{1}{4}\right)\Gamma\left(s+\frac{1}{2}\right)\Gamma\left(s+\frac{3}{4}\right)\Gamma(-s)^2\Gamma(u-s).
\end{equation}
We could close the contour to the left to avoid the second order poles, but it actually turns out to be better to separate the second order poles and close the contour to the right. For this purpose we introduce a small parameter $\epsilon$ so that the integral has first order poles at $s=k$, $s=k+\epsilon$ and $s=k+u$ for $s\in\mathbb{Z}_{\geq 0}$. At the end of the calculation we take the limit $\epsilon\rightarrow 0$. We get three contributions and apply the Thomae relations (\ref{thomae1}) and (\ref{thomae2}) to factor out poles in $u$.
\begin{align}
I_2^{(1)}&=\frac{\pi^2}{\sin\pi u\sin\pi\epsilon}\frac{\Gamma\left(\frac{1}{4}\right)\Gamma\left(\frac{3}{4}\right)\Gamma\left(\frac{1}{2}-u-\epsilon\right)}{\Gamma\left(\frac{3}{4}-u-\epsilon\right)\Gamma\left(\frac{5}{4}-u-\epsilon\right)}{}_3F_2\left(\begin{array}{c}\frac{1}{2}-\epsilon,\frac{1}{2}-u,\frac{1}{2}-u-\epsilon\\\frac{3}{4}-u-\epsilon,\frac{5}{4}-u-\epsilon\end{array};1\right)\\
I_2^{(2)}&=\frac{\pi^2 e^{-i\pi\epsilon}}{\sin\pi(-\epsilon)\sin\pi(u-\epsilon)}\frac{\Gamma\left(\frac{1}{4}+\epsilon\right)\Gamma\left(\frac{3}{4}+\epsilon\right)\Gamma\left(\frac{1}{2}-u-\epsilon\right)}{\Gamma\left(\frac{3}{4}-u\right)\Gamma\left(\frac{5}{4}-u\right)}{}_3F_2\left(\begin{array}{c}\frac{1}{2}-u,\frac{1}{2},\frac{1}{2}-u-\epsilon\\\frac{3}{4}-u,\frac{5}{4}-u\end{array};1\right)\\
I_2^{(3)}&=\frac{\pi^2 e^{-i\pi u}}{\sin\pi(-u)\sin\pi(-u+\epsilon)}\frac{\Gamma\left(\frac{1}{4}+u\right)\Gamma\left(\frac{1}{2}+u\right)\Gamma\left(\frac{3}{4}+u\right)\Gamma\left(\frac{1}{2}-u-\epsilon\right)}{\Gamma(1+u)\Gamma\left(\frac{5}{4}-\epsilon\right)\Gamma\left(\frac{1}{4}-\epsilon\right)}{}_3F_2\left(\begin{array}{c}\frac{3}{4},\frac{1}{2},\frac{3}{4}+u\\\frac{5}{4}-\epsilon,1+u\end{array};1\right)
\end{align}
Expanding these expressions in $\epsilon$ one immediately sees that the $\frac{1}{\epsilon}$-contributions originating from $I_2^{(1)}$ and $I_2^{(2)}$ cancel end we can safely take the limit $\epsilon\rightarrow 0$. The expansion in particular introduces derivative terms of ${}_3F_2$ with respect to the parameters. Since the result is quite ugly and can easily be obtained by feeding the expressions above into Mathematica we do not give details here. After this we have to perform the $u$-integration by closing the integration contour to the right. There are second order poles at $u=k$ and first order poles at $u=\frac{1}{2}+k$ for $k\in\mathbb{Z}_{\geq 0}$. By another set of highly non-trivial identities, similar to (\ref{newid1}) and (\ref{newid2}) but with derivatives on the hypergeometric functions, the terms with $\log y$ and $\sqrt{y}$ cancel in accordance with the discussion in section \ref{sec-method2}. Since we have not found a way to simplify these lengthy expressions we omit them here. One can show that the final result is a linear combination of the two periods $\Phi_{1,11}$ and $\Phi_{1,12}$. Numerical comparison to the results of method 1 shows agreement and we seem to have uncovered a further set of non-trivial identities between generalized hypergeometric functions and their derivatives.
\subsection{The Quintic}
\subsubsection{Method 1}
For the quintic we have $n=4$ and $\beta_4=1$. The analytic continuation of the holomorphic solution yields
\begin{equation}
 \Phi_{0,11} = l_0 \Phi_{1,11} + w_0\Phi_{1,12} + q_0
  \Phi_{1,13} + \left(w_1-\tfrac{7}{10}w_0\right) \Phi_{1,14}.
\end{equation}
The left-hand side has been identified with $Z_{D^2}^{\zeta\gg0}(\mathcal{O}_p)$ in (\ref{quinticd0}). With
\begin{equation}
A^{(4)}(k) =\frac{(\frac{3}{5})_k (\frac{2}{5})_k}{\Gamma(1+k)} \,{}_3F_2
\left(\genfrac{}{}{0pt}{}{\frac{1}{5},\frac{1}{5},-k}{\frac{3}{5},
    \frac{3}{5}-k};1\right)
\end{equation}
the values of the coefficients are
\begin{align}
\label{eq:lqw}
  l_0 &= \frac{\Gamma(\frac{1}{5})}{\Gamma(\frac{3}{5})} \sum_{k=0}^\infty
  \frac{\Gamma(\frac{3}{5}+k)\Gamma(\frac{2}{5}+k)}{\Gamma(\frac{6}{5}+k)\Gamma(\frac{7}{5}+k)}
  \,{}_3F_2\left(\genfrac{}{}{0pt}{}{\frac{1}{5},\frac{1}{5},-k}{\frac{3}{5},\frac{3}{5}-k};1\right)\\
    q_0 &= 1\\
    w_0 & =- \psi(1) - \psi(2) + \psi(\tfrac{6}{5}) + \psi(\tfrac{7}{5})
    -\sum_{k=1}^\infty \frac{(\frac{3}{5})_k
      (\frac{2}{5})_k}{k\,(\frac{6}{5})_k(\frac{7}{5})_k}\,{}_3F_2
    \left(\genfrac{}{}{0pt}{}{\frac{1}{5},\frac{1}{5},-k}{\frac{3}{5},
        \frac{3}{5}-k};1\right)\\
    w_1 &= -\tfrac{21}{25}\left( \vphantom{\sum_{k=2}^\infty} \psi(2) +
      \psi(3) - \psi(\tfrac{11}{5}) - \psi(\tfrac{12}{5})-\tfrac{1}{6} \left( \psi(1) + \psi(3) - \psi(\tfrac{11}{5}) -
        \psi(\tfrac{12}{5}) \right) \right.\notag\\
    &\phantom{=} \left. - \sum_{k=2}^\infty \frac{(\frac{3}{5})_k
        (\frac{2}{5})_k}{k(k-1)(\frac{6}{5})_k(\frac{7}{5})_k}
      \,{}_3F_2
      \left(\genfrac{}{}{0pt}{}{\frac{1}{5},\frac{1}{5},-k}{\frac{3}{5},
          \frac{3}{5}-k};1\right) \right).
\end{align}
The analytic continuation of $\Phi_{0,12}$ becomes
\begin{align}
    \Phi_{0,12} = -
      \frac{h_0}{2\pi i} \Phi_{1,11} - \frac{h_1}{2\pi i}
      \Phi_{1,12} -
      \left(\frac{h_2}{2\pi i} - \frac{7}{10} \frac{h_1}{2\pi i} \right) \Phi_{1,14}.
  \end{align}
This describes the analytic continuation of $Z_{D^2}^{\zeta\gg0}(\mathcal{O}_l)$ (\ref{quinticd2}). The coefficients are
\begin{align}
    \label{eq:hm}
    h_0 &=
    \Gamma(\tfrac{1}{5})^2\Gamma(\tfrac{2}{5})^2\Gamma(\tfrac{4}{5})\sum_{\ell=0}^\infty
    \frac{(\frac{1}{5})_\ell(\frac{2}{5})_\ell}{(\frac{3}{5})_\ell
      \ell!}\Hyp{-\ell,\frac{3}{5},\frac{4}{5}}{1,1}{1}\\
    h_1 &=
    \tfrac{2}{15}\Gamma(\tfrac{1}{5})^2\Gamma(\tfrac{2}{5})^2\Gamma(\tfrac{4}{5})
    \sum_{\ell=0}^\infty
    \frac{(\frac{1}{5})_\ell(\frac{2}{5})_\ell}{(\frac{8}{5})_\ell
      \ell!}\Hyp{-\ell,\frac{3}{5},\frac{4}{5}}{1,1}{1}\\
    h_2 &=
    \tfrac{7}{100}\Gamma(\tfrac{1}{5})^2\Gamma(\tfrac{2}{5})^2\Gamma(\tfrac{4}{5})\sum_{\ell=0}^\infty
    \frac{(\frac{1}{5})_\ell(\frac{2}{5})_\ell}{(\frac{13}{5})_\ell
      \ell!}\Hyp{-\ell,\frac{3}{5},\frac{4}{5}}{1,1}{1}.
  \end{align}
The analytic continuation of the  $\Phi_{0,13}$ is
\begin{align}
  \Phi_{0,13}
    &= \frac {5\,k_0}{(2\pi i)^2}  \Phi_{1,11} + \frac {5\,k_1}{(2\pi i)^2} \Phi_{1,12} + \left(\frac {5\,k_2}{(2\pi i)^2} -\frac{7}{10} \frac {5\,k_1}{(2\pi i)^2}\right) \Phi_{1,14}
\end{align}
Recall that this is not the analytic continuation of the ``canonical'' D4 brane (\ref{quinticd4}) in the GLSM. The coefficients are
\begin{align}
 \label{eq:km2}
    k_0 &= \Gamma(\tfrac{2}{5}) \Gamma(\tfrac{3}{5})
    \Gamma(\tfrac{4}{5}) \sum_{\ell=0}^\infty
    \frac{\Gamma(\frac{1}{5}+\ell)^2 }{\Gamma(\ell+1)^2} \notag\\
    &\phantom{=} \cdot \left(
      \e{-i\pi\frac{2}{5}}
      \frac{\Gamma(\frac{1}{5})\Gamma(\frac{2}{5})\Gamma(\frac{2}{5}+\ell)}{\Gamma(\frac{3}{5})\Gamma(\frac{3}{5}+\ell)}
      {}_3F_2\left(\genfrac{}{}{0pt}{0}{\frac{2}{5},\frac{2}{5}+\ell,\frac{2}{5}}{\frac{4}{5},\frac{3}{5}+\ell};1\right)
    \right. \left. +\, \e{-i\pi\frac{3}{5}}
      \frac{\Gamma(-\frac{1}{5})\Gamma(\frac{3}{5})\Gamma(\frac{3}{5}+\ell)}{\Gamma(\frac{2}{5})\Gamma(\frac{4}{5}+\ell)}
      {}_3F_2\left(\genfrac{}{}{0pt}{0}{\frac{3}{5},\frac{3}{5}+\ell,\frac{3}{5}}{\frac{6}{5},\frac{4}{5}+\ell};1\right)
    \right)  \\
    k_1 &= \Gamma(\tfrac{4}{5}) \Gamma(\tfrac{7}{5})
    \Gamma(\tfrac{8}{5}) \sum_{\ell=0}^\infty
    \frac{\Gamma(\frac{1}{5}+\ell)^2 }{\Gamma(\ell+1)^2} \notag\\
    &\phantom{=} \cdot \left(
      \e{-i\pi\frac{2}{5}}
      \frac{\Gamma(\frac{1}{5})\Gamma(\frac{2}{5})\Gamma(\frac{2}{5}+\ell)}{\Gamma(\frac{8}{5})\Gamma(\frac{3}{5}+\ell)}
      {}_3F_2\left(\genfrac{}{}{0pt}{0}{\frac{2}{5},\frac{2}{5}+\ell,-\frac{3}{5}}{\frac{4}{5},\frac{3}{5}+\ell};1\right)
    \right. \left. +\, \e{-i\pi\frac{3}{5}}
      \frac{\Gamma(-\frac{1}{5})\Gamma(\frac{3}{5})\Gamma(\frac{3}{5}+\ell)}{\Gamma(\frac{7}{5})\Gamma(\frac{4}{5}+\ell)}
      {}_3F_2\left(\genfrac{}{}{0pt}{0}{\frac{3}{5},\frac{3}{5}+\ell,-\frac{2}{5}}{\frac{6}{5},\frac{4}{5}+\ell};1\right)
    \right) \\
    k_2 &= \tfrac{1}{2}\Gamma(\tfrac{4}{5})\Gamma(\tfrac{12}{5})
      \Gamma(\tfrac{13}{5}) \sum_{\ell=0}^\infty
    \frac{\Gamma(\frac{1}{5}+\ell)^2 }{\Gamma(\ell+1)^2} \notag\\
    &\phantom{=} \cdot \left(
      \e{-i\pi\frac{2}{5}}
      \frac{\Gamma(\frac{1}{5})\Gamma(\frac{2}{5})\Gamma(\frac{2}{5}+\ell)}{\Gamma(\frac{13}{5})\Gamma(\frac{3}{5}+\ell)}
      {}_3F_2\left(\genfrac{}{}{0pt}{0}{\frac{2}{5},\frac{2}{5}+\ell,-\frac{8}{5}}{\frac{4}{5},\frac{3}{5}+\ell};1\right)
     \right. \left. +\, \e{-i\pi\frac{3}{5}}
      \frac{\Gamma(-\frac{1}{5})\Gamma(\frac{3}{5})\Gamma(\frac{3}{5}+\ell)}{\Gamma(\frac{12}{5})\Gamma(\frac{4}{5}+\ell)}
      {}_3F_2\left(\genfrac{}{}{0pt}{0}{\frac{3}{5},\frac{3}{5}+\ell,-\frac{7}{5}}{\frac{6}{5},\frac{4}{5}+\ell};1\right)
    \right)
\end{align}
Finally, the analytic continuation of $Z_{D^2}^{\zeta\gg0}(\mathcal{O}_X)$ (\ref{quinticd6}) can be determined using~\eqref{eq:xi_n} as
\begin{equation}
\Phi_{1,12}(y) = \tfrac{5}{2\pi i} \left( y^*_4(z) -2\, y^*_3(z) + 2\,y^*_2(z) - y_1^*(z) \right)=\frac{1}{2\pi i}\Phi_{0,14}(z).
\end{equation}
Since $\Phi_{1,12}(0)=0$ we confirm that this D6 branes becomes massless at the conifold point. This result has first been obtained through a monodromy argument in~\cite{Candelas:1991rm}.

In summary, the following analytic continuation matrix reads
\begin{equation}
    M_{10} =
    \begin{pmatrix}
     \medskip
      l_0 & -\frac{h_0}{2\pi i} & \frac{5\,k_0}{(2\pi i)^2} & 0\\
     \medskip
     w_0 & -\frac{h_1}{2\pi i} &\frac{5\,k_1}{(2\pi i)^2} & 2\pi i \\
     \medskip
     1 & 0 & 0 & 0 \\
     w_1 - \frac{7}{10}w_0 & -\frac{h_2}{2\pi i} + \frac{7}{10}\frac{h_1}{2\pi i}
     &\frac{5\,k_2}{(2\pi i)^2} - \frac{7}{10}\frac{5\,k_1}{(2\pi
       i)^2} & 0
    \end{pmatrix}
\end{equation}
At present, we are not aware of any identities that help evaluating the infinite sums in $h_m$ and $k_m$. We can however evaluate them to high precision and find perfect agreement with numerical analytic continuation. We observe that the following identity should hold: $\Im k_m = \pi i h_m, m=0,1,2.$ This can be absorbed into the following change of basis in~\eqref{eq:PhiBasis}: $\Phi_{0,12} \to \Phi_{0,13} - \frac{5}{2} \Phi_{0,11}$. Then all the constants in $M_{10}$ are real up to the displayed factors of $2\pi i$. 
\subsubsection{Outline of Method 2}
We have already seen for the quartic that the alternative application of analytic continuation crucially depends on hypergeometric identities that help to regularize the divergent sums. The same difficulties arise for the quintic. Let us demonstrate this by discussing the D0-brane. The double integral we have to solve is
\begin{equation}
y_1^{\ast}(z)=\int\frac{du}{2\pi i}(-y)^u\Gamma(-u)\int\frac{ds}{2\pi i}\frac{\Gamma\left(s+\frac{1}{5}\right)\Gamma\left(s+\frac{2}{5}\right)\Gamma\left(s+\frac{3}{5}\right)\Gamma\left(s+\frac{4}{5}\right)\Gamma(u-s)}{\Gamma(1+s)^3}e^{i\pi s}.
\end{equation}
Closing the contour to the right the result for the $s$-integral is
\begin{equation}
e^{i\pi u} \frac{\Gamma\left(u+\frac{1}{5}\right)\Gamma\left(u+\frac{2}{5}\right)\Gamma\left(u+\frac{3}{5}\right)\Gamma\left(u+\frac{4}{5}\right)}{\Gamma(1+u)^3}{}_4 F_3\left(\begin{array}{c}u+\frac{1}{5},u+\frac{2}{5},u+\frac{3}{5},u+\frac{4}{5}\\ 1+u,1+u,1+u\end{array};1\right),
\end{equation}
which is divergent for positive $u$. The generalized hypergeometric functions of type ${}_4F_3$ are not well-studied. In particular we do not know of any identities that may help us to isolate the poles. One way out of this dilemma is to use B\"uhring's recursion (\ref{buehring}) to express ${}_4F_3$ in terms of ${}_3F_2$. However, this comes at the cost of a further integral:
\begin{align}
y_1^{\ast}(z)=&\frac{\sqrt{5}}{(2\pi)^2}\int\frac{du}{2\pi i}(z-1)^u\Gamma(-u)e^{i\pi u}\frac{\Gamma\left(u+\frac{1}{5}\right)\Gamma\left(u+\frac{2}{5}\right)\Gamma\left(u+\frac{3}{5}\right)}{\Gamma(1+u)}\nonumber\\
  &\cdot \frac{1}{\Gamma\left(\frac{1}{5}\right)^2}\int\frac{dt}{2\pi i}e^{\pm i\pi t}\frac{\Gamma(-t)\Gamma\left(\frac{1}{5}+t\right)^2}{\Gamma\left(\frac{6}{5}+u+t\right)} {}_3 F_2\left(\begin{array}{c}u+\frac{1}{5},u+\frac{2}{5},u+\frac{3}{5},\\ 1+u,\frac{6}{5}+u+t\end{array};1\right).
\end{align}
Keeping $u$ as a parameter, we can evaluate the $t$-integral by closing the contour to the right and enclosing the poles at $t=k$. Then we get
\begin{align}
y_1^{\ast}(z)=&\int\frac{du}{2\pi i}(-y)^u\Gamma(-u)e^{i\pi u}\frac{\Gamma\left(u+\frac{1}{5}\right)\Gamma\left(u+\frac{2}{5}\right)\Gamma\left(u+\frac{3}{5}\right)}{\Gamma(1+u)}\nonumber\\
  &\cdot \frac{1}{\Gamma\left(\frac{1}{5}\right)^2}\sum_{k=0}^{\infty}\frac{\Gamma\left(k+\frac{1}{5}\right)^2}{\Gamma(k+1)\Gamma\left(u+k+\frac{6}{5}\right)}{}_3 F_2\left(\begin{array}{c}u+\frac{1}{5},u+\frac{2}{5},u+\frac{3}{5},\\ 1+u,\frac{6}{5}+u+k\end{array};1\right)
\end{align}
Using (\ref{thomae1}) we can rewrite
\begin{equation}
{}_3 F_2\left(\begin{array}{c}u+\frac{1}{5},u+\frac{2}{5},u+\frac{3}{5},\\ 1+u,\frac{6}{5}+u+k\end{array};1\right)=\frac{\Gamma\left(\frac{6}{5}+k+u\right)\Gamma(1+k-u)}{\Gamma\left(\frac{8}{5}+k\right)\Gamma\left(\frac{3}{5}+k\right)}{}_3F_2\left(\begin{array}{c}\frac{4}{5},\frac{3}{5},\frac{3}{5}+u\\\frac{8}{5}+k,1+u \end{array};1\right).
\end{equation}
This can be evaluated for $\Re(u)>0$, where the integral has at most double poles. This gives the expected behavior of the periods around the conifold point. Numerical analysis shows that the result is convergent (however, badly) and that we find agreement with B\"uhring's method. This shows that also this alternative method leads to coefficients in the analytic continuation matrix that are infinite sums and the result does not have a simpler form than with the other approach. Since also the convergence issue is not as clear as for the first method we will not discuss any more examples on the quintic.
\section{Conclusions}
There are several obvious directions in which one can generalize our results. Application of our methods to one-parameter CY hypersurfaces in weighted projective space is straightforward since these cases are already included in the methods of analytic continuation discussed in section \ref{sec-math} and \cite{Scheidegger:2016ab}. What is more challenging is the generalization to CY hypersurfaces in toric varieties with more than one K\"ahler parameter. Here the most difficult problem is how to find suitable recurrences which are necessary for analytic continuation to the singular locus. A further possibility is to study exotic CYs related to abelian and non-abelian GLSMs \cite{Hori:2006dk,Caldararu:2007tc,Hori:2011pd,Jockers:2012zr,Addington:2012zv,Sharpe:2012ji,Halverson:2013eua,Sharpe:2013bwa,Hori:2013gga}. Of particular interest are the GLSMs associated to the one-parameter CYs found by R{\o}dland CY \cite{rodland98} and Hosono and Takagi \cite{Hosono:2011np}. The main difference between these one-parameter examples and our discussion is that they have more than one conifold point, which complicates matters. The corresponding differential equation will have more than three regular singular points. So far, much less is known about the analytic properties of the solutions to these differential equations.

Another important question is what one can learn from our results about the physics near the conifold point, where we do not have a well-understood low-energy effective description. Most of what we know comes from monodromy considerations and the space-time picture \cite{Huang:2006hq}. Our methods shed light on the behavior of the central charge of the brane near the conifold point. One hint of a possible approach comes from the cubic curve where we observed in (\ref{dual1}) and (\ref{dual2}) that the analytic continuation of the hemisphere partition function of a D0 and D2 to the singular point looks like a hemisphere partition function for a D2 and D0 on the cubic, respectively, where $e^{-t}$ has to be identified with $y=1-z$, instead of $z$. As for one of the phases, one can evaluate the hemiphere partition function at the conifold point by closing the contour in such a way that one obtains a convergent series in $y$. This is due to the fact that for the cubic, the monodromy behaviour around the large radius limit and the singular point is the same. More generally, we observe that one can write the double Mellin--Barnes integrals that appear in the analytic continuation in the form of a hemisphere partition function by separating bulk and brane contributions. One might then try to find a low energy theory which reproduces the bulk contribution.

While we have been working mostly with the relation to the geometric phase, our arguments can be extended in a straightforward manner to the Landau--Ginzburg phase. In this way, one can obtain a generalization of~\cite{Chiodo:2014ab} to the GLSM setting. This also allows us to address further mathematical aspects such as the variation of Hodge structure in the A--model, or Bridgeland's stability conditions for D--branes. Assuming the conjecture that the hemisphere partition function computes the exact central charge in a phase, the contour integral formula can be taken as a definition for the central charge, and one can try and verify whether this contour integral satisfies the axioms of such a stability condition. This would provide the first description which is intrinsic to the A--model and does not rely on mirror symmetry.

Beyond CYs and GLSMs one may ask if our results can be useful in other, not directly related contexts. Mellin-Barnes integrals play an important role in the computation of amplitudes in string theory and quantum field theory. It would be interesting to see if the Mellin-Barnes integrals we encounter also play a role in this context. Moreover, our different approaches to the analytic continuation lead to nontrivial identities between generalized hypergeometric functions. It might be interesting from a mathematical point of view to study these further.

\appendix
\section{Identities and Contour Integrals}
\label{app-identities}
Here we collect some essential identities for the Gamma function, for generalized hypergeometric functions at $z=1$ and certain contour integrals.
\subsection{Gamma Function Identities}
The most important identity is the reflection formula
\begin{align}
\label{reflection}
\Gamma(z)\Gamma(1-z)=\frac{\pi}{\sin\pi z}.
\end{align}
A further useful identity is the multiplication theorem
\begin{align}
\label{multiplication}
\prod_{k=0}^{N-1}\Gamma\left(z+\frac{k}{N}\right)=(2\pi)^{\frac{N-1}{2}}N^{\frac{1}{2}-Nz}\Gamma(Nz).
\end{align}
Furthermore recall that
\begin{align}
\label{plusone}
\Gamma(z+1)=z\Gamma(z).
\end{align}
The digamma function is defined as
\begin{align}
\psi(z)=\frac{d}{dz}\log\Gamma(z)=\frac{\Gamma'(z)}{\Gamma(z)}.
\end{align}
\subsection{Identities of generalized hypergeometric functions at $z=1$}
Using the Pochhammer symbol
\begin{align}
(a)_n=\frac{\Gamma(a+n)}{\Gamma(a)},
\end{align}
the generalized hypergeometric function is defined as
\begin{align}
{}_pF_q\left(\begin{array}{c}a_1,\ldots,a_p\\ b_1,\ldots,b_q\end{array};z\right)=\sum_{n=0}^{\infty}\frac{(a_1)_n\cdot\ldots\cdot(a_p)_n}{(b_1)_n\cdot\ldots\cdot(b_q)_n}\frac{z^n}{n!}.
\end{align}
This implies in particular:
\begin{align}
{}_pF_q\left(\begin{array}{c}a_1,\ldots,a_{p-1},c\\ b_1,\ldots,b_{q-1},c\end{array};z\right)={}_{p-1}F_{q-1}\left(\begin{array}{c}a_1,\ldots,a_{p-1}\\ b_1,\ldots,b_{q-1}\end{array};z\right).
\end{align}
Via Gau{\ss}'s theorem the hypergeometric function at $z=1$ can be written in terms of a quotient of Gamma functions:
\begin{align}
\label{gauss}
{}_2 F_1\left(\begin{array}{c}a,b\\c\end{array};1\right)=\frac{\Gamma(c)\Gamma(c-a-b)}{\Gamma(c-a)\Gamma(c-b)}\qquad \mathrm{Re}c>\mathrm{Re}(a+b).
\end{align}
The Thomae relations connect different generalized hypergeometric functions of type ${}_3F_2$ at $z=1$. The two identities we use are
\begin{align}
\label{thomae1}
{}_3F_2\left(\begin{array}{c}a,b,c\\d,e\end{array};1\right)&=\frac{\Gamma(e)\Gamma(d+e-a-b-c)}{\Gamma(e-a)\Gamma(d+e-b-c)}{}_3F_2\left(\begin{array}{c}a,d-b,d-c\\d,d+e-b-c\end{array};1\right)\\
{}_3F_2\left(\begin{array}{c}a,b,c\\d,e\end{array};1\right)&=\frac{\Gamma(d)\Gamma(e)\Gamma(d+e-a-b-c)}{\Gamma(b)\Gamma(d+e-a-b)\Gamma(d+e-b-c)}{}_3F_2\left(\begin{array}{c}d-b,e-b,d+e-a-b-c\\d+e-a-b,d+e-b-c\end{array};1\right).\label{thomae2}
\end{align}
Using the symmetries of the generalized hypergeometric function, further identities of this type can be generated. We also make use of Dixon's identity \cite{Slater:1966ab}
\begin{multline}
\label{dixon}
  \,{}_3 F_{2}\left(\genfrac{}{}{0pt}{}{a_1,a_2,
      a_3}{1+a_1-a_2,1+a_1-a_3};1 \right) \\=
  \frac{\Gamma(1+\frac{a_1}{2})\Gamma(1+\frac{a_1}{2}-a_2-a_3)\Gamma(1+a_1-a_2)\Gamma(1+a_1-a_3)}{\Gamma(1+a_1)\Gamma(1+a_1-a_2-a_3)\Gamma(1+\frac{a_1}{2}-a_2)\Gamma(1+\frac{a_1}{2}-a_3)}
\end{multline}
and a generalization due to Lavoie \cite{Lavoie:1994ab}, where we need the following special cases:
\begin{align}
\label{lavoie}
  &\Hyp{a_1,a_2,a_3}{a_1-a_2,1+a_1-a_3}{1} = \frac
  {{2}^{-2\,a_{{3}}}\Gamma ( a_{{1}}-a_{{2}} ) \Gamma (
    a_{{1}}-a_{{3}}+1 ) }{\Gamma ( a_{{1}}-2\,a
    _{{3}}+1 ) \Gamma  ( a_{{1}}-a_{{2}}-a_{{3}}+1 ) }\nonumber\\
  \cdot& \left( {\frac {\Gamma ( \frac{a_1}{2}-a_{{3}}+\frac{1}{2} )
        \Gamma ( \frac{a_1}{2}-a_{{2}}-a_{{3}}+1 ) }{\Gamma (
        \frac{a_1}{2}+\frac{1}{2} ) \Gamma ( \frac{a_1}{2}-a_{{2}} )
      }}+{\frac {\Gamma ( \frac{a_1}{2}-a_{{3}}+1 ) \Gamma (
        \frac{a_1}{2}-a_{{2}}-a_{{3}}+\frac{1}{2} ) }{\Gamma (
        \frac{a_1}{2} ) \Gamma (
        \frac{a_1}{2}-a_{{2}}+\frac{1}{2} ) }} \right)\\
&\Hyp{a_1,a_2,a_3}{2+a_1-a_2,1+a_1-a_3}{1} = \frac
  {{2}^{1-2\,a_{{2}}}\Gamma ( a_{{1}}-a_{{3}}+1 ) \Gamma
    ( a_{{1}}-a_{{2}}+2 ) \Gamma ( a_{{2}}-1 )
  }{\Gamma ( a_{{1}}-2\,a_{{2}}+2 ) \Gamma (
      a_{{1}}-a_{{2}}-a_{{3}}+2 ) \Gamma ( a_{{2}} )
  } \nonumber \\
&  \cdot \left( -{\frac {\Gamma ( \frac{a_1}{2}-a_{{2}}+\frac{3}{2} )
        \Gamma ( \frac{a_1}{2}-a_{{3}}-a_{{2}}+2 ) }{\Gamma
        ( \frac{a_1}{2}+\frac{1}{2} ) \Gamma (
          \frac{a_1}{2}-a_{{3}}+1 ) }}+{\frac {\Gamma (
          \frac{a_1}{2}-a_{{2}}+1 ) \Gamma (
          \frac{a_1}{2}-a_{{3}}-a_{{2}}+\frac{3}{2} ) }{\Gamma (
          \frac{a_1}{2} ) \Gamma ( \frac{a_1}{2}-a_{{3}}+\frac{1}{2}
        ) }} \right).\label{lavoie2}
\end{align}
There is a large number of further identities of ${}_3F_2$ for special values of the arguments. For $n\in\mathbb{Z}_{\geq 0}$ we use the identity
\begin{align}
\label{id3}
{}_3F_2\left(\begin{array}{c}a,b,-n\\d,e\end{array};1\right)&=\frac{\Gamma(e-b+n)\Gamma(e)}{\Gamma(e-b)\Gamma(e+n)}{}_3F_2\left(\begin{array}{c}-n,b,d-a\\d,1+b-e-n\end{array};1\right).
\end{align}
\subsection{Contour Integrals}
One of the key identities to analytically continue to the conifold point is the following:
\begin{align}
\label{identityMB}
\frac{1}{2\pi i}\int ds\Gamma(s-u)\Gamma(-s)(-z)^s=\Gamma(-u)(1-z)^u.
\end{align}
A well-known series of identities are the Barnes lemmas. The first Barnes lemma is
\begin{align}
\label{barnes1}
\int\frac{ds}{2\pi i}\Gamma(a+s)\Gamma(b+s)\Gamma(c-s)\Gamma(d-s)=\frac{\Gamma(a+c)\Gamma(a+d)\Gamma(b+c)\Gamma(b+d)}{\Gamma(a+b+c+d)}.
\end{align}
The second Barnes lemma is
\begin{align}
\label{barnes2}
&\int\frac{dz}{2\pi i}\frac{\Gamma(\alpha_1-z)\Gamma(\alpha_2-z)\Gamma(\beta_1+z)\Gamma(\beta_2+z)\Gamma(\beta_3+z)}{\Gamma(\alpha_1+\alpha_2+\beta_1+\beta_2+\beta_3+z)}\nonumber\\
&=\frac{\Gamma(\alpha_1+\beta_1)\Gamma(\alpha_2+\beta_2)\Gamma(\alpha_1+\beta_3)}{\Gamma(\alpha_1+\alpha_2+\beta_1+\beta_2)\Gamma(\alpha_1+\alpha_2+\beta_1+\beta_3)}\frac{\Gamma(\alpha_1+\beta_1)\Gamma(\alpha_2+\beta_2)\Gamma(\alpha_2+\beta_3)}{\Gamma(\alpha_1+\alpha_2+\beta_2+\beta_3)}.
\end{align}
Another useful formula is Jantzen's additional identity \cite{Jantzen:2012cb}:
\begin{align}
\label{jantzen}
\int \frac{ds}{2\pi i}\frac{\Gamma\left(s+\beta_1\right)\Gamma\left(s+\beta_2\right)\Gamma\left(u-s\right)}{\Gamma(\gamma+s)}e^{i\pi s}=e^{\pm i\pi u}\frac{\Gamma\left(u+\beta_1\right)\Gamma\left(u+\beta_2\right)\Gamma(-u+\gamma-\beta_1-\beta_2)}{\Gamma\left(\gamma-\beta_1\right)\Gamma\left(\gamma-\beta_2\right)}.
\end{align}
\section{Differential Equations and Bases of Solutions}
\label{app-pf}
Depending on the problem, we make use of various bases of periods. In this appendix we discuss how they are connected. In the large radius limit we use two types of bases: $\Phi_0$ and $\varpi_i$. The former appears naturally in the context of analytic continuation of GLSM branes, the latter is a standard basis of solutions to the Picard-Fuchs equations associated to the mirror CY. We give power series expansions of periods at large radius (coordinate $z$) in these two bases. Near the conifold point we choose a basis $\Phi_1$ of periods whose power series expansion we can obtain from solving the hypergeometric differential equation in the variable $y=1-z$. Finally we also define a basis $y^{\ast}$ of Mellin-Barnes representations of periods, following N{\o}rlund:
\begin{align}
y^*_q(z) = \int_{-i\infty}^{i\infty} ds\, \frac{\prod_{j=1}^{N-1}\Gamma\left(s+\frac{j}{N}\right)}{\Gamma(s+1)^{N-1}}\frac{z^s}{(1-e^{2\pi i s})^{q}}.
\end{align}
Since we are concerned with analytic continuation the relative normalization of the bases of periods at large radius and at the conifold point is important. We have chosen it in such a way that our results have a relatively simple form. Recall that we use the following differential operator:
\begin{align}
 \left( \theta\, \prod_{j=1}^{N-2} \left( \theta - \gamma_j\right) - z \prod_{j=1}^{N-1} \left(\theta - \alpha_j\right) \right),
\end{align}
where $\theta=z\frac{d}{dz}$. Note that the variable $z$ is chosen such that the singular points are at $\{0,1,\infty\}$. This differs from the standard convention in the mirror symmetry literature where the large radius variable is typically chosen as $z'=\frac{z}{5^5}$ so that the conifold point is at $z=N^{-N}$.
\subsection{Quintic}
For the quintic we choose $N=5$ and
\begin{align}
\alpha_i=\frac{i}{5}\quad\gamma_i=0\qquad  i=1,2,3,4.
\end{align}
The topological characteristics of the quintic are
\begin{align}
H^3=5\qquad c_2\cdot H=50\qquad c_3=-200.
\end{align}
The basis $\Phi_0$ in the large radius limit $z=0$ can be determined by the Frobenius method as described in the main text. The matrix of solutions can be written as
\begin{align}
\Phi_0(z) = S_0(z)z^{R_0}5^{-5R_0}C_0,
\end{align}
where
\begin{align}
\begin{aligned}
  R_0 &=
  \begin{pmatrix}
    0 & 1 & 0 & 0\\
    0 & 0 & 1 & 0\\
    0 & 0 & 0 & 1\\
    0 & 0 & 0 & 0
  \end{pmatrix},
  &
  C_0 &=
  \begin{pmatrix}
    1 & 0 & 0 & 0\\
    0 & \frac{1}{2\pi i} & 0 & 0\\
    0 & 0 &\frac{1}{(2\pi i)^2} &  0\\
    0 & 0 & 0 & \frac{1}{(2\pi i)^3} \\
  \end{pmatrix}
  \cdot
  \begin{pmatrix}
    1 & 0 & -\frac{25}{12} & \frac{200}{(2\pi i)^3} \zeta(3)\\
    0 & 1 & \frac{5}{2} & -\frac{25}{12}\\
    0 & 0 & 5 & 0\\
    0 & 0 & 0 & -5
  \end{pmatrix}.
\end{aligned}
\end{align}
The choice of $C_0$ follows from~\cite{Hosono:2000eb}.
To construct the matrix $S_{0}(z)$ we only need to know the first row, which is given in terms of the following power series expansions:
\begin{align}
S_{0,11}&=1+\frac{24 z}{625}+\frac{4536 z^2}{390625}+\frac{1345344 z^3}{244140625}+\frac{488864376z^4}{152587890625}+O\left(z^5\right) \\
S_{0,12}&=\frac{154 z}{625}+\frac{32409 z^2}{390625}+\frac{29965432 z^3}{732421875}+\frac{296135721
   z^4}{12207031250}+O\left(z^5\right) \\
S_{0,13}&= \frac{23 z}{125}+\frac{168327 z^2}{1562500}+\frac{135716176 z^3}{2197265625}+\frac{57606926969 z^4}{1464843750000}+O\left(z^5\right)\\
S_{0,14}&=-\frac{46 z}{125}-\frac{26387 z^2}{312500}-\frac{373292959 z^3}{13183593750}-\frac{104105463971z^4}{8789062500000}+O\left(z^5\right)
\end{align}
Evaluating the Mellin-Barnes representation $y^{\ast}_j$ and the solution $\xi_4$ in the large radius phase we find the following relation to the basis $\Phi_0(z)$:
\begin{align}
y_1^{\ast}(z)&=\frac{4\pi^2}{\sqrt{5}}\Phi_{0,11}\\
y_2^{\ast}(z)-y_1^{\ast}(z)&=-\frac{4\pi^2}{\sqrt{5}}\Phi_{0,12}\\
y_3^{\ast}(z)-2y_2^{\ast}(z)+y_1^{\ast}(z)&=\frac{4\pi^2}{\sqrt{5}}\Phi_{0,13}\\
\frac{\sqrt{5}}{4\pi^2}\xi_4(z)=y_4^{\ast}-2y_3^{\ast}+2y_2^{\ast}-y_1^{\ast}&=\frac{1}{5}\frac{1}{2\pi i}\Phi_{0,14}
\end{align}
Note that $\frac{4\pi^2}{\sqrt{5}}=\Gamma\left(\frac{1}{5}\right)\Gamma\left(\frac{2}{5}\right)\Gamma\left(\frac{3}{5}\right)\Gamma\left(\frac{4}{5}\right)$. In the context of D-branes in the GLSM it is convenient to use another standard large radius basis $\varpi_i$ ($i=0,1,2,3$) which is related to the basis $\Phi_0$ in the following way:
\begin{align}
\label{eq:PhiBasis}
\Phi_{0,11}&=\varpi_0\left(\frac{z}{5^5}\right)\\
\Phi_{0,12}&=\varpi_1\left(\frac{z}{5^5}\right)\\
\Phi_{0,13}&=\frac{H^3}{2}\varpi_2\left(\frac{z}{5^5}\right)+\frac{H^3}{2}\varpi_1\left(\frac{z}{5^5}\right)-\frac{c_2\cdot H}{24}\varpi_0\left(\frac{z}{5^5}\right)\\
\Phi_{0,14}&=-\left[\frac{H^3}{6}\varpi_3\left(\frac{z}{5^5}\right)+\frac{c_2\cdot H}{24}\varpi_1\left(\frac{z}{5^5}\right)+\frac{c_3\zeta(3)}{(2\pi i)^3}\varpi_0\left(\frac{z}{5^5}\right) \right].
\end{align}
Near the singular point the basis of solutions can be written as power series in $y=1-z$ with
\begin{align}
\Phi_1(y) = S_1(y)y^{R_1}C_1,
\end{align}
where
\begin{align}
\begin{aligned}
  R_1 &=
  \begin{pmatrix}
    0 & 0 & 0 & 0 \\
    0 & 1 & 1 & 0\\
    0 & 0 & 1 & 0\\
    0 & 0 & 0 & 2\\
  \end{pmatrix},
  &
  C_1 &= \frac{\sqrt{5}}{4\pi^2}
  \begin{pmatrix}
    1 & 0 & 0 & 0 \\
    0 & 1 & 0 & 0\\
    0 & 0 & 1 & 0\\
    0 & 0 & 0 & 1\\
  \end{pmatrix}.
\end{aligned}
\end{align}
The first line in the matrix $S_1(y)$ is
\begin{align}
S_{1,11}&=1+\frac{2 y^3}{625}+\frac{97 y^4}{18750}+\frac{2971
   y^5}{468750}+O\left(y^6\right)\\
S_{1,12}&=1+\frac{7 y}{10}+\frac{41 y^2}{75}+\frac{1133 y^3}{2500}+\frac{6089y^4}{15625}+\frac{160979 y^5}{468750}+O\left(y^6\right)\\
S_{1,13}&=-\frac{23 y^3}{360}-\frac{6397 y^4}{60000}-\frac{333323y^5}{2500000}+O\left(y^6\right)\\
S_{1,14}&=1+\frac{37 y}{30}+\frac{2309 y^2}{1800}+\frac{286471
   y^3}{225000}+\frac{41932661 y^4}{33750000}+\frac{237108737
   y^5}{196875000}+O\left(y^6\right)
\end{align}
\subsection{Quartic}
\label{app-quartic-periods}
The quartic corresponds to $N=4$ with
\begin{align}
\alpha_i=\frac{i}{4}\quad\gamma_i=0\qquad i=1,2,3.
\end{align}
The topological characteristics of the quartic are
\begin{align}
H^2=4\qquad c_2\cdot H=6.
\end{align}
The basis $\Phi_0(z)$ in the large radius limit is
\begin{align}
\Phi_0(z) = S_0(z) z^{R_0} 4^{-4R_0}C_0
\end{align}
with
\begin{align}
\begin{aligned}
  R_0 &=
  \begin{pmatrix}
    0 & 1 & 0 \\
    0 & 0 & 1 \\
    0 & 0 & 0
  \end{pmatrix},
  &
  C_0 &=
  \begin{pmatrix}
    1 & 0 & \frac{1}{4} \\
    0 & \frac{1}{2\pi i} & 0 \\
    0 & 0 & \frac{1}{(2\pi i)^2}
  \end{pmatrix}.
\end{aligned}
\end{align}
The first line in the matrix $S_0(z)$ is given by the following power series expansions:
\begin{align}
S_{0,11}&=1+\frac{3 z}{32}+\frac{315 z^2}{8192}+\frac{5775 z^3}{262144}+\frac{7882875 z^4}{536870912}+\frac{183324141z^5}{17179869184}+O\left(z^6\right)\\
S_{0,12}&=\frac{13 z}{32}+\frac{3069 z^2}{16384}+\frac{176005 z^3}{1572864}+\frac{163635325 z^4}{2147483648}+\frac{19276992819z^5}{343597383680}+O\left(z^6\right) \\
S_{0,13}&=\frac{169 z^2}{2048}+\frac{35841 z^3}{524288}+\frac{86041595 z^4}{1610612736}+\frac{2917954325 z^5}{68719476736}+O\left(z^6\right)
\end{align}
The relation to the basis of Mellin-Barnes integrals is
\begin{align}
y_1^{\ast}&=\sqrt{2}\pi^{\frac{3}{2}}\Phi_{0,11}\\
y_2^{\ast}-y_1^{\ast}&=-\sqrt{2}\pi^{\frac{3}{2}}\Phi_{0,12}\\
-\frac{\pi}{\Gamma\left(\frac{3}{2}\right)}\xi_3=2y_1^{\ast}-3y_2^{\ast}+2y_3^{\ast}&=2\sqrt{2}\pi^\frac{3}{2}\Phi_{0,13}.
\end{align}
Note that $\sqrt{2}\pi^{\frac{3}{2}}=\Gamma\left(\frac{1}{4}\right)\Gamma\left(\frac{1}{2}\right)\Gamma\left(\frac{3}{4}\right)$. When we consider a basis of GLSM branes, we also use the basis $\varpi_i$ ($i=0,1,2$) where
\begin{align}
\Phi_{0,11}&=\varpi_0\left(\frac{z}{4^4}\right)\\
\Phi_{0,12}&=\varpi_1\left(\frac{z}{4^4}\right)\\
\Phi_{0,13}&=\frac{1}{2}\varpi_2\left(\frac{z}{4^4}\right)+\frac{1}{4}\varpi_0\left(\frac{z}{4^4}\right)
\end{align}
Near the conifold point the solutions are
\begin{align}
\Phi_1(y) = S_1(y) y^{R_1}
\end{align}
with
\begin{align}
 R_1 =
  \begin{pmatrix}
    0 & 0 & 0 \\
    0 & 1 & 0 \\
    0 & 0 & \frac{1}{2}
  \end{pmatrix}
\end{align}
and
\begin{align}
S_{1,11}&=1-\frac{y^2}{32}-\frac{131 y^3}{3840}-\frac{9407
   y^4}{286720}-\frac{211711 y^5}{6881280}+O\left(y^6\right) \\
S_{1,12}&=1+\frac{35 y}{48}+\frac{665 y^2}{1152}+\frac{5915
   y^3}{12288}+\frac{122395 y^4}{294912}+\frac{57015413
   y^5}{155713536}+O\left(y^6\right) \\
S_{1,13}&=1+\frac{11 y}{24}+\frac{39 y^2}{128}+\frac{1181
   y^3}{5120}+\frac{385397 y^4}{2064384}+\frac{1361519
   y^5}{8650752}+O\left(y^{6}\right) .
\end{align}
\subsection{Cubic}
For completeness we also discuss the cubic with $N=3$, where
\begin{align}
\alpha_1=\frac{i}{3}\quad\gamma_i=0\qquad i=1,2.
\end{align}
The solutions around $z=0$ can be written as $\Phi_0=S_0(z)z^{R_0}3^{-3R_0}C_0$ with
\begin{align}
R_0=\left(\begin{array}{cc}0&1\\0&0\end{array}\right),\qquad C_0=\left(\begin{array}{cc}1&0\\0&\frac{3}{2\pi i} \end{array}\right).
\end{align}
The power series expansions in the first line of the matrix $S_0(z)$ are
\begin{align}
S_{0,11}&=1+\frac{8 z}{9}+\frac{280 z^2}{81}+\frac{123200 z^3}{6561}+\frac{7007000 z^4}{59049}+\frac{144848704  z^5}{177147}+O\left(z^6\right)\\
S_{0,12}&=\frac{104 z}{27}+\frac{1364 z^2}{81}+\frac{5632160 z^3}{59049}+\frac{327270650 z^4}{531441}+\frac{11423403152 z^5}{2657205}+O\left(z^6\right)
\end{align}
The relation to the basis of Mellin-Barnes integrals is
\begin{align}
y_1^{\ast}(z)&=\frac{2\pi}{\sqrt{3}}\Phi_{0,11}\\
\frac{1}{3}\xi_2=y_2^{\ast}(z)-y_1^{\ast}(z)&=-\frac{2\pi}{3\sqrt{3}}\Phi_{0,12},
\end{align}
where $\frac{2\pi}{\sqrt{3}}=\Gamma\left(\frac{1}{3}\right)\Gamma\left(\frac{2}{3}\right)$. The relation to the other large radius basis $\varpi_i$ ($i=0,1$) is: $\Phi_{0,11}=\varpi_0\left(\frac{z}{3^3}\right)$ and $\Phi_{0,12}=3\varpi_1\left(\frac{z}{3^3}\right)$.

One can easily show that the differential operator transforms into itself under the coordinate change $z\rightarrow y=1-z$. Therefore the solutions near the the conifold point look the same as at large radius and we choose $\Phi_1=\Phi_0(y)$.
\section{Further  GLSM branes on the Quintic and Quartic}
\label{app-grr}
In the following we discuss a set of $(D0,D2,D4)$-branes on the quartic which we can analytically continue to the conifold point. Before that we also discuss further examples of geometric branes on the quintic. These examples are necessary for understanding the D4 branes we encounter in the GLSM and in the mathematical approach for analytic continuation. They further show how the algorithmic approach to grade restriction works in non-trivial examples.
\subsection{Quintic}
\subsubsection{D0}
We have identified the D0 brane with minimal charge as a permutation-type GLSM matrix factorization at the Fermat point. Another example on the quintic is a complete intersection of the generic quintic $G_5(x)=0$ with three divisors $h_i=\sum_{j=1}^5\alpha_j^ix_i$ ($i=1,2,3$). This information can easily be encoded in the matrix factorization
\begin{align}
Q=h_1\eta_1+h_2\eta_2+h_3\eta_3+G_5\eta_4+p\bar{\eta}_4.
\end{align}
With a suitable normalization of $\rho$ and $\mathbf{r}_{*}$ this defines the following GLSM brane: 
\begin{align}
\widehat{\mathcal{W}}_{D0}:\quad\xymatrix{\mathcal{W}(0)_+\ar@<2pt>[r] & \ar@<2pt>[l]{\begin{array}{c}\mathcal{W}(1)_-^{\oplus 3}\\\oplus\\\mathcal{W}(5)_-\end{array}}\ar@<2pt>[r] & \ar@<2pt>[l]{\begin{array}{c}\mathcal{W}(2)_+^{\oplus 3}\\\oplus\\\mathcal{W}(6)_+^{\oplus 3}\end{array}}\ar@<2pt>[r] & \ar@<2pt>[l]{\begin{array}{c}\mathcal{W}(3)_-\\\oplus\\\mathcal{W}(7)_-^{\oplus 3}\end{array}}\ar@<2pt>[r] & \ar@<2pt>[l]\mathcal{W}(8)_+
}.
\end{align}
The brane is clearly not grade restricted. The brane factor is
\begin{align}
f_{\widehat{\mathcal{W}}_{D0}}=1-3e^{2\pi\sigma}+3e^{4\pi\sigma}-e^{6\pi\sigma}-e^{10\pi\sigma}+3e^{12\pi\sigma}-3e^{14\pi\sigma}+e^{16\pi\sigma}.
\end{align}
The hemisphere partition evaluated in the large radius phase is $Z_{D^2}(\widehat{\mathcal{W}}_{D0})=5\varpi_0=5\Phi_{0,11}$. To grade restrict this example to the charge window $q\in\{0,1,2,3,4\}$ we have to get rid of the unwanted Wilson line branes $\mathcal{W}(5)$, $\mathcal{W}(6)^{\oplus 3}$, $\mathcal{W}(5)^{\oplus 3}$ and $\mathcal{W}(8)$. This leads to a whole cascade of empty branes that needs to be bound to grade restrict. For the readers' amusement and as a means to demonstrate that the algorithmic approach to grade restriction indeed works, we collect the necessary steps in the following table.
\begin{align}
\label{d0grr}
\begin{array}{ccccccccc|c}
-&+&-&+&-&+&-&+&-&\#\\
\hline
&&&\mathcal{W}(0)&\mathcal{W}(1)^{\oplus 3}&\mathcal{W}(2)^{\oplus 3}&\mathcal{W}(3)&&&-\\
&&&&{\color{color4}\mathcal{W}(5)}&{\color{color3}\mathcal{W}(6)^{\oplus 3}}&{\color{color2}\mathcal{W}(7)^{\oplus 3}}&{\color{color1}\mathcal{W}(8)}&&-\\
\hline
&&&\mathcal{W}(3)&\mathcal{W}(4)^{\oplus 5}&{\color{color7}\mathcal{W}(5)^{\oplus 10}}&{\color{color6}\mathcal{W}(6)^{\oplus 10}}&{\color{color5}\mathcal{W}(7)^{\oplus 5}}&{\color{color1}\mathcal{W}(8)}&1\\
&&\mathcal{W}(2)&\mathcal{W}(3)^{\oplus 5}&\mathcal{W}(4)^{\oplus 10}&{\color{color9}\mathcal{W}(5)^{\oplus 10}}&{\color{color8}\mathcal{W}(6)^{\oplus 5}}&{\color{color2}\mathcal{W}(7)}&&3\\
&\mathcal{W}(1)&\mathcal{W}(2)^{\oplus 5}&\mathcal{W}(3)^{\oplus 10}&\mathcal{W}(4)^{\oplus 10}&{\color{color10}\mathcal{W}(5)^{\oplus 5}}&{\color{color3}\mathcal{W}(6)}&&&3\\
\mathcal{W}(0)&\mathcal{W}(1)^{\oplus 5}&\mathcal{W}(2)^{\oplus 10}&\mathcal{W}(3)^{\oplus 10}&\mathcal{W}(4)^{\oplus 5}&{\color{color4}\mathcal{W}(5)}&&&&1\\
&&&\mathcal{W}(2)&\mathcal{W}(3)^{\oplus 5}&\mathcal{W}(4)^{\oplus 10}&{\color{color12}\mathcal{W}(5)^{\oplus 10}}&{\color{color11}\mathcal{W}(6)^{\oplus 5}}&{\color{color5}\mathcal{W}(7)}&5\\
&&\mathcal{W}(1)&\mathcal{W}(2)^{\oplus 5}&\mathcal{W}(3)^{\oplus 10}&\mathcal{W}(4)^{\oplus 10}&{\color{color13}\mathcal{W}(5)^{\oplus 5}}&{\color{color6}\mathcal{W}(6)}&&10\\
&\mathcal{W}(0)&\mathcal{W}(1)^{\oplus 5}&\mathcal{W}(2)^{\oplus 10}&\mathcal{W}(3)^{\oplus 10}&\mathcal{W}(4)^{\oplus 5}&{\color{color7}\mathcal{W}(5)}&&&10\\
&&\mathcal{W}(1)&\mathcal{W}(2)^{\oplus 5}&\mathcal{W}(3)^{\oplus 10}&\mathcal{W}(4)^{\oplus 10}&{\color{color14}\mathcal{W}(5)^{\oplus 5}}&{\color{color8}\mathcal{W}(6)}&&15\\
&\mathcal{W}(0)&\mathcal{W}(1)^{\oplus 5}&\mathcal{W}(2)^{\oplus 10}&\mathcal{W}(3)^{\oplus 10}&\mathcal{W}(4)^{\oplus 5}&{\color{color9}\mathcal{W}(5)}&&&30\\
&\mathcal{W}(0)&\mathcal{W}(1)^{\oplus 5}&\mathcal{W}(2)^{\oplus 10}&\mathcal{W}(3)^{\oplus 10}&\mathcal{W}(4)^{\oplus 5}&{\color{color10}\mathcal{W}(5)}&&&15\\
&&&\mathcal{W}(1)&\mathcal{W}(2)^{\oplus 5}&\mathcal{W}(3)^{\oplus 10}&\mathcal{W}(4)^{\oplus 10}&{\color{color15}\mathcal{W}(5)^{\oplus 5}}&{\color{color11}\mathcal{W}(6)}&25\\
&&\mathcal{W}(0)&\mathcal{W}(1)^{\oplus 5}&\mathcal{W}(2)^{\oplus 10}&\mathcal{W}(3)^{\oplus 10}&\mathcal{W}(4)^{\oplus 5}&{\color{color12}\mathcal{W}(5)}&&50\\
&&\mathcal{W}(0)&\mathcal{W}(1)^{\oplus 5}&\mathcal{W}(2)^{\oplus 10}&\mathcal{W}(3)^{\oplus 10}&\mathcal{W}(4)^{\oplus 5}&{\color{color13}\mathcal{W}(5)}&&50\\
&&\mathcal{W}(0)&\mathcal{W}(1)^{\oplus 5}&\mathcal{W}(2)^{\oplus 10}&\mathcal{W}(3)^{\oplus 10}&\mathcal{W}(4)^{\oplus 5}&{\color{color14}\mathcal{W}(5)}&&75\\
&&&\mathcal{W}(0)&\mathcal{W}(1)^{\oplus 5}&\mathcal{W}(2)^{\oplus 10}&\mathcal{W}(3)^{\oplus 10}&\mathcal{W}(4)^{\oplus 5}&{\color{color15}\mathcal{W}(5)}&125
\end{array}
\end{align}
Calling this brane $\mathcal{W}_{D0}$ and extracting the brane factor out of this we obtain
\begin{align}
\label{geomd0}
f_{\mathcal{W}_{D0}}=5(1-e^{2\pi\sigma})^4,
\end{align}
which is, as expected, five times the brane factor of the permutation-type D0-brane.
\subsubsection{D2}
Analogously we can construct a D2-brane on the quintic. We take two divisors $h_1$ and $h_2$ as above and intersect it with $G_5(x)=0$. This lifts to a matrix factorization
\begin{align}
Q=h_1\eta_1+h_2\eta_2+G_5\eta_3+p\bar{\eta}_3,
\end{align}
which we can associate with the following non-grade-restricted GLSM brane:
\begin{align}
\widehat{\mathcal{W}}_{D2}:\quad \xymatrix{\mathcal{W}(-1)_-\ar@<2pt>[r] & \ar@<2pt>[l]{\begin{array}{c}\mathcal{W}(0)_+^{\oplus 2}\\\oplus\\\mathcal{W}(4)_+\end{array}}\ar@<2pt>[r] & \ar@<2pt>[l]{\begin{array}{c}\mathcal{W}(1)_-\\\oplus\\\mathcal{W}(5)_-^{\oplus 2}\end{array}}\ar@<2pt>[r] & \ar@<2pt>[l]\mathcal{W}(6)_+
}.
\end{align}
The brane factor is
\begin{align}
f_{\widehat{\mathcal{W}}_{D2}}=-e^{-2\pi\sigma}+2+e^{8\pi\sigma}-e^{2\pi\sigma}-2e^{10\pi\sigma}+e^{12\pi\sigma}.
\end{align}
From this we get the hemisphere partition function $Z_{D^2}^{\zeta\gg0}(\widehat{\mathcal{W}}_{D2})=5\varpi_1=5\Phi_{0,12}$ in the large radius phase.
The grade restriction procedure is simpler than for the D0 brane:
\begin{align}
\begin{array}{ccccccc|c}
-&+&-&+&-&+&-&\#\\
\hline
&&{\color{color1}\mathcal{W}(-1)}&\mathcal{W}(0)^{\oplus 2}&\mathcal{W}(1)&&&-\\
&&&\mathcal{W}(4)&{\color{color3}\mathcal{W}(5)^{\oplus 2}}&{\color{color2}\mathcal{W}(6)}&&-\\
\hline
&{\color{color1}\mathcal{W}(-1)}&\mathcal{W}(0)^{\oplus 5}&\mathcal{W}(1)^{\oplus 10}&\mathcal{W}(2)^{\oplus 10}&\mathcal{W}(3)^{\oplus 5}&\mathcal{W}(4)&1\\
&\mathcal{W}(1)&\mathcal{W}(2)^{\oplus 5}&\mathcal{W}(3)^{\oplus 10}&\mathcal{W}(4)^{\oplus 10}&{\color{color4}\mathcal{W}(5)^{\oplus 5}}&{\color{color2}\mathcal{W}(6)}&1\\
\mathcal{W}(0)&\mathcal{W}(1)^{\oplus 5}&\mathcal{W}(2)^{\oplus 10}&\mathcal{W}(3)^{\oplus 10}&\mathcal{W}(4)^{\oplus 5}&{\color{color3}\mathcal{W}(5)}&&2\\
&\mathcal{W}(0)&\mathcal{W}(1)^{\oplus 5}&\mathcal{W}(2)^{\oplus 10}&\mathcal{W}(3)^{\oplus 10}&\mathcal{W}(4)^{\oplus 5}&{\color{color4}\mathcal{W}(5)}&5
\end{array}
\end{align}
The grade-restricted brane factor of the resulting brane $\mathcal{W}_{D2}$ is
\begin{align}
\label{geomd2}
f_{\mathcal{W}_{D2}}=-5e^{2\pi\sigma}(1-e^{2\pi\sigma})^3.
\end{align}
\subsection{Quartic}
We have not discussed any GLSM branes on the quartic in the main text since their construction is completely analogous to the examples discussed for the cubic and the quintic. However, since we have been successful in analytically continuing a basis of solutions of the hypergeometric equation on the quartic to the conifold point with two different methods we find it necessary to complete the discussion by explicitly giving the GLSM branes related to these Mellin-Barnes integrals.
\subsubsection{D0}
As for the quintic, there are two canonical ways to construct the D0 brane: the object of minimal charge is the GLSM-lift of a permutation brane on the Fermat quartic. Alternatively we can also construct a ``geometric'' D0 as a complete intersection of two linear divisors with the hypersurface equation. Let us start with the permutation brane. In a Landau-Ginzburg framework this has already been discussed in \cite{Brunner:2006tc}. To discuss the particular GLSM-lift of the matrix factorization on the Fermat quartic we are interested in, we define:
\begin{align}
f_1&=x_1-\kappa_1 x_2\\
g_1&=x_1^3+\kappa_1x_1^2x_2+\kappa_1^2 x_1x_2^2+\kappa_1^3x_2^3,
\end{align}
with $\kappa_1^4=-1$. Choosing a $2^3$-dimensional Clifford basis, we consider the matrix factorization
\begin{align}
Q=f_1\eta_1+x_3\eta_2+x_4\eta_3+p g_1\bar{\eta}_1 +p x_3^3\bar{\eta}_2+ px_4^3\bar{\eta}_3.
\end{align}
To this, we can associate the Koszul brane $K((f_1,x_3,x_4),\mathcal{W}(0)_-)$:
\begin{align}
\mathcal{W}_{\mathrm{pt}}:\quad \xymatrix{\mathcal{W}(0)_-\ar@<2pt>[r] & \ar@<2pt>[l]\mathcal{W}(1)_+^{\oplus 3}\ar@<2pt>[r] & \ar@<2pt>[l]\mathcal{W}(2)_-^{\oplus 3}\ar@<2pt>[r] & \ar@<2pt>[l]\mathcal{W}(2)_+.
}
\end{align}
This is automatically grade restricted with respect to the window $\{0,1,2,3\}$. The brane factor is
\begin{align}
f_{\mathcal{W}_{\mathrm{pt}}}=(1-e^{2\pi\sigma})^3.
\end{align}
The hemisphere partition function, evaluated in the large radius phase is $Z_{D^2}^{\zeta\gg0}(\mathcal{W}_{\mathrm{pt}})=\varpi_0=\Phi_{0,11}$. Comparing with the basis $y_j^{\ast}$ of Mellin-Barnes integrals, the correspondence is
\begin{align}
\label{quarticd0}
Z_{D^2}(\mathcal{W}_{\mathrm{pt}})=\frac{1}{\sqrt{2}\pi^{\frac{3}{2}}} y_1^{\ast}.
\end{align}
Geometrically we can construct the D0-brane by intersecting two divisors $h_1,h_2$ with the hypersurface equation $G_4(x)=0$ with the generic quartic. The matrix factorization is exactly the same as the matrix factorization of a D2 on the quintic, with $G_5\rightarrow G_4$. The associated GLSM brane is
\begin{align}
\widehat{\mathcal{W}}_{D0}:\quad \xymatrix{\mathcal{W}(0)_+\ar@<2pt>[r] & \ar@<2pt>[l]{\begin{array}{c}\mathcal{W}(1)_-^{\oplus 2}\\\oplus\\\mathcal{W}(4)_-\end{array}}\ar@<2pt>[r] & \ar@<2pt>[l]{\begin{array}{c}\mathcal{W}(5)_+\\\oplus\\\mathcal{W}(5)_+^{\oplus 2}\end{array}}\ar@<2pt>[r] & \ar@<2pt>[l]\mathcal{W}(6)_-
}.
\end{align}
This is not grade restricted and the brane factor is
\begin{align}
f_{\widehat{\mathcal{W}}_{D0}}=1-2e^{2\pi \sigma}+e^{4\pi\sigma}-e^{8\pi\sigma}+2e^{10\pi\sigma}-e^{12\pi\sigma}.
\end{align}
The hemisphere partition in the large radius phase is $Z_{D^2}^{\zeta\gg0}(\widehat{\mathcal{W}}_{D0})=4\varpi_0=4\Phi_{0,11}$, which reflects that this construction leads to four point-like objects on the quartic. In order to make contact with the Mellin-Barnes integrals we have to grade-restrict to obtain a brane $\mathcal{W}_{D0}$. The steps are summarized in the table below.
\begin{align}
\begin{array}{ccccccc|c}
+&-&+&-&+&-&+&\#\\
\hline
&&\mathcal{W}(0)&\mathcal{W}(1)^{\oplus 2}&\mathcal{W}(2)&&&-\\
&&&{\color{color3}\mathcal{W}(4)}&{\color{color2}\mathcal{W}(5)^{\oplus 2}}&{\color{color1}\mathcal{W}(6)}&&-\\
\hline
&&\mathcal{W}(2)&\mathcal{W}(3)^{\oplus 4}&{\color{color5}\mathcal{W}(4)^{\oplus 6}}&{\color{color4}\mathcal{W}(5)^{\oplus 4}}&{\color{color1}\mathcal{W}(6)}&1\\
&\mathcal{W}(1)&\mathcal{W}(2)^{\oplus 4}&\mathcal{W}(3)^{\oplus 6}&{\color{color6}\mathcal{W}(4)^{\oplus 4}}&{\color{color2}\mathcal{W}(5)}&&2\\
\mathcal{W}(0)&\mathcal{W}(1)^{\oplus 4}&\mathcal{W}(2)^{\oplus 6}&\mathcal{W}(3)^{\oplus 4}&{\color{color3}\mathcal{W}(4)}&&&1\\
&&\mathcal{W}(1)&\mathcal{W}(2)^{\oplus 4}&\mathcal{W}(3)^{\oplus 6}&{\color{color7}\mathcal{W}(4)^{\oplus 4}}&{\color{color4}\mathcal{W}(5)}&4\\
&\mathcal{W}(0)&\mathcal{W}(1)^{\oplus 4}&\mathcal{W}(2)^{\oplus 6}&\mathcal{W}(3)^{\oplus 4}&{\color{color5}\mathcal{W}(4)}&&6\\
&\mathcal{W}(0)&\mathcal{W}(1)^{\oplus 4}&\mathcal{W}(2)^{\oplus 6}&\mathcal{W}(3)^{\oplus 4}&{\color{color6}\mathcal{W}(4)}&&8\\
&&\mathcal{W}(0)&\mathcal{W}(1)^{\oplus 4}&\mathcal{W}(2)^{\oplus 6}&\mathcal{W}(3)^{\oplus 4}&{\color{color7}\mathcal{W}(4)}&16
\end{array}
\end{align}
This results in the following grade-restricted brane factor
\begin{align}
f_{\mathcal{W}_{D0}}=4(1-e^{2\pi\sigma})^3,
\end{align}
which we can identify with our basis of Mellin-Barnes integrals:
\begin{align}
Z_{D^{2}}(\mathcal{W}_{D0})=\frac{4}{\sqrt{2}\pi^{\frac{3}{2}}} y_1^{\ast}.
\end{align}
\subsubsection{D2}
The D2-brane with minimal charge is also described by the GLSM lift of a specific permutation brane of the Fermat quartic to the GLSM whose matrix factorization is
\begin{align}
Q=f_1\eta_1+f_2\eta_2+p g_1\bar{\eta}_1+p g_2\bar{\eta}_2
\end{align}
with
\begin{align}
f_1&=x_1-\kappa_1 x_2\\
f_2&=x_3-\kappa_2 x_4\\
g_1&=x_1^3+\kappa_1x_1^2x_2+\kappa_1^2 x_1x_2^2+\kappa_1^3x_2^3\\
g_2&= x_3^3+\kappa_2x_3^2x_4+\kappa_2^2 x_3x_4^2+\kappa_2^3x_4^3,
\end{align}
where $\kappa_1^4=\kappa_2^4=-1$.
The associated Koszul brane $K((f_1,f_2),\mathcal{W}(1)_-)$ is grade restricted:
\begin{align}
\mathcal{W}_{\ell}:\quad\xymatrix{\mathcal{W}(1)_-\ar@<2pt>[r] & \ar@<2pt>[l]\mathcal{W}(2)_+^{\oplus 2}\ar@<2pt>[r] & \ar@<2pt>[l]\mathcal{W}(3)_-
}
\end{align}
and the brane factor is
\begin{align}
f_{\mathcal{W}_{\ell}}=-e^{2\pi\sigma}(1-e^{2\pi\sigma})^2.
\end{align}
Evaluated in the large radius phase we get
\begin{align}
Z_{D^2}^{\zeta\gg0}(\mathcal{W}_{\ell})=\varpi_1=\Phi_{0,12}.
\end{align}
We match this with the following expression in terms of the Mellin-Barnes basis:
\begin{align}
\label{quarticd2}
Z_{D^2}(\mathcal{W}_{\ell})=-\frac{1}{\sqrt{2}\pi^{\frac{3}{2}}}(y_2^{\ast}-y_1^{\ast}).
\end{align}
For completeness, we also give an example of a D2-brane which is constructed as an intersection of a divisor $h$ with the hypersurface equation $G_4(x)$. The matrix factorization $Q=h\eta_1+G_4\eta_2+p\bar{\eta}_2$ is the same as the one for the D4 brane on the quintic discussed in the main text. The associated GLSM brane
\begin{align}
\widehat{\mathcal{W}}_{D2}:\quad\xymatrix{\mathcal{W}(-1)_-\ar@<2pt>[r]& \ar@<2pt>[l]{\begin{array}{c}\mathcal{W}(0)_+\\\oplus\\\mathcal{W}(3)_+\end{array}}\ar@<2pt>[r] & \ar@<2pt>[l]\mathcal{W}(4)_-
}.
\end{align}
The brane is not grade restricted and the brane factor is
\begin{align}
f_{\widehat{\mathcal{W}}_{D2}}=-e^{-2\pi\sigma}+1+e^{6\pi\sigma}-e^{8\pi\sigma}.
\end{align}
The hemisphere partition function in the large radius phase is
\begin{align}
Z_{D^2}^{\zeta\gg0}(\widehat{\mathcal{W}}_{D2})=4\varpi_1-2\varpi_0.
\end{align}
Grade-restriction is achieved by binding two empty branes:
\begin{align}
\mathcal{W}_{D2}:\quad \begin{array}{ccccc|c}
+&-&+&-&+&\#\\
\hline
&{\color{color2}\mathcal{W}(-1)}&\mathcal{W}(0)&&&-\\
&&\mathcal{W}(3)&{\color{color1}\mathcal{W}(4)}&&-\\
\hline
\mathcal{W}(0)&\mathcal{W}(1)^{\oplus 4}&\mathcal{W}(2)^{\oplus 6}&\mathcal{W}(3)^{\oplus 4}&{\color{color1}\mathcal{W}(4)}&1\\
{\color{color2}\mathcal{W}(-1)}&\mathcal{W}(0)^{\oplus 4}&\mathcal{W}(1)^{\oplus 6}&\mathcal{W}(2)^{\oplus 4}&\mathcal{W}(3)&1
\end{array}
\end{align}
The associated brane factor is
\begin{align}
f_{\mathcal{W}_{D2}}=-2(1-e^{2\pi\sigma}-e^{4\pi\sigma}+e^{6\pi\sigma}).
\end{align}
The hemisphere partition function for this brane factor can be expressed as the following combination of Mellin-Barnes integrals:
\begin{align}
Z_{D^2}(\mathcal{W}_{D2})=\frac{2}{\sqrt{2}\pi^{\frac{3}{2}}}(-2y_2^{\ast}+y_1^{\ast}).
\end{align}
\subsubsection{Structure Sheaf}
As usual, the GLSM lift $\widehat{\mathcal{W}}_X$ of structure sheaf is given by the matrix factorization $Q=G_4\eta+p\bar{\eta}$ with $f_{\widehat{\mathcal{W}}_X}=1-e^{8\pi\sigma}$. This is not grade restricted. In the large radius phase the hemisphere partition function is
\begin{align}
Z_{D^2}^{\zeta\gg0}(\widehat{\mathcal{W}}_X)=(2\varpi_2+\varpi_0)=2\Phi_{0,13}.
\end{align}
Grade restriction is almost trivial and leads to the following GLSM brane:
\begin{align}
\mathcal{W}_X:\quad \xymatrix{\mathcal{W}(0)_+\ar@<2pt>[r] & \ar@<2pt>[l]\mathcal{W}(1)_-^{\oplus 4}\ar@<2pt>[r] & \ar@<2pt>[l]\mathcal{W}(2)_+^{\oplus 6}\ar@<2pt>[r] & \ar@<2pt>[l]\mathcal{W}(3)_-^{\oplus 4}\ar@<2pt>[r] & \ar@<2pt>[l]\mathcal{W}(0)_+
}.
\end{align}
The brane factor is
\begin{align}
f_{\mathcal{W}_X}=2-4e^{2\pi\sigma}+6e^{4\pi\sigma}-4e^{6\pi\sigma},
\end{align}
which yields
\begin{align}
\label{quarticd4}
Z_{D^2}(\mathcal{W}_X)=\frac{1}{\sqrt{2}\pi^{\frac{3}{2}}}(2y_1^{\ast}-3y_2^{\ast}+2y_3^{\ast})=-\frac{\sqrt{2}}{\pi}\xi_3.
\end{align}

\bibliographystyle{fullsort}
\bibliography{bibliography}
\end{document}